\documentclass[manuscript, a4paper]{aastex}

\slugcomment{To appear in ApJ.}

\shorttitle{BAHAMAS: SNIa re-analysis }
\shortauthors{Shariff et al.}

\title{BAHAMAS: new SNIa analysis reveals inconsistencies with standard cosmology }
\date{\today}

\usepackage{graphicx,psfrag}
\usepackage{morefloats}
\usepackage{amsmath}
\usepackage{color}
\usepackage{enumerate}
\usepackage{multirow}
\usepackage{epsfig}
\usepackage{subfigure}
\usepackage{theorem}
\usepackage{tikz}
\usetikzlibrary{fit,positioning,calc}
\usepackage{mathrsfs}
\usepackage{scalefnt}

\newcommand{\D}{{\rm d}}
\newcommand{\N}{\mathcal{N}}

\newcommand{\be}{\begin{equation}}
\newcommand{\ee}{\end{equation}}

\newcommand{\salt}{\textsc{SALT2}}
\newcommand{\lcdm}{$\Lambda$CDM}

\newcommand{\indep}{\buildrel{\rm indep}\over\sim}
\newcommand{\dL}{d_{\rm L}}

\newcommand{\Cparams}{\mathscr{C}} 
\newcommand{\OmK}{\Omega_\kappa}
\newcommand{\OmM}{\Omega_{\rm m}}
\newcommand{\OmDE}{\Omega_{\rm DE}}
\newcommand{\OmL}{\Omega_{\Lambda}}
\newcommand{\saltdata}{\mathscr{\widehat D}}
\newcommand{\Mgaldata}{\mathscr{\widehat D}_\text{g}}
\newcommand{\zhat}{\hat{z}}
\newcommand{\mB}[1]{{m}^\star_{B#1}}
\newcommand{\xone}[1]{{x}_{1#1}}

\newcommand{\mBhat}[1]{\hat{m}^\star_{B#1}}
\newcommand{\mBbar}[1]{\bar{m}^\star_{B#1}}
\newcommand{\xhat}[1]{\hat{x}_{1#1}}
\newcommand{\chat}{\hat{c}}

\newcommand{\tc}{c_i}
\newcommand{\covhat}{\hat{C}}
\newcommand{\Mnot}{M_0}
\newcommand{\Mlow}{M_0^{\rm lo}}
\newcommand{\Mhigh}{M_0^{\rm hi}}

\newcommand{\sigmares}{\sigma_{\rm res}}
\newcommand{\sigmareslow}{\sigma_{\rm res}^{\rm lo}}
\newcommand{\sigmareshigh}{\sigma_{\rm res}^{\rm hi}}
\newcommand{\sigmaint}{\sigma_{\rm int}}
\newcommand{\xstar}{x_{1\star}}
\newcommand{\Rx}{R_{x_1}}
\newcommand{\Mgal}{{M_{\text{g} \, i}}}
\newcommand{\Mgalobs}{{\widehat M_{\text{g} \, i}}}
\newcommand{\sigMgal}{\sigma_{\text{g} \, i}}
\newcommand{\Mgalstar}{{M_{\text{g} \star}}}
\newcommand{\Rgal}{R_{\rm g}}
\newcommand{\cstar}{c_\star}
\newcommand{\Rc}{R_c}

\newcommand{\half}{\frac{1}{2}}
\newcommand{\E}{{\rm exp}}

\newcommand{\diag}{{\rm diag}}
\newcommand{\unif}{{\hbox{\sc Uniform}}}
\newcommand{\covariates}{X}
\newcommand{\regcoeff}{\mathscr{B}}
\newcommand{\nonlin}{\tau}
\newcommand{\betalin}{$z$-Linear color Correction}
\newcommand{\betastep}{$z$-Jump color Correction}
\newcommand{\HC}{Hard Classification}
\newcommand{\SC}{Soft Classification}
\newcommand{\CA}{Covariate Adjustment}
\newcommand{\NSN}{n}

\newcommand{\remu}{{\Delta\mu}_i}
\newcommand{\mremu}{\Delta\bar{\mu}}
\newcommand{\omb}{\hat{m}_{Bi}}
\newcommand{\ox}{\hat{x}_{i}}
\newcommand{\oc}{\hat{c}_{i}}
\newcommand{\sremu}{\sigma_{\Delta\mu}^2}
\newcommand{\ep}{^\epsilon}

\newcommand{\tk}{\mathcal{M}}

\newcommand{\logunif}{{\hbox{\sc LogUniform}}}
\newcommand{\normal}{\N}
\newcommand{\invgamma}{{\hbox{\sc InvGamma}}}
\newcommand{\betanot}{\beta_0}
\newcommand{\deltabeta}{\Delta \beta}
\newcommand{\ztrans}{z_t}

\newcommand{\chatn}[1]{\hat{c}_{#1}}
\newcommand{\cbar}[1]{\bar{c}_{#1}}
\newcommand{\hatz}[1]{\hat{z}_{#1}}
\newcommand{\remun}[1]{{\Delta\mu}_{#1}}
\newcommand{\cumuremun}[1]{s_{#1}}
\newcommand{\dobs}{{\hat D}(\Cparams)}
\newcommand{\dmis}{D}
\newcommand{\tdmis}{\tilde{\dmis}}
\newcommand{\dmean}{D_\star}
\newcommand{\dprior}{D_{\star\star}}

\newcommand{\sigsalt}{\Sigma_\saltdata}
\newcommand{\sigobs}{\Sigma_{\hat D}}
\newcommand{\sigmis}{\Sigma_D}
\newcommand{\sigprior}{\Sigma_{D_\star}}
\newcommand{\Jmat}{J}
\newcommand{\Amat}{A}
\newcommand{\sigA}{\Sigma_A}
\newcommand{\sigK}{\Sigma_K}
\newcommand{\sigdelta}{\Delta}
\newcommand{\kvec}{{k_\star}}
\newcommand{\gapara}{\lambda}
\newcommand{\dimmat}{{(3n \times 3n)}}
\newcommand{\txone}[1]{\tilde{x}_{1#1}}
\newcommand{\ttc}{\tilde{c}_i}
\newcommand{\ttM}{\tilde{M}}
\newcommand{\vone}{V_m}
\newcommand{\rvone}{V_{-m}}
\newcommand{\pvone}{V_{m,-m}}
\newcommand{\pxobs}{\hat{\xi}_m}
\newcommand{\rxobs}{\hat{\xi}_{-m}}
\newcommand{\pxmis}{\xi_m}
\newcommand{\tpxmis}{\tilde{\xi}_m}
\newcommand{\rxmis}{\xi_{-m}}
\newcommand{\ximnot}{\xi_{M_0}}
\newcommand{\delxi}{\Delta\xi}
\newcommand{\xmat}{E}
\newcommand{\txmat}{\tilde{E}}
\newcommand{\sigmaB}{\Sigma_B}
\newcommand{\zB}{\zeta_B}
\newcommand{\tsigmaB}{\tilde{\Sigma}_B}
\newcommand{\tzB}{\tilde{\zeta}_B}
\newcommand{\rth}{{\rm th}}
\newcommand{\yobs}{Y_{\rm obs}}
\newcommand{\ymis}{Y_{\rm mis}}
\newcommand{\symis}{Y_{{\rm mis},S}}
\newcommand{\aymis}{Y_{{\rm mis},A}}

\newcommand{\cutt}{(t)}
\newcommand{\nextt}{(t+1)}

\newcommand{\logr}{\mathrm{log}}

\newcommand{\BAHAMAS}{{\it BAHAMAS}}
\newcommand{\LCDM}{$\Lambda$CDM}
\newcommand{\wCDM}{$w$CDM}

\def\spacingset#1{\renewcommand{\baselinestretch}%
{#1}\small\normalsize}

\begin{document}

\author{Hikmatali Shariff}
\affil{Astrophysics Group, Physics Department, Imperial College London, Prince Consort Rd, London SW7 2AZ\altaffilmark{1}}

\author{Xiyun Jiao}
\affil{Statistics Section, Mathematics Department, 
Huxley Building, South Kensington Campus, Imperial College London,  London SW7 2AZ\altaffilmark{1}}
\author{Roberto Trotta}
\affil{Astrophysics Group, Physics Department, Imperial College London, Prince Consort Rd, London SW7 2AZ\altaffilmark{1}}
\author{David A. van Dyk}
\affil{Statistics Section, Mathematics Department,  Huxley Building, South Kensington Campus, Imperial College London,  London SW7 2AZ\altaffilmark{1}}

\altaffiltext{1}{Imperial Centre for Inference and Cosmology, Blackett Laboratory, Prince Consort Rd, London SW7 2AZ}

\begin{abstract}
We present results obtained by applying our BAyesian HierArchical Modeling for the Analysis of Supernova cosmology (\BAHAMAS) software package to the 740 spectroscopically confirmed supernovae type Ia (SNIa) from the ``Joint Light-curve Analysis'' (JLA) dataset. We simultaneously determine cosmological parameters and standardization parameters, including host galaxy mass corrections, residual scatter and object-by-object intrinsic magnitudes. 
Combining JLA and {\em Planck} Cosmic Microwave Background data, we find significant discrepancies in cosmological parameter constraints with respect to the standard analysis:  we find $\OmM = 0.399 \pm 0.027$, $2.8\sigma$ higher than previously reported and $w = -0.910 \pm 0.045$,  $1.6\sigma$ higher than the standard analysis. We determine the residual scatter to be $\sigmares = 0.104\pm 0.005$. 
We confirm (at the 95\% probability level) the existence of two sub-populations segregated by host galaxy mass, separated at $\log_{10}(M/M_\odot) = 10$, differing in mean intrinsic magnitude by $0.055 \pm 0.022$~mag, lower than previously reported. Cosmological parameter constraints are however unaffected by inclusion of host galaxy mass corrections.
We find $\sim 4\sigma$ evidence for a sharp drop in the value of the color correction parameter, $\beta(z)$,  at a redshift $\ztrans = 0.662 \pm 0.055$. We rule out some possible explanations for this behaviour, which remains unexplained. 
\end{abstract}


\section{Introduction}

Supernovae type Ia (SNIa) have been instrumental in establishing the accelerated expansion of the Universe, starting with the momentous discovery of the Supernova Cosmology Project and the High-Z Supernova Search Team in the late 90's~\citep{Riess:1998cb,Perlmutter:1998np}. 
The accelerated expansion is currently widely attributed to the existence of a ``dark energy'' component, which is compatible with Einstein's cosmological constant. Over the last decade, the SNIa sample has increased dramatically~\citep[e.g.,][]{Astier:2005qq, WoodVassey2007,Amanullah2010Spectra,Kowalski2008Improved, Kessler2009Firstyear,Freedman:2009vv,Contreras2009Carnegie,Balland2009ESOVLT,Bailey2008Initial,Hicken2009CfA3,2012ApJ...746...85S,Rest:2013mwz,Betoule:2014frx}, and it now comprises several hundred spectroscopically confirmed SNIa's. Since SNIa's probe the low-redshift Universe, they are ideal tools to measure the properties of dark energy. Two of the most important tasks required to shed light on the origin of dark energy are to establish whether or not the dark energy equation of state is evolving with time and  whether modified gravity scenarios might provide a viable alternative explanation.

SNIa's are observationally characterized by an absence of H in their spectrum, and by the presence of strong SiII lines. They occur when material from a companion
accreting onto a white dwarf  triggers carbon fusion, which proceeds until a core of typical mass $~0.7 \text{M}_\odot$ of $^{56}$Ni is created. The radioactive decay of $^{56}$Ni to $^{56}$Co and, subsequently to $^{56}$Fe, produces $\gamma$-rays that heat up the ejecta, thus powering the light curve (LC). While it is believed that this happens when the mass of the white dwarf approaches (without reaching) the Chandrasekhar limit of $1.4 \text{M}_\odot$, the debate about progenitor scenarios is not settled. There is strong evidence that some systems are likely single degenerate~\citep{2011Natur.480..344N} (where a white dwarf accretes mass from a large, perhaps a main sequence, companion star~\citep{2011Natur.480..348L}), but studies of SNIa rates point to the existence of two classes of progenitors~\citep{2006MNRAS.370..773M}. Furthermore, single-degenerate models have been ruled out for the supernova remnant SNR 0509-67 by the lack of an ex-companion star~\citep{2012arXiv1201.2195S}, and pre-explosion X-ray and optical data for SN2007on are compatible with a single-degenerate model~\citep{2008Natur.451..802V}. Multiple progenitor channels would help explain the observed variability within the type Ia category~\citep{2011MNRAS.412.1441L}. 

Within the more restricted sub-class of so-called ``normal'' SNIa's, the fundamental assumption underlying their use to measure expansion history is that they can be standardized so that their intrinsic magnitudes (after empirical corrections) are sufficiently homogeneous. This makes them into ``standard candles'', i.e. object of almost uniform intrinsic luminosity (within $\sim 0.1$ mag) that can be used to determine the distance-redshift relation. This relies on the empirical observation that intrinsic magnitudes are correlated with decay times of light-curves: intrinsically brighter SNIa's are slower to fade~\citep{Phillips:1993ng,Phillips:1999vh}. It also appears that fainter SNIa's are redder in color~\citep{Riess:1996pa}. Therefore, multi-wavelength observations of light curves can be used to exploit this correlations and reduce the residual scatter in the intrinsic magnitude to typically $\sim 0.10-0.15$ mag. Near infra-red light curve data can significantly reduce residual scatter still further~\citep{2011ApJ...731..120M}, as does selecting SNIa in young star-forming environments~\citep{Kelly:2014hla}.

One of the most widely-used frameworks for determining an estimate of the distance modulus from LC data is the \salt{} method~\citep{Guy:2005me, Guy:2007dv}, 
which derives color and stretch corrections for the magnitude from the LC fit, and then uses the corrected distance modulus to fit the underlying cosmological parameters. By contrast, the Multi-color Lightcurve Shape~\citep{Riess:1996pa,JhaRiess2007} 
approach simultaneously infers the Phillips corrections and the cosmological parameters of interest, 
while explicitly modeling the dust absorption and reddening in the host galaxy. Recently, a fully Bayesian, hierarchical model approach to LC fitting has emerged~\citep{2009ApJ...704..629M,2011ApJ...731..120M}, but this so-called BAYESN algorithm has not yet been applied for cosmological parameter inference. 
 
As the SNIa sample size grows, so does the importance of systematic errors relative to statistical errors, to the point that current measurements of the cosmological parameters (including dark energy properties) are limited by systematics~\citep{Betoule:2014frx}. A better understanding of how SNIa properties correlate with their environment (such as host galaxy properties) will help in improving their usage as standard candles. 

In this paper, we introduce \BAHAMAS\ (BAyesian HierArchical Modeling for the Analysis of Supernova cosmology), an extention of the method first introduced by~\cite{2011MNRAS.418.2308M}, and apply it to the SNIa sample from the ``joint light-curve analysis'' (JLA) \citep{Betoule:2014frx}. \cite{Betoule:2014frx} re-analysed 740 spectroscopically confirmed SNIa's obtained by the SDSS-II and SNLS collaboration. \cite{2011MNRAS.418.2308M} demonstrated with simulated data that a Bayesian hierarchical model approach of the kind developed here has a reduced posterior uncertainties, smaller mean squared error and better coverage properties than the standard approach (see also~\cite{2014MNRAS.437.3298M,Karpenka:2015vva} for further detailed comparisons). More recently, \cite{Rubin:2015rza} applied a similar method to analyse Union2.1 SNIa data, extending it to deal with selection effect and non-Gaussian distribution. \cite{Nielsen:2015pga} adopted the effective likelihood introduced in \cite{2011MNRAS.418.2308M} but interpreted the results in terms of profile likelihood (rather than posterior distributions), showing that the profile likelihood in the $\Omega_\Lambda, \Omega_m$ plane obtained from JLA data is much wider than what is recovered with the usual $\chi^2$ approach.

This paper re-evaluates the JLA data in the light of the principled statistical analysis made possible by \BAHAMAS. As demonstrated in~\cite{2011MNRAS.418.2308M}, the standard $\chi^2$ fitting is an approximation to the Bayesian result in a particular regime, which is usually violated by \salt{}\ outputs. Therefore we address the question of whether the cosmological constraints obtained from the standard analysis remain unchanged when using a principled likelihood function within a fully Bayesian analysis, as in \BAHAMAS.  We use our framework to test for evolution with redshift in the SNIa properties, and in particular in their color correction. Finally, we investigate whether the residual scatter around the Hubble law can be further reduced by exploiting correlations between SNIa intrinsic magnitudes and their host galaxy mass.

This paper is organized as follows: in Section~\ref{sec:model} we introduce our notation, the parameters of interest, and our Bayesian hierarchical model. In Section~\ref{sec:results} we present results obtained when our approach is applied to the JLA sample; conclusions appear in Section~\ref{sec:conclusions}. In Appendix~\ref{apsec:algo} we review our statistical algorithms; in Appendix~\ref{apsec:pos} we present the full posterior distributions, and in Appendix~\ref{apsec:sampler} give details of the Gibbs-type samplers that we use to fit our Bayesian models.   


\section{BAHAMAS: Bayesian Hierarchical Modeling for the Analysis of Supernova Cosmology}
\label{sec:model}

In this section, we review \BAHAMAS, an extension of the method introduced by \cite{2011MNRAS.418.2308M} for estimating cosmological parameters using SNIa peak magnitudes adjusted for the stretch and color of their LCs via \salt. We then discuss features of  the model and methods that allow us to adjust for systematic errors, host galaxy mass, and a possible dependence of the color correction on redshift.  We also provide a new estimate of the residual scatter in SNIa absolute magnitudes. An outline of our statistical models and methods is presented here. Details of the statistical posterior distributions and the computational techniques we use to explore them appear in Appendix~\ref{apsec:pos}. 

\subsection{Distance Modulus in an FRW Cosmology} 
\label{subsec:dismodu}

Our overall modeling strategy leverages the homogeneity of SNIa absolute magnitudes to allow us to estimate their distance modulus from their apparent magnitudes and thereby estimate the underlying cosmological parameters that govern the relationship between distance modulus and redshift, $z$.  Consider, for example, a sample of $n$ SNIa's with apparent B-band peak magnitudes, $m^\star_i$. The distance modulus in any pass band, $\mu(z; \Cparams)$, is the difference between the apparent and the intrinsic magnitudes in that band. Ignoring measurement error for the moment, we can express this relationship statistically via the regression model
\be
m^\star_i = \mu(z_i; \Cparams) + M_i, \ \hbox{ for } \ i=1,\ldots, n
\label{eq:basicmodel}
\ee
where $M_i \sim \N(M_0, \sigmaint^2)$ is the absolute magnitude of SNIa $i$ with $M_0$ and $\sigmaint$ the mean and intrinsic standard deviation of SNIa absolute magnitudes in the underlying population\footnote{We use $\N(\mu, \Sigma)$ to denote a (multivariate) Gaussian distribution of mean $\mu$ and variance-covariance matrix $\Sigma$. For the 1-dimensional case, $\Sigma$ reduces to the variance, $\sigma^2$.}.  Clearly the smaller $\sigmaint$ the better we can estimate $\mu(z; \Cparams)$. In Section~\ref{subsec:model} we discuss the inclusion of correlates in Eq.~\eqref{eq:basicmodel} that aim to reduce its residual variance, i.e. to make the SNIa better standard candles.

The distance modulus is given by 
\be
\mu(z; \Cparams)  =25+{5}\log  {\dL(z; \Cparams) \over \text{Mpc}}  , 
\ee
where  $\Cparams$ represents a set of underlying cosmological parameters and $\dL(z; \Cparams)$ is the luminosity distance to redshift $z$. In the case of the \lcdm\ cosmological model (based on a Friedman-Robertson-Walker (FRW) metric), the luminosity distance is 
\be
\dL(z; \Cparams)=
\frac{c}{H_0}\frac{(1+z)}{\sqrt{|\OmK|}}{\rm sinn}_{\OmK}
\left\{
\sqrt{|\OmK|}\int_{0}^{z}{\D}z^{\prime}{
\left[{(1+z^\prime)}^3\OmM+\OmDE(z^\prime)+{(1+z^\prime)}^2\OmK
\right]}^{-1/2}
\right\},
\ee 
where 
\be
{\rm sinn}_{\OmK}(x)=
\begin{cases}
x, &\mbox{if } \Omega_{k}=0 \\
{\rm sin}(x), & \mbox{if } \Omega_{k}<0 \\
{\rm sinh}(x), & \mbox{if } \Omega_{k}>0
\end{cases}
\ee
and $\Cparams= \{ \OmK, \OmM, H_0, w \}^T$,  with
$\OmK$  the curvature parameter and $\OmM$ the total (both baryonic and dark) matter density (in units of the critical density); $c$ is the speed of light, and $H_0 = 100h$~km/s/Mpc is the Hubble parameter today depending on the  dimensionless quantity $h$. For a general dark energy equation of state as a function of redshift, $w(z)$, we can express 
\be
\OmDE(z)=\OmL \exp \left[3\int_0^z\frac{1+w(x)}{1+x}\D x\right],
\ee
where $\OmL$ is the dark energy density parameter. In our analyses, we either assume a flat Universe (i.e., $\OmK = 0$) with $w(z)$ equal to a constant other than  $-1$ or a curved Universe with a cosmological constant (i.e., $w(z)=-1$).   In either case, $w(z) = w$ becomes a time-independent constant, and thus
\be
\OmL= 1-\OmK-\OmM.
\ee

\subsection{\salt{} Output and Standardization of SNIa}
\label{subsec:model}

\subsubsection{Baseline Model}
\label{sec:baseline}

As described in~\cite{Guy:2007dv}, the \salt{} fit of the multi-color LC observation of SNIa $i$ produces measured quantities: $\zhat_i$ is the measured heliocentric redshift,  $\mBhat{i}$ the measured B-band apparent magnitude, $\xhat{i}$ the measured stretch correction parameter, $\chat_i$ the measured color correction parameter, and $\covhat_i$ a $(3\times3)$ variance-covariance matrix describing the measurement error of $\mBhat{i}$, $\xhat{i}$, and $\chat_i$.  As shown in \cite{2011MNRAS.418.2308M}, accounting for observational error of spectroscopically determined redshifts does not lead to any appreciable difference in the results. Thus, after correcting for the translation from heliocentric redshift to the frame of reference of the Cosmic Microwave Background, we ignore measurement error in the observed redshift and set $\zhat_i = z_i$ throughout. Each $\covhat_i$ is treated as a known constant,  and we denote the \salt\ data by
\be
\saltdata_i = \{\mBhat{i}, \xhat{i}, \chat_i\}^T,  \ \mbox{ for } \ i=1,\ldots, n.
\ee

\begin{figure}[tbh]
  \begin{tikzpicture}
\tikzstyle{epic}=[circle, minimum size = 10mm, ultra thick, draw =black!80, node distance = 10mm]
\tikzstyle{circ}=[circle, minimum size = 11mm, ultra thick, draw =black!80, node distance = 0mm]
\tikzstyle{func}=[circle, double, double distance=1pt, minimum size = 10mm, thick, draw =black!80, node distance = 15mm]
\tikzstyle{fun}=[circle, double, double distance=1pt, minimum size = 8mm, thick, draw =black!80, node distance = 10mm]
\tikzstyle{main}=[circle, minimum size = 11mm, ultra thick, draw =black!80, node distance = 12mm]
\tikzstyle{side}=[rectangle, minimum size = 8mm, ultra thick, draw =black!80, node distance = 14mm]
\tikzstyle{sid}=[rectangle, minimum size = 8mm, ultra thick, draw =black!80, node distance = 6mm]
\tikzstyle{connect}=[-latex, thick]
\tikzstyle{box}=[rectangle, draw=black!100,rounded corners]
  \node[side, fill = gray!30] (hmi){ $\mBhat{i}$};
  \node[side, fill = gray!30] (hxi) [right=of hmi] { $\xhat{i}$ };
  \node[side, fill = gray!30] (hci) [right=of hxi] {$\chat_i$};
  \node[side, fill = gray!30] (hgi) [right=of hci] {\footnotesize{$\Mgalobs$}};
  \node[sid, fill = gray!30] (sig) [below=of hxi] {$\covhat_i$};
  \node[sid, fill = gray!30] (siggi) [below=of hgi] {$\sigMgal^2$};
  \node(num)[left=of sig] {$i=1,\dots,n$};
  \node[func] (mi) [above=of hmi] {$\mB{i}$ };
  \node[main] (Mi) [above=of mi] {$M_i\ep$ };
  \node[func] (mui) [left=of Mi] {$\mu_i$ };
  \node[side, fill = gray!30] (hzi) [below=of mui] { $\zhat_i$ };
  \node[main] (xi) [right=of Mi] {$\xone{i}$ };
  \node[main] (ci) [right=of xi] {$c_i$ };
  \node[main] (gi) [right=of ci] {\footnotesize{$\Mgal$}};
  \node[main] (alpha) [above=of mui] {$\regcoeff$};
  \node[epic] (m0) [right=of alpha] {$\Mnot\ep$};
  \node[circ] (sigmaint) [right=of m0] {\footnotesize{$\sigmares^2$}};
  \node[main] (rx) [right=of sigmaint] {\footnotesize{$\Rx^2$}};
  \node[circ] (xstar) [left=of rx] {$\xstar$};
  \node[main] (rc) [right=of rx] {$\Rc^2$};
  \node[circ] (cstar) [left=of rc] {$\cstar$};
  \node[main] (rg) [right=of rc] {$\Rgal^2$};
  \node[circ] (gstar) [left=of rg] {\footnotesize{$\Mgalstar$}};
  \node[main] (wk) [left=of alpha] {$\Cparams$};
  \node[sid, fill = gray!30] (a) [right=of gi] {};
  \node[node distance=1mm] (b) [right=of a] {\footnotesize{known/observed}};
  \node[node distance=0mm] (b1) [below=of b] {\footnotesize{quantities}};
  \node[epic] (c) [below=of a] {};
  \node[node distance=1mm] (d) [right=of c] {\footnotesize{latent variable}};
  \node[fun] (e) [below=of c] {};
  \node[node distance=1mm] (f) [right=of e] {\footnotesize{function of data}};
  \node[node distance=0mm] (g) [below=of f] {\footnotesize{and parameters}};
  \node[box,thick, inner sep=0.13in, fit = (hzi)(mui)(hmi) (hxi) (hci) (hgi) (mi) (Mi) (xi) (ci) (gi) (num) (sig) (siggi)](obs){};
  \node[side, fill = gray!30] (sys) [below=of obs] {$C_{\rm syst}$};
\path (mi) edge [connect] (hmi)
         (xi) edge [connect] (hxi)
         (ci) edge [connect] (hci)
         (gi) edge [connect] (hgi)
         (sig) edge [connect] (hmi)
         (sig) edge [connect] (hxi)
         (sig) edge [connect] (hci)
         (siggi) edge [connect] (hgi)
         (Mi) edge [connect] (mi)
         (mui) edge [connect] (mi)
         (hzi) edge [connect] (mui)
         (xi) edge [connect] (mi)
         (ci) edge [connect] (mi)
         (gi) edge [connect] (mi)
         (alpha) edge [connect] (mi)
         (wk) edge [connect] (mui)
         (sigmaint) edge [connect] (Mi)
         (rx) edge [connect] (xi)
         (rc) edge [connect] (ci)
         (rg) edge [connect] (gi)
        (m0) edge [connect] (Mi)
        (xstar) edge [connect] (xi)
        (cstar) edge [connect] (ci)
        (gstar) edge [connect] (gi)
        (sys) edge [connect] (obs);
\end{tikzpicture}
   \caption{Graphical representation of \BAHAMAS. The meaning of the symbols is given in Table~\ref{table:main_params}.}
\label{fig:hierarchicalmodel}
\end{figure}
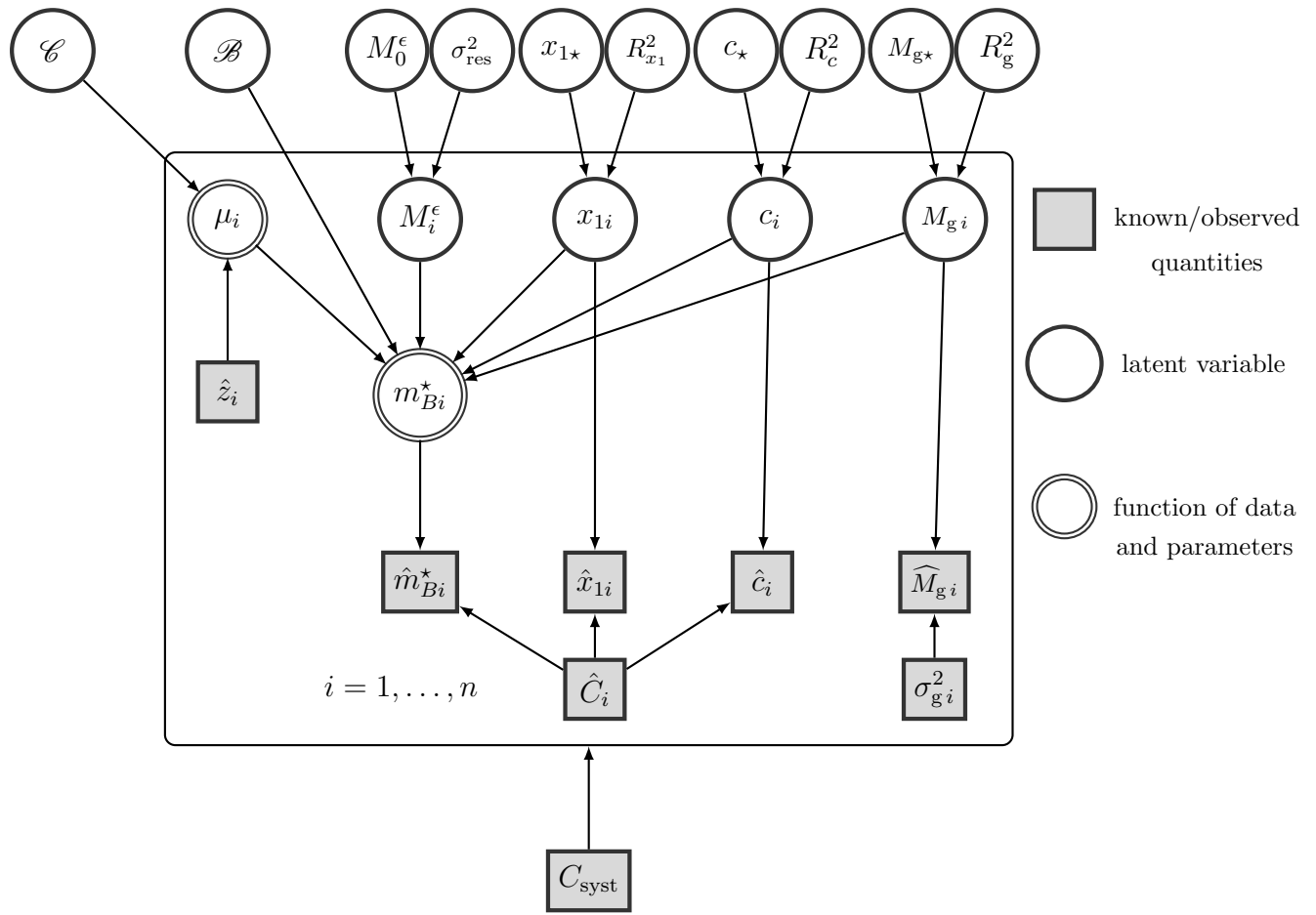

Here we review our Baseline Model that was first introduced by \cite{2011MNRAS.418.2308M}; extensions appear in Sections~\ref{sec:system} and \ref{subsec:extend}. We model $\saltdata = \{\saltdata_1^T, \ldots, \saltdata_n^T\}^T$ via a Bayesian hierarchical model~\citep{Kelly:2007jy}, see Fig.~\ref{fig:hierarchicalmodel}. At the observation level, we model the measured \salt{} fits as independent Gaussian variables centered at their true values,
\be
\left(\begin{array}{c}
\mBhat{i}\\ 
\xhat{i}\\
\chat_i 
\end{array}\right)
\indep
\N\left[\left(\begin{array}{c}\mB{i}\\ \xone{i}\\ c_{i}\end{array}\right),{\covhat}_{i}\right], \ \hbox{ for } \ i=1,\ldots, n.
\label{eq:obslevel}
\ee
The true (but unobserved) values, $\mB{i}$, $\xone{i}$, and $c_i$, are treated as latent variables, with $\xone{i}$ and $c_i$ used to predict the intrinsic (absolute) magnitude $M_i$ via the linear regression
\be 
M_i =  -\alpha \xone{i}+\beta \tc + M_i\ep,
\label{eq:linear-reg}
\ee
where $M_{i}\ep 
\sim {\N}( \Mnot\ep, \sigmares^{2})$. Here $\xone{i}$ and $\tc$  represent the Phillips stretch and color corrections, respectively, whose predictive strength is controlled by the unknown parameters, $\alpha$ and $\beta$, which must be inferred from $\saltdata$. Whereas $M_i$ appearing in Eq.~\eqref{eq:basicmodel} is the physical absolute magnitude of SNIa $i$ and  $M_i\ep$ is the empirically corrected absolute magnitude, after application of the Phillips relations.  Substituting Eq.~\eqref{eq:linear-reg} into Eq.~\eqref{eq:basicmodel} yields
\be 
\mB{i} =\mu_{i}(\zhat_i, \Cparams)- \alpha \xone{i}+\beta \tc + M_i\ep.
\label{eq:covariates_relation}
\ee 
From a statistical point of view Eq.~\eqref{eq:linear-reg} is a linear regression model with residuals $M_i\ep$. In principle, including the stretch and color corrections in Eqs.~\eqref{eq:linear-reg} and \eqref{eq:covariates_relation} should reduce the residual variance, i.e., $\sigmares^2 \leq \sigmaint^2$, and improve the precision of the estimates of $\Cparams$.\footnote{This intuition stems from standard linear regression where the dependent variables (here the $\mB{i}$) and independent variables (here the $\xone{i}$ and $\tc$) are observed directly. The situation is more complicated when these variables are observed with error.}  In Section~\ref{subsec:extend} we investigate whether introducing either host galaxy mass or an interaction\footnote{In statistical terms, an {\it interaction} between two variables means the effect of one variable depends on the values of the second. In Section~\ref{subsec:extend} we allow the effect of the color correction to vary with redshift.} between redshift and the color correction as additional correlated variables in Eq.~\eqref{eq:linear-reg} can further reduce the residual variance and increase the precision of the estimate of $\Cparams$. 

The population distributions of the latent variables $M_i\ep$, $\xone{i}$ and $\tc$ are modelled as Gaussian\footnote{We assume a single underlying population, but it would be simple to extend our model to multiple populations by drawing $M_{i}\ep$ from a mixture of Gaussians, for example to account for different progenitor scenarios, or contamination from non-Ia's.}, with unknown hyperparameters controlling the mean and variance of each population: 
\begin{align}
M_{i}\ep | \Mnot\ep, \sigmares & \sim {\N}( \Mnot\ep, \sigmares^{2}), 
\label{eq:Mdist} \\ 
\xone{i} | \xstar, \Rx& \sim {\N}(\xstar, \Rx^2), 
\label{eq:xdist}
\\ 
\tc| \cstar, \Rc  & \sim {\N}(\cstar, \Rc^2)
\label{eq:cdist}
\end{align} 
The distribution in Eq.~\eqref{eq:Mdist} is the model for the residuals in Eq.~\eqref{eq:linear-reg}.

The prior distributions used for the model parameters are given in Table~\ref{table:main_params} (along with those for parameters introduced in extensions to the model in Section~\ref{subsec:extend}).  We adopt non-informative proper prior distributions for $\alpha$, $\beta$, and the parameters in $\Cparams$. The value of the Hubble parameter is fixed to $H_{0}/{\rm km/s/Mpc} = 67.3$ from Planck\footnote{The Hubble parameter is perfectly degenerate with the mean absolute magnitude $\Mnot$, hence SNIa data only constraints the degenerate combination $\Mnot - 5\log h$. Therefore changing the value of $h$ amounts to a shift in the mean absolute magnitude.}.
Among the population level parameters, the choice of prior distribution for $\sigmares^2$ is the most subtle. The simple choice of a log-uniform prior, as adopted in~\cite{2011MNRAS.418.2308M}, requires specification of arbitrary bounds to make it proper. Because this might lead to difficulties in interpreting the posterior distribution, we instead adopt a proper inverse Gamma\footnote{We parameterize the inverse Gamma distribution so that $X\sim\invgamma(u,v)$ means that ${2v/X}$ follows a $\chi^2$ distribution with $2u$ degrees of freedom.} prior distribution, $\sigmares^2 \sim  \invgamma(0.003,0.003)$. We perform a sensitivity analysis for the choice of scale for this  distribution and demonstrate that our results (including the posterior distribution of $\sigmares$) are robust to this choice, see Fig.~\ref{fig:res_prior_post}.

\subsubsection{Systematics Covariance Matrix and Selection Effects}
\label{sec:system}

In the Baseline Model described in Section~\ref{sec:baseline}, we assume that the \salt\ measurements for each SNIa are conditionally independent (given their means and variances, see Eq.~\eqref{eq:obslevel}), i.e., the $(3n \times 3n)$ variance-covariance matrix $C_{\rm stat} \equiv \text{diag}(\covhat_1, \dots, \covhat_n)$ is block diagonal. 
\cite{Betoule:2014frx} derive a systematic variance-covariance matrix, $C_{\rm syst}$,  
with correlations among the SNIa's. The systematic covariance matrix includes contributions from calibration, model uncertainty, bias correction, host, dust, peculiar velocities and contamination. 
We account for these systematics by replacing the matrix $C_{\rm stat}$ with $\sigsalt = C_{\rm stat} +C_{\rm syst}$ in the full posterior distribution; see Appendix~\ref{apsec:pos}.

\cite{Betoule:2014frx} use SNANA simulations to model observational selection effects and correct for them by shifting the value of $\mB{i}$ accordingly. We adopt the bias-corrected values of $\mB{i}$ and thus do not need to separately account for selection effects. A fully Bayesian approach to forward-modelling of such effects appears in~\cite{Rubin:2015rza}.

\begin{table}[htb]\footnotesize
\begin{center}
\begin{tabular}{ll} 
\hline\hline 
Parameter  & Notation and Prior Distribution\\ 
\hline 
\multicolumn{2}{c}{Cosmological parameters} \\\hline
Matter density parameter & $\OmM \sim \unif(0,2)$ \\ 
Cosmological constant density parameter  & $\OmL \sim \unif(0,2)$ \\ 
Dark energy EOS  & $w \sim \unif(-2,0)$ \\ 
Hubble parameter & $H_{0}/{\rm km/s/Mpc} = 67.3$ \\ \hline 
\multicolumn{2}{c}{Covariates} \\\hline
Coefficient of stretch covariate  & $\alpha \sim \unif(0,1)$ \\ 
Coefficient of color covariate & $\beta\ (\text{or }\beta_0) \sim \unif(0,4)$ \\ 
Coefficient of interaction of color correction and $z$ & $\beta_1 \sim \unif(-4, 4)$ \\  
Jump in coefficient of color covariate  & $\deltabeta \sim \unif(-1.5,1.5)$\\ 
Redshift of jump in color covariate  & $\ztrans \sim \unif(0.2,1)$ \\ 
Coefficient of host galaxy mass covariate   & $\gamma \sim \unif(-4,4)$ \\ \hline
\multicolumn{2}{c}{Population-level distributions} \\\hline
Mean of absolute magnitude  & $\Mnot\ep \sim \normal(-19.3,2^2)$ \\ 
Residual scatter after corrections  & $\sigmares^2 \sim \invgamma(0.003,0.003)$ \\ 
Mean of absolute magnitude, low galaxy mass  & $\Mlow \sim \normal(-19.3,2^2)$ \\ 
SD of absolute magnitude, low galaxy mass  & ${\sigmareslow}^2 \sim \invgamma(0.003,0.003)$ \\ 
Mean  of absolute magnitude, high galaxy mass  & $\Mhigh \sim \normal(-19.3,2^2)$ \\ 
SD of absolute magnitude, high galaxy mass  & ${\sigmareshigh}^2 \sim \invgamma(0.003,0.003)$ \\ 
Mean of stretch & $\xstar \sim \normal(0,10^2)$\\ 
SD of stretch & $\Rx \sim \logunif(-5,2)$ \\ 
Mean of color  & $\cstar \sim \normal(0,1^2)$  \\ 
SD of color  & $\Rc \sim \logunif(-5,2)$ \\ 
Mean of host galaxy mass & $\Mgalstar \sim \normal(10,100^2)$ \\ 
SD of host galaxy mass & $\Rgal \sim \logunif(-5,2)$ \\
\hline \hline
\end{tabular}
\caption{Summary of the parameters, notation, and prior distributions used in our hierarchical model.  These include parameters in the Baseline Model described in Sections~\ref{subsec:dismodu}-\ref{subsec:model} and its extensions described in Section~\ref{subsec:extend}. ``SD'' stands for ``standard deviation''. \label{table:main_params}}
\end{center}
\end{table}%

\subsection{Generalizing the Phillips Corrections}
\label{subsec:extend}

The advantage of the Phillips corrections is that they are expected to reduce the residual variance in Eq.~\eqref{eq:covariates_relation} and thus increase the precision in the estimates of $\Cparams$. Introducing additional correlates may further improve precision. In the context of \BAHAMAS, it is straightforward to generalize the Phillips corrections to include additional 
covariates. To formalize this, we replace
$\xone{i}$ and $\tc$ in  Eq.~\eqref{eq:linear-reg} with a set of $p$ covariates and substitute into Eq.~\eqref{eq:basicmodel} to obtain
\be \label{eq:covariates_relation_generalized}
\mB{i} =\mu_{i}(\zhat_i, \Cparams) + \covariates_i^T \regcoeff  +M_i\ep
\ee 
where $\covariates_i$ is a $(p\times 1)$ vector of covariates and $\regcoeff$ is a $(p \times 1)$ vector of  regression coefficients. The usual case, given in Eq.~\eqref{eq:covariates_relation}, is a special case of Eq.~\eqref{eq:covariates_relation_generalized} in which only the stretch and color covariates are included ($p=2$) and  can be recovered by setting $\covariates_i = \{\xone{i}, \tc \}^T$ and $\regcoeff = \{-\alpha, \beta\}^T$.
If the covariate vector depends non-linearly on a set of parameters $\nonlin$, Eq.~\eqref{eq:covariates_relation_generalized} can be further generalized to
\begin{equation}
\label{eq:covariates_relation_nonlinear}
\mB{i} =\mu_{i}(\zhat_i, \Cparams)+ \covariates_i(\nonlin)^T \regcoeff +M_i \ep.
\ee
Eq. (\ref{eq:covariates_relation_nonlinear}) allows for both linear and non-linear covariate adjustment.

We consider various instances of Eq.~\eqref{eq:covariates_relation_nonlinear}. First, we investigate the effect of the environment by including the host galaxy mass as a covariate in the correction. The host mass is  a (relatively easy to measure) proxy for more fundamental changes in the environment, such as evolution of metallicity. Second, we are interested in testing for possible redshift-dependence of the color correction. This could have a physical origin (e.g., dust environments in high-redshift galaxy being different) or be a reflection of systematic differences between low- and high-redshift survey. 

Future work will aim at investigating the dependency on environmental properties, such as star formation rates and metallicities, a topic of active investigation~\citep{Childress:2013wna,Rigault:2013gux,Rigault:2014kaa,Kelly:2014hla,Jones:2015uaa}.
 
 \subsubsection{Dependency on Host Galaxy Mass}

There is strong evidence that the absolute magnitude (after corrections) of SNIa correlates with host galaxy mass (e.g,~\cite{Sullivan:2006ah, 2012ApJ...750....1M}). Current results indicate that more massive galaxies ($\log_{10} (M/M_\odot) > 10$) host brighter SNIa's, with their average absolute magnitude being of order $\sim 0.1$~mag smaller than in less massive hosts~\citep{Kelly:2009iy,2010MNRAS.406..782S,Campbell2015}. This could be a reflection of dust, age and/or metallicity in the progenitor systems~\citep{Childress:2013xna}.

We investigate three formulations that incorporate host galaxy mass as a covariate in Eq. (\ref{eq:covariates_relation_nonlinear}) and study how they affect inference for $\Cparams$. In particular, we consider models that (i) divide the SNIa's into two populations using a hard host galactic mass threshold (``\HC\ Model''), (ii) divide the SNIa's into two populations using soft probabilistic classification (``\SC\ Model''), and  (iii) adjust for host galaxy mass as a covariate in the regression, analogously to the stretch and color corrections (``\CA\ Model''). Specifically, we model the observed host mass galaxies (on the  $\log_{10}$ scale) as
\be
\Mgalobs
\indep 
\N \left(\Mgal, \sigMgal^2\right), 
 \ \hbox{ for } \ i=1,\ldots, n,
\label{eq:hostmasslike}
\ee
where $\Mgal$ is the (true) host galaxy mass of SNIa $i$  (in $\log_{10}$ solar masses) and
$\sigMgal$ is the (known) standard deviation of the measurement error.

In the ``\HC\ Model'', we divide the SNIa's into two classes using the observed mass: high host galaxy mass if $\Mgalobs \geq 10$ and low host galaxy mass if $\Mgalobs < 10$. (In this way, we ignore measurement errors in $\Mgalobs$.) The two classes are allowed to have their own population-level values for the mean absolute SNIa magnitude and residual standard deviation, i.e., $(\Mhigh, \sigmareshigh)$ for high mass hosts and $(\Mlow, \sigmareslow)$ for low mass hosts. Common values are used for $\alpha$ and  $\beta$ (and of course for $\Cparams$)  for both classes. We do not assume a redshift dependency for the color correction. 
{We fix the host galaxy mass classification at $10^{10}$ solar masses, analogous to location of the step function used for the host galaxy mass by ~\cite{Betoule:2014frx} to enable a direct comparison with their results.}

The ``\SC\ Model'' is identical to the \HC\ Model except that measurement errors in the observed masses are accounted for by probabilistically classifying each SNIa; these errors can be quite significant.  Specifically, we let $Z_i$ be an indicator variable that equals one for high host galaxy masses and equals zero for low host galaxy masses, that is,
\be
Z_i = 
\begin{cases}
0, &\mbox{if } \Mgal< 10 \\
1, & \mbox{if } \Mgal \geq 10.
\end{cases}
\ee
We treat $\{Z_1, \ldots, Z_\NSN\}^T$ as a vector of unknown latent variables that are estimated along with the other model parameters and latent variables via Bayesian model fitting.  This requires specification of a prior distribution on each $\Mgal$. We choose a flat prior so that $\Mgal | \Mgalobs \indep \N(\Mgalobs, \sigMgal^2)$; details appear in Appendix~\ref{apsec:pos}.
	
The ``\CA\ Model'' introduces $\Mgal$ as a covariate in the regression in Eq.~\eqref{eq:covariates_relation_generalized} rather than by classifying the SNIa's on galactic mass. In particular, we use Eq.~\eqref{eq:covariates_relation_generalized}, but with $p=3$, $\covariates_i= \{\xone{i}, \tc,  \Mgal\}^T$, and $\regcoeff = \{-\alpha, \beta, \gamma \}^T$ with $\regcoeff $ being estimated from the data.  The population model for the latent variables $M_i\ep$, $\xone{i}$ and $\tc$ given in Eq. (\ref{eq:Mdist})--(\ref{eq:cdist}) is also expanded to include host galaxy mass:
 \begin{equation}	
 \Mgal | \Mgalstar, \Rgal  \sim {\N}(\Mgalstar, \Rgal^2),
 \label{eq:gdist}
 \end{equation}
where $\Mgalstar$ and $\Rgal$ are hyperparameters analogous e.g., to  $\xstar$ and $\Rx$; their prior distributions are given in Table~\ref{table:main_params}.

%
%

\subsubsection{Redshift Evolution of the Color Correction} 

The \salt{} color correction gives the offset with respect to the average color at maximum B-band luminosity, $c_i = (B-V)_i - \langle B - V \rangle$. This time-independent color variation encompasses both intrinsic color differences and those due to host galaxy dust. It is possible that the color correction varies with redshift, as a consequence of evolution of the progenitor and/or changes in the environment, for example, variation in the dust composition with galactic evolution~\citep{Childress:2013xna}. Redshift-dependent dust extinction can lead to biased estimates of cosmological parameters \citep{Menard:2009wy,Menard:2009yb}.
This is not captured by the \salt{} fits, since they use a training sample that is distributed over a large redshift range ($0.002 \leq z \lesssim 1$) \citep{Guy:2007dv} and thus the training sample color correction is averaged across redshift. It is therefore important to check for a redshift dependence in the color correction by allowing $\beta$, which controls the amplitude of the linear correction to the distance modulus, to vary with $z$. 

We consider two phenomenological models that allow the color correction to depend on $z$. In the first, the dependence is linear: we replace the constant $\beta$ in Eq.~\eqref{eq:covariates_relation} with the $z$-dependent $\beta_0+\beta_1 \zhat_i$. This is expressed by setting  $\covariates_i = \{\xone{i}, \tc , \tc \zhat_i\}^T$ and $\regcoeff = \{-\alpha, \beta_0, \beta_1 \}^T$ in Eq.~\eqref{eq:covariates_relation_generalized}, leading to
\be
M_i =  -\alpha \xone{i}+\beta_0 \tc + \beta_1 \tc \zhat_i + M_i\ep.
\label{eq:linear-color}
\ee
We refer to this as the ``\betalin\ Model''. 

The second model allows for a sharp transition from a high-redshift to a low-redshift regime:  we replace the constant $\beta$ in Eq.~\eqref{eq:covariates_relation} with 
$\beta_0 + \Delta\beta \left({1\over 2} + \frac{1}{\pi} \arctan\left(\frac{\zhat_i-\ztrans}{0.01}\right)\right)$, where $\beta_0, \, \Delta\beta$, and $z_t$ are parameters. This can be viewed as a smoothed step function in that it approaches $\beta_0$ as $z\rightarrow 0$ and approaches $\beta_0+\Delta\beta$ as $z\rightarrow \infty$, with a smooth monotone local transition centered at $z=\ztrans$. Substituting into Eq.~\eqref{eq:covariates_relation}, we have
\be
M_i =  -\alpha \xone{i}+\beta_0\tc + \Delta\beta \left({1\over 2} + \frac{1}{\pi} \arctan\left(\frac{\zhat_i-\ztrans}{0.01}\right)\right)\tc+ M_i\ep,
\label{eq:step-color}
\ee
where the covariate associated with $\Delta\beta$ depends nonlinearly on $\ztrans$ as described in Eq.~\eqref{eq:covariates_relation_nonlinear}.  We refer to this model as the  ``\betastep\ Model''.

The several model extensions we consider are summarized in Table~\ref{tab:model_extensions}. 

\begin{table}[t]
\begin{center} \small
\begin{tabular}{lll}
\hline\hline
\multicolumn{3}{l}{\bf Models that adjust for host galaxy mass} \\

\HC && $(\Mnot, \sigmares)$ split for low/high host galaxy mass at $\Mgalobs = 10$. \\ 
\SC && $(\Mnot, \sigmares)$ split for low/high host galaxy mass at $\Mgal = 10$. \\
\CA && Host galaxy mass included in linear regression with\\[-8pt]
&& coefficient, $\gamma$, see Eq.~\eqref{eq:covariates_relation_generalized}. \\

\multicolumn{3}{l}{\bf Models that allow the color adjustment to depend on redshift} \\
\betalin && color correction given by $\beta + \beta_1 z$, see Eq.~\eqref{eq:linear-color}. \\ 
\betastep &&   color correction changes smoothly by $\Delta\beta$ near $z=\ztrans$, see Eq.~\eqref{eq:step-color}. \\

\hline\hline
\end{tabular}
\end{center}
\caption{Summary of extensions to the Baseline Model.}
\label{tab:model_extensions}
\end{table}%

\subsection{Posterior Sampling} 
\label{sec:sampling}

To significantly reduce the dimension of the parameter space under the Baseline Model, \cite{2011MNRAS.418.2308M} marginalizes out the $3n$ latent variables, $\{M_i\ep, \xone{i}, c_i\}$, from the posterior distribution. This relies on the Gaussian population distributions for analytic tractability. A consequence is that the  posterior distributions of the latent variables of each SNIa are inaccessible. 

\BAHAMAS\ improves on~\cite{2011MNRAS.418.2308M} by using a Gibbs-type sampler to sample from the joint posterior distribution of the parameters and latent variables. This has the advantage that we can present object-by-object posterior distributions for the latent color, stretch and intrinsic magnitude values. These can also be mapped onto posterior distributions for the residuals of the Hubble diagram, see Fig.~\ref{fig:1D_shrinkage}.

Furthermore, \BAHAMAS\ does not require Gaussian population distributions, as the posterior sampling is fully numerical; \cite{Rubin:2015rza} take a similar approach.  Although we do not use non-Gaussian distributions here, \BAHAMAS\ opens the door to a fully Bayesian treatment of non-Gaussianities and selection effects. This will be investigated in a future work. 

We present the algorithmic details of our Gibbs-type sampler in Appendix~\ref{apsec:sampler}. We have cross-checked the results obtained with Gibbs-type sampler with those obtained with the Metropolis-Hastings sampler of \cite{2011MNRAS.418.2308M} and with  the  {\sc MultiNest} sampler (a nested sampling algorithm, see~\cite{Feroz:2008xx}). 
{This comparison was carried out for the baseline model as well as the extensions in Table \ref{tab:model_extensions}. The main difference being that the Gibbs sampler directly simulates the latent variables while the other two algorithms do not.}
The marginal distributions obtainable with the latter two methods match within the numerical sampling margin of error with the Gibbs-type sampler output.  
We use the ~\cite{Gelman:1992} stopping criterion and require their scale reduction factor, $\hat R$, to be less than 1.1 for all the components of $\Cparams$ and $\regcoeff$. This leads to a chain of typically $\sim 3300$ samples, with an effective sample size\footnote{The effective sample size of the parameter $\psi$ is defined as
\begin{equation}
\mathrm{ESS}(\psi)=\frac{T}{1+2\sum_{t=1}^\infty \rho_{t}(\psi)},
\label{eq:ess}
\end{equation}
where $T$ is the total posterior sample size and $\rho_{t}(\psi)$ is the lag-$t$ autocorrelation of $\psi$ in the MCMC sample. ESS$(\psi)$ approximates the size of an independent posterior sample that would be required to obtain the same Monte Carlo variance of the posterior mean of $\psi$,  see~\cite{kass:etal:98} and \cite{liu:01:book}. ESS$(\psi)$ is an indicator of how well the MCMC chain for $\psi$ mixes; ESS$(\psi)$ is necessarily less than $T$ and larger values of are preferred.} of around 200 for $\Cparams$ components, and 400 for $\regcoeff$ components. This requires a CPU time of order $2.0\times 10^5$ seconds, where the cost of evaluating a single likelihood is of the order 5--10 seconds on a single CPU.

\section{Results} \label{sec:results}

Here we present  the \BAHAMAS\ fits to the JLA data, as well as in combination with {\em Planck} cosmic microwave background (CMB) temperature data, complemented by WMAP9 polarisation data~\citep{2015arXiv150201589P}. 
 
\subsection{Baseline Model}

We begin by displaying in Fig.~\ref{fig:Baseline_triangle} the 1D and 2D marginal posterior distributions for the cosmological parameters, and color and stretch correction parameters from the JLA SNIa sample analysed with \BAHAMAS\ (black contours). We also show the combination with {\em Planck} CMB data, which we obtained via importance sampling (red contours).  We consider either a Universe containing a cosmological constant, $w=-1$ (\LCDM), or a flat Universe with a dark energy component with redshift-independent $w\neq-1$ (\wCDM).

Table~\ref{tab:1d_constraints_wm1} (\LCDM) and Table~\ref{tab:1d_constraints_flat} (\wCDM) report the corresponding marginal posterior credible intervals. For the $w=-1$ case (i.e., \LCDM), we find $\Omega_m = 0.340 \pm 0.101$ and $\Omega_\kappa = 0.119 \pm 0.249$ (JLA alone). Including {\em Planck} data results in\footnote{We summarize marginal posterior distributions with their posterior mean and approximate 68\% ($1\sigma$) posterior credible intervals. We report highest posterior density (HPD) posterior intervals, which are the shortest intervals that capture 68\% of the posterior probability.  In most cases, the marginal posterior distributions are symmetric and approximately Gaussian, in which case the reported error bar is the posterior standard deviation.  The exceptions are the intervals reported for $w$ which are reported with asymmetric positive and negative errors due to the non-Gaussian shape of the posterior distribution.} $\Omega_m = 0.399 \pm 0.027$, a significantly higher value of the matter content than reported in the standard analysis. (More detailed comparisons are given below.)   The curvature parameter is $\Omega_\kappa = -0.024 \pm 0.008$, excluding a flat Universe, $\OmK = 0$, at the $\sim 3\sigma$ level. For the case of a flat Universe (i.e., \wCDM, Table~\ref{tab:1d_constraints_flat}), we find from JLA and {\em Planck}, $\Omega_m = 0.343 \pm 0.019$, and $w=-0.910 \pm 0.050$. 
The contours of the posterior distribution of $\OmM$ and $\OmL$ based on the JLA data only are similar to those obtained by~\cite{Nielsen:2015pga}. These authors marginalized latent variables out of their effective likelihood, in an approach similar to our own, although with a number of detailed differences\footnote{\cite{Nielsen:2015pga}, adopted implicit uniform priors on the population variances, as well as on $\sigmaint$. They also maximised the likelihood to obtain confidence intervals on cosmological parameters (after marginalisation of the latent variables), rather than integrating the posterior to obtain marginalised credible regions (as in this work). Because \BAHAMAS\ is a non-linear, non-Gaussian model there is no reason to expect {\em a priori} that our results ought to be similar to those obtained by \cite{Nielsen:2015pga}.}. In particular, the $1\sigma$ (marginal posterior) contour obtained with \BAHAMAS\ overlaps closely with the $1\sigma$ (profile likelihood) contours in~\cite{Nielsen:2015pga}, while the $2\sigma$ contour from \BAHAMAS\ shows a degree of asymmetry that is not present in~\cite{Nielsen:2015pga}. (Recall that the analysis of ~\cite{Nielsen:2015pga} relies on approximating the confidence regions using Gaussians, while the numerical sampling of \BAHAMAS\ does not.) 

The residual intrinsic dispersion is in all cases close to $\sigmares = 0.104 \pm 0.005$. This value is to be understood as the average residual scatter in the (post-correction) intrinsic magnitudes across all surveys that make up the JLA data set. This value should be compared with the parameter $\sigma_\text{coh}$ in~\cite{Betoule:2014frx} (which in that work has an approximately equivalent meaning as our $\sigmares$)  ranging from 0.08 (for SNLS) to 0.12 (low-$z$).
It would be easy to extend our analysis to allow for a different value of $\sigmares$ for each of the data sets (SNSL, SDSS, low-$z$, and HST) comprised in JLA. 

The posterior constraints on the residual intrinsic dispersion are, in principle, sensitive to the choice of scale in its inverse Gamma prior distribution. To test the robustness of our posterior inference on $\sigmares$ with respect to its prior specification, we have compared the posterior distributions obtained with three very different prior distributions; each is an inverse Gamma distribution, but with parameters  $u=v=0.003, 0.03, 0.1$. The resulting posterior distributions (alongside their prior distributions) are shown in Fig.~\ref{fig:res_prior_post}. Despite the widely differing prior distributions, the posteriors are  nearly identical, demonstrating the prior-independence of our result. We have verified that all constraints on the other parameters are similarly insensitive to the choice of prior for $\sigmares$. 

\begin{figure}[t!]
\centering
 \includegraphics[width=0.85 \linewidth]{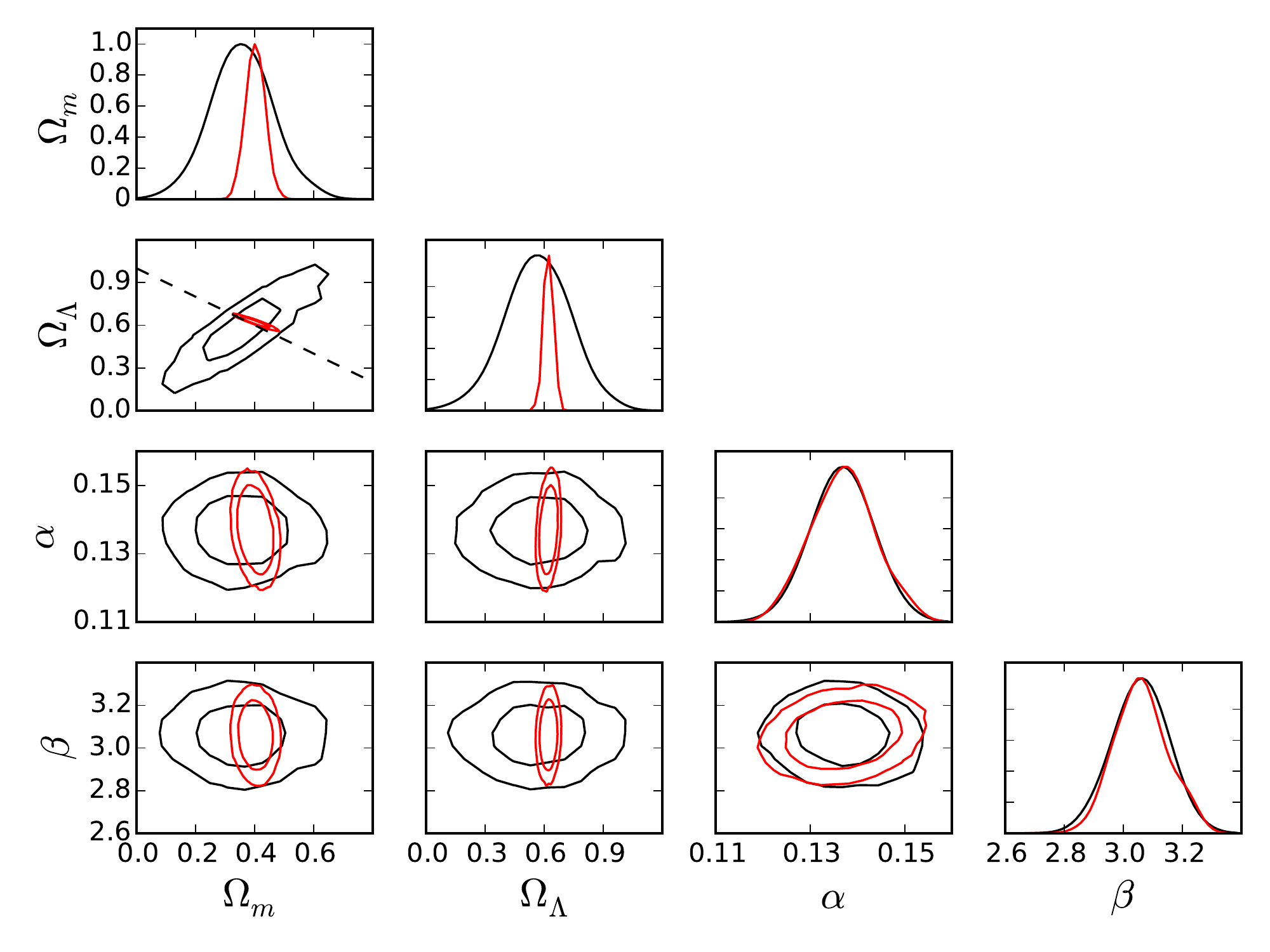}
 \includegraphics[width=0.85 \linewidth]{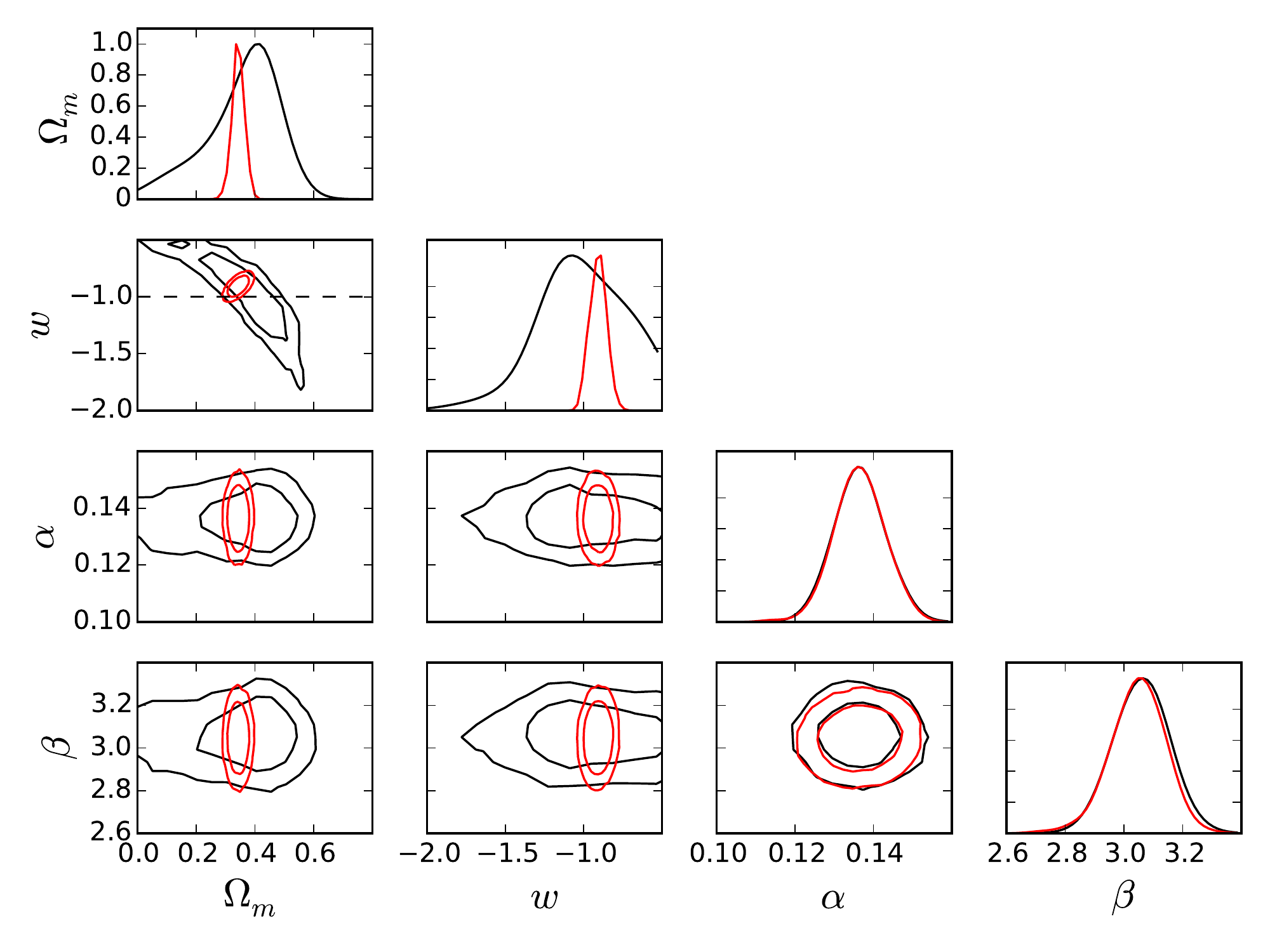}
  \caption{1D and 2D marginal posterior distributions for the cosmological parameters, and the color and stretch correction parameters. Black (red) contours show 68\% and 95\% highest posterior density regions for JLA SNIa data only (JLA combined with {\em Planck}). The top (bottom) panels display results for the \LCDM\ (\wCDM) model. 
\label{fig:Baseline_triangle}}
\end{figure}

\begin{figure}[tbh]
\centering
 \includegraphics[width=0.85\linewidth]{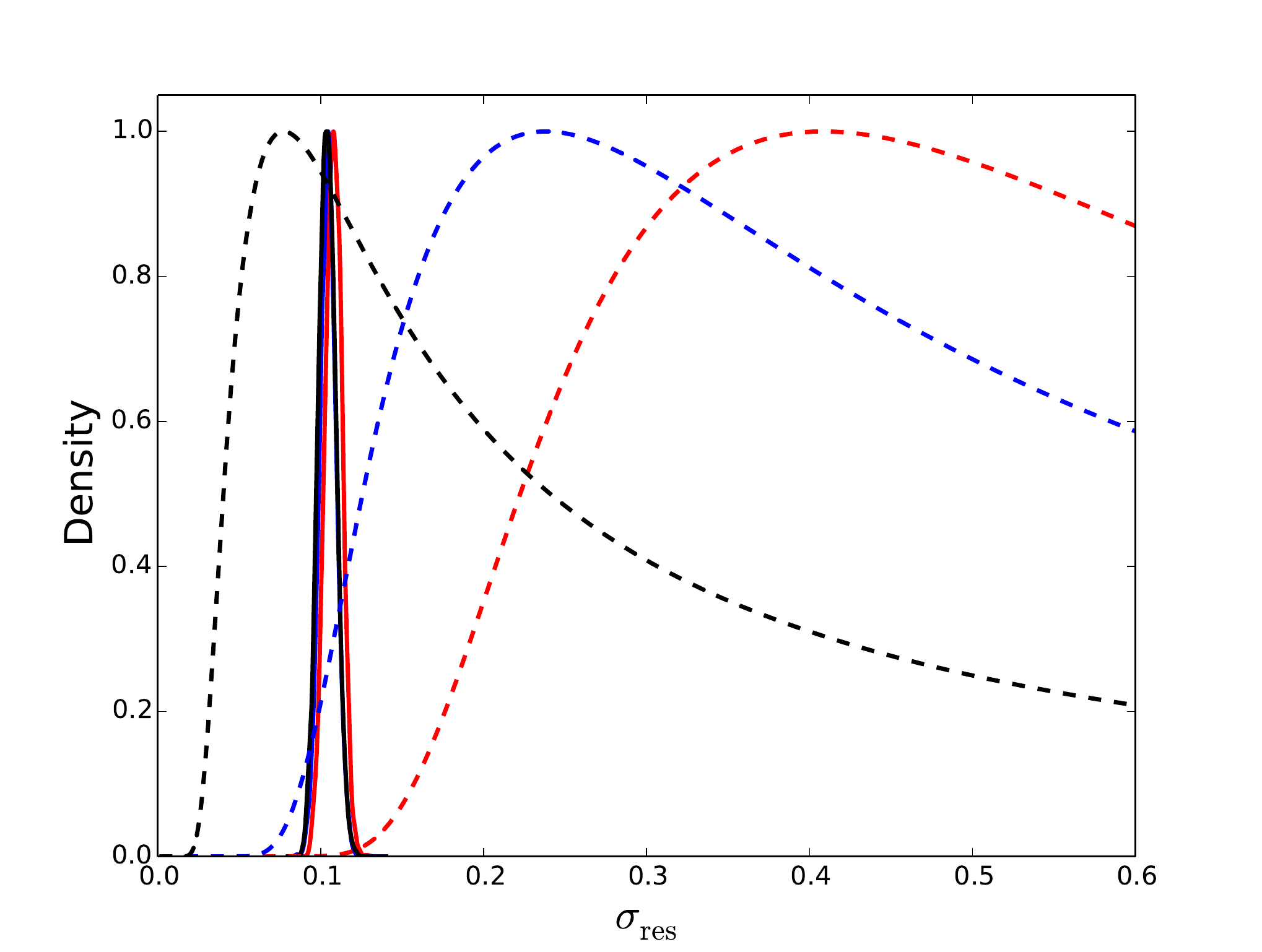}
  \caption{Robustness of the posterior distribution for $\sigmares$ (solid lines) with respect to three different prior specifications (dashed lines) Black: $\sigmares^2\sim \invgamma(0.003,0.003)$; blue: $\sigmares^2\sim \invgamma(0.03,0.03)$; red: $\sigmares^2\sim \invgamma(0.1,0.1)$.  Since the three posterior distributions are very similar, we conclude that the posterior distribution of $\sigmares$ is largely insensitive to its prior specification (assuming \LCDM). Densities have been normalized to their peak for ease of comparison. In the rest of this paper, we  use $\sigmares^2\sim \invgamma(0.003,0.003)$.}
\label{fig:res_prior_post}
\end{figure}

\BAHAMAS\ allows us to compute the posterior distribution for all latent variables, and for the Hubble residuals. It is instructive to compare the posterior distribution to the standard best fit estimate to illustrate the phenomenon of ``shrinkage'': the hierarchical regression structure of the Bayesian model allows estimators to ``borrow strength'' across the SNIa's and thus reduces their residual scatter around the regression plane. 

We illustrate the shrinkage effect using the Baseline Model. We divide the SNIa's into four bins using the quartiles of $\xhat{}$. For each bin, in the four panels of the first row of Fig.~\ref{fig:1D_shrinkage}, we plot in blue $\hat{M}_i \equiv \mBhat{i}-\mu_i(\zhat_i, \overline{\Cparams})$ versus $\chatn{i}$. Here, $\overline{\Cparams}$ is the posterior mean of the cosmological parameters, and $\hat{M}_i$ is a plug-in estimate of the intrinsic magnitude of SNIa $i$ before corrections. This is equivalent to the standard ``best-fit'' estimate of the intrinsic magnitude. In red we plot the posterior means, i.e. $\mBbar{i}-\mu_i(\zhat_i, \overline{\Cparams})$ vs $\cbar{i}$. The regression line in each bin (black) has slope $\bar{\beta}$ and intercept $\bar{M}_0-\bar{\alpha} x_1$, where the bar represents the average with respect to the posterior distribution while $x_1$ is the mean of $\xhat{}$ in that bin. 

In each bin, we observe the expected positive correlation between intrinsic magnitude and color (top panels), and negative correlation between intrinsic magnitude and stretch (bottom panels) The most striking feature is that the posterior estimates are dramatically shrunk towards the regression line, when compared with the plug-in estimates. This is because \BAHAMAS\ accounts for the uncertainty in the measured value of $\{\mBhat{i}, \ox, \oc \}$, and adjusts their fitted values (i.e., their posterior distributions) by ``shrinking'' them towards their fitted population means and the fitted regression line.   

\begin{figure}[tbh]
\centering
 \includegraphics[width=0.995\linewidth]{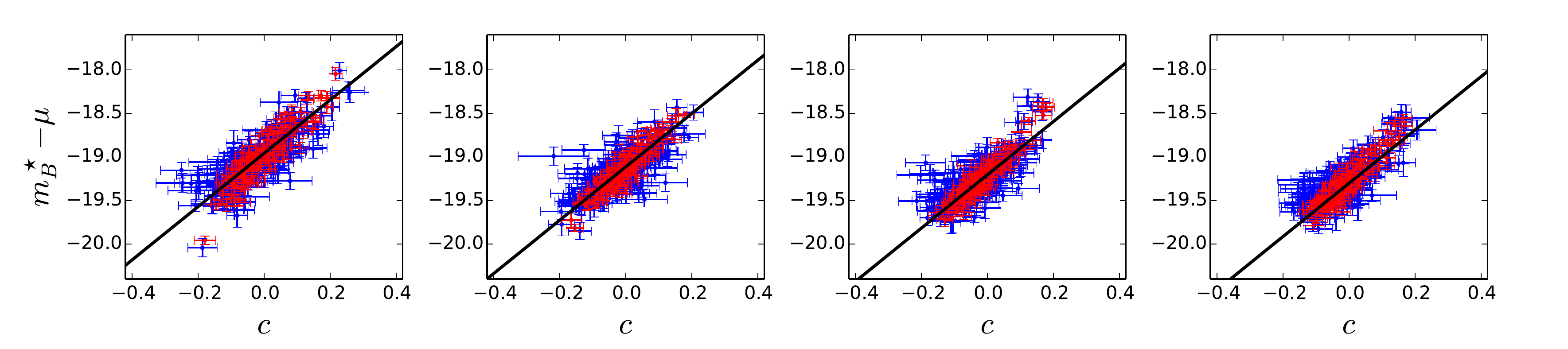}
  \includegraphics[width=0.995\linewidth]{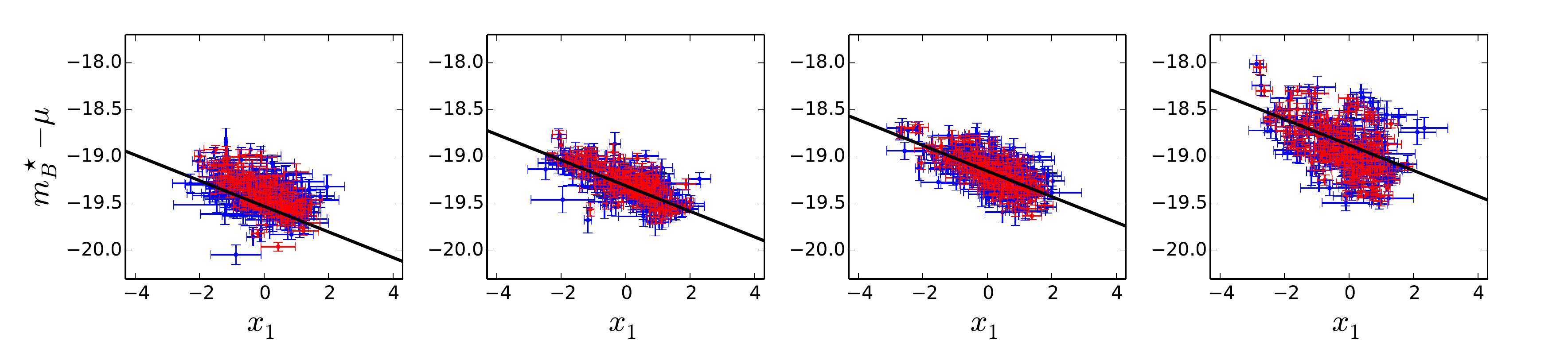}
  \caption{Shrinkage of posterior estimates in \BAHAMAS: Intrinsic magnitude plug-in estimates (blue) and posterior mean (red). The four panels in the first row correspond to quartiles of $\xhat{}$;  we plot the regression line as a function of the color parameter in each. The horizontal axis plots $\chat_{i}$ (blue) and the posterior mean of $c_i$ (red). 
The four panels in the bottom row correspond to quartiles of $\chat{}$; we plot the regression as a function of the stretch parameter $\xhat{}$ in each.
The horizontal axis plots $\xhat{i}$ (blue) and and the posterior mean of $x_{1i}$ (red). 
The regression lines use the posterior means of the parameters and the mean of the observed covariates in each quartile. 
The posterior estimates shrink from the plug-in estimates toward the regression line and thus reduce scatter around the regression plane. This is a consequence of the hierarchical regression in the model (this plot is for the \LCDM\ case).\label{fig:1D_shrinkage}}
\end{figure}


\begin{table}[t!]
\scriptsize
\begin{center}
\begin{tabular}{lccccccc}
\hline\hline
& \multicolumn{3}{c}{JLA SNIa only} && \multicolumn{3}{c}{JLA SNIa + Planck 2015}\\
 &  & $z$-Linear & $z$-Jump &    & & $z$-Linear & $z$-Jump \\[-5pt]
 & Baseline & color Corr & color Corr  &&   Baseline & color Corr  & color Corr \\ \cline{2-4} \cline{6-8}
&\multicolumn{7}{c}{\underline{ \it Baseline Model parameters}}\\
$\OmM$  & $0.340 \pm  0.101$ & $0.362 \pm  0.094$ & $0.429 \pm  0.097$ &&$0.399 \pm  0.027$ & $0.420 \pm  0.031$&  $0.425 \pm  0.025$ \\ 
$\OmL$  &$0.542 \pm  0.157$ &$0.557 \pm  0.145$ & $0.632 \pm  0.155$ &&$0.625 \pm  0.020$ & $0.609 \pm  0.025$&  $0.604 \pm  0.019$  \\
$\OmK$ & $0.119 \pm  0.249$ &$0.081 \pm  0.230$ & $ -0.061 \pm 0.244$ &&  $-0.024 \pm  0.008$ & $-0.028 \pm  0.008$ &  $-0.029 \pm  0.007$\\ 
$\alpha$& $0.137 \pm  0.006$  & $0.136 \pm  0.006$ & $0.136 \pm  0.006$&& $0.137 \pm  0.006$ &$0.135 \pm  0.007$ &  $0.136 \pm  0.006$ \\ 
$\beta$ & $3.058 \pm  0.085$ & n/a & n/a && $3.068 \pm  0.097$ & n/a&  n/a \\ 
&\multicolumn{7}{c}{\underline{ \it Redshift evolution of color correction parameters}} \\ 
$\beta_0$ & n/a & $3.211 \pm  0.120$ & $3.137 \pm  0.092$ && n/a & $3.219 \pm  0.119$&  $3.136 \pm  0.096$ \\ 
$\beta_1$ & n/a & $-0.622 \pm  0.342$ &n/a && n/a & $-0.732 \pm  0.360$ &  n/a \\ 
$\Delta\beta$ & n/a & n/a &  $-1.120 \pm  0.240$ && n/a & n/a & $-1.145 \pm  0.243$  \\ 
$\ztrans$ & n/a & n/a &  $0.662 \pm  0.055$ && n/a & n/a & $0.670 \pm 0.056$  \\ 
&\multicolumn{7}{c}{\underline{ \it Intrinsic magnitude and residual dispersion parameters}} \\ 
$\Mnot$ &$-19.140 \pm  0.022$ & $-19.140 \pm  0.020$ & $-19.144 \pm  0.021$ &&$-19.140 \pm  0.018$ &  $-19.138 \pm  0.018$ & $-19.140 \pm  0.016$\\ 
$\sigmares$ & $0.104 \pm 0.005$& $0.104 \pm 0.005$& $0.103 \pm 0.005$  && $0.105 \pm 0.005$ &  $0.105 \pm 0.004$ & $0.103 \pm 0.005$\\ \hline\hline
\end{tabular}
\caption{Marginalized posterior constraints on cosmological and SNIa correction parameters for the \LCDM\ model, assuming $H_0 = 67.3$ km/s/Mpc. 
\label{tab:1d_constraints_wm1}}
\end{center} 
\end{table}%

\begin{table}[htb]
\scriptsize
\begin{center}
\begin{tabular}{lccccccc}
\hline\hline
& \multicolumn{3}{c}{JLA SNIa only} && \multicolumn{3}{c}{JLA SNIa + Planck 2015}\\
 &  & $z$-Linear & $z$-Jump &    & & $z$-Linear & $z$-Jump \\[-5pt]
 & Baseline & color Corr & color Corr  &&   Baseline & color Corr  & color Corr \\ \cline{2-4} \cline{6-8}
$\OmM$ & $0.355 \pm  0.117$ &  $0.366 \pm  0.119$ & $0.422 \pm  0.097$  && $0.343 \pm  0.019$ & $0.349 \pm  0.015$ & $0.353 \pm  0.018$  \\ 
$\OmL$ & $0.645 \pm  0.117$ &  $0.634 \pm  0.119$ &  $0.578 \pm  0.097$  && $0.657 \pm  0.019$ & $0.651 \pm  0.015$ & $0.647 \pm  0.018$ \\
$w$ &$-0.995 ^{+0.418}_{-0.275}$ & $-1.022 ^{+0.425}_{-0.227}$ & $-1.145 ^{+0.394}_{-0.293}$ && $-0.910 \pm  0.045$ & $-0.905 \pm  0.050$ &   $-0.883 \pm  0.043$\\ 
$\alpha$ &$0.136 \pm  0.006$ &  $0.136 \pm  0.006$ & $0.136 \pm  0.006$  &&$0.136 \pm  0.006$ & $0.137 \pm  0.006$ & $0.136 \pm  0.005$  \\ 
$\beta$ & $3.060 \pm  0.088$ &  n/a & n/a && $3.047 \pm  0.087$ & n/a & n/a  \\ 
\cline{2-4} \cline{6-8}
$\beta_0$ & n/a &  $3.206 \pm  0.358$ & $3.137 \pm  0.090$ && n/a & $3.199 \pm  0.109$ & $3.128 \pm  0.082$ \\ 
$\beta_1$ & n/a &  $-0.629 \pm  0.358$& n/a && n/a &  $-0.603 \pm  0.320$& n/a \\ 
$\Delta\beta$ & n/a & n/a & $-1.116 \pm  0.240$   && n/a & n/a &  $-1.083 \pm  0.237$ \\
$\ztrans$ & n/a & n/a &  $0.661 \pm  0.055$ && n/a & n/a & $0.655 \pm 0.055$  \\ 
\cline{2-4} \cline{6-8}
$\Mnot$ &$-19.146 \pm  0.024$ & $-19.142 \pm  0.022$ & $-19.145 \pm  0.021$ && $-19.148 \pm  0.024$ & $-19.144 \pm  0.020$& $-19.143 \pm  0.020$\\ 
$\sigmares$ & $0.103 \pm 0.005$ &  $0.104 \pm 0.005$ & $0.103 \pm 0.005$  && $0.103 \pm 0.007$ &$0.104 \pm 0.005$ & $0.102 \pm 0.005$ \\
\hline\hline
\end{tabular}\caption{As in Table~\ref{tab:1d_constraints_wm1}, but for \wCDM. \label{tab:1d_constraints_flat}}
\end{center}
\end{table}%


\subsection{Including Host Galaxy Mass Corrections} 
\label{sec:galaxymass}

We now investigate the impact of including information on the host galaxy mass. Marginalized posterior constraints on our model parameters when the host galaxy mass is used as a predictor or a covariate are reported in Tables~\ref{tab:1d_constraints_Mgal_wm1} (\LCDM) and \ref{tab:1d_constraints_Mgal_flat} (\wCDM). The posterior distributions are shown in Fig.~\ref{fig:comp_baseline_mgalprobclass}, where they are compared with the case when no mass correction is used. 

\begin{figure}[tbh]
\centering
\begin{tabular}{cc}
  \includegraphics[width=0.48\linewidth]{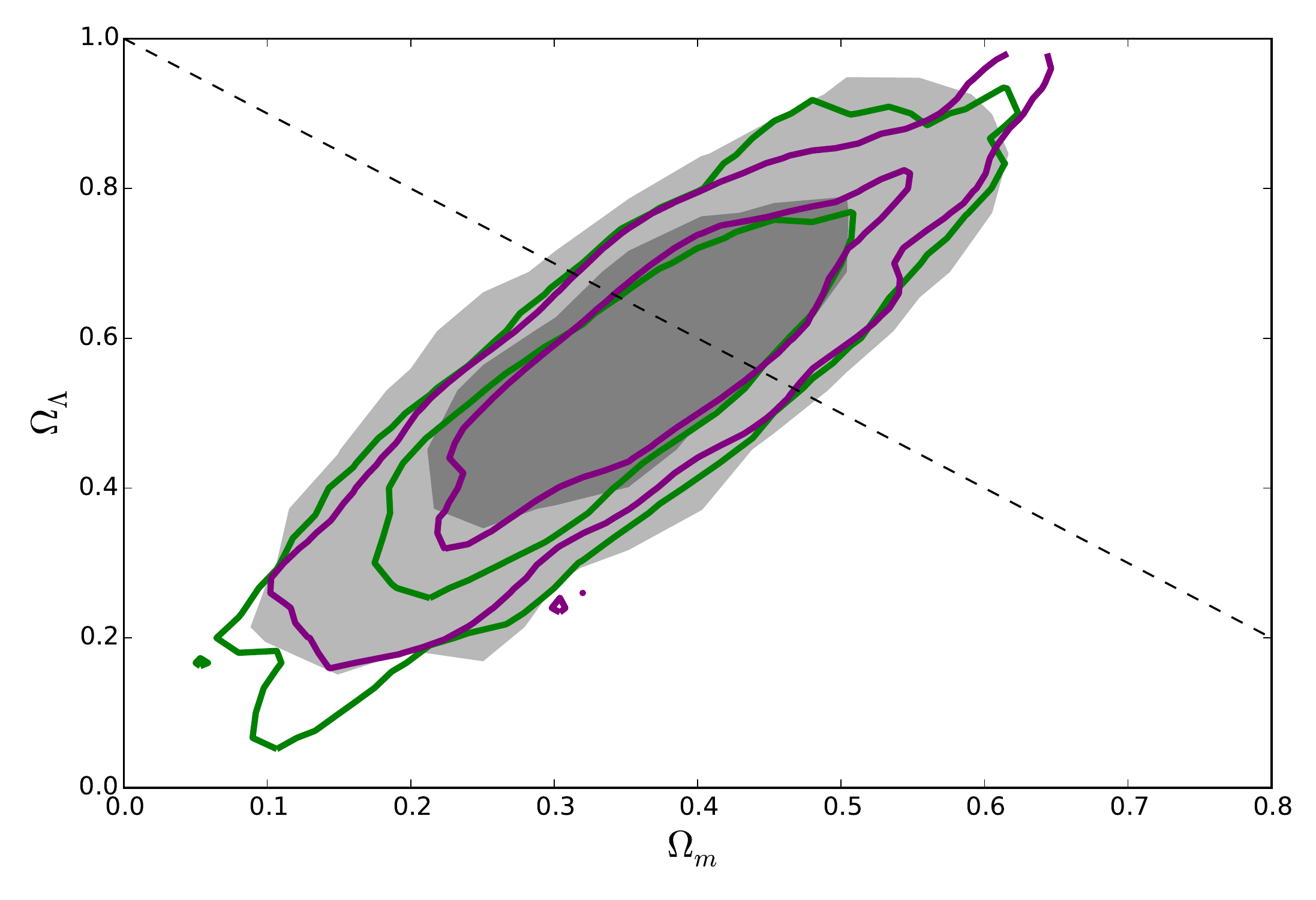} &   \includegraphics[width=0.48\linewidth]{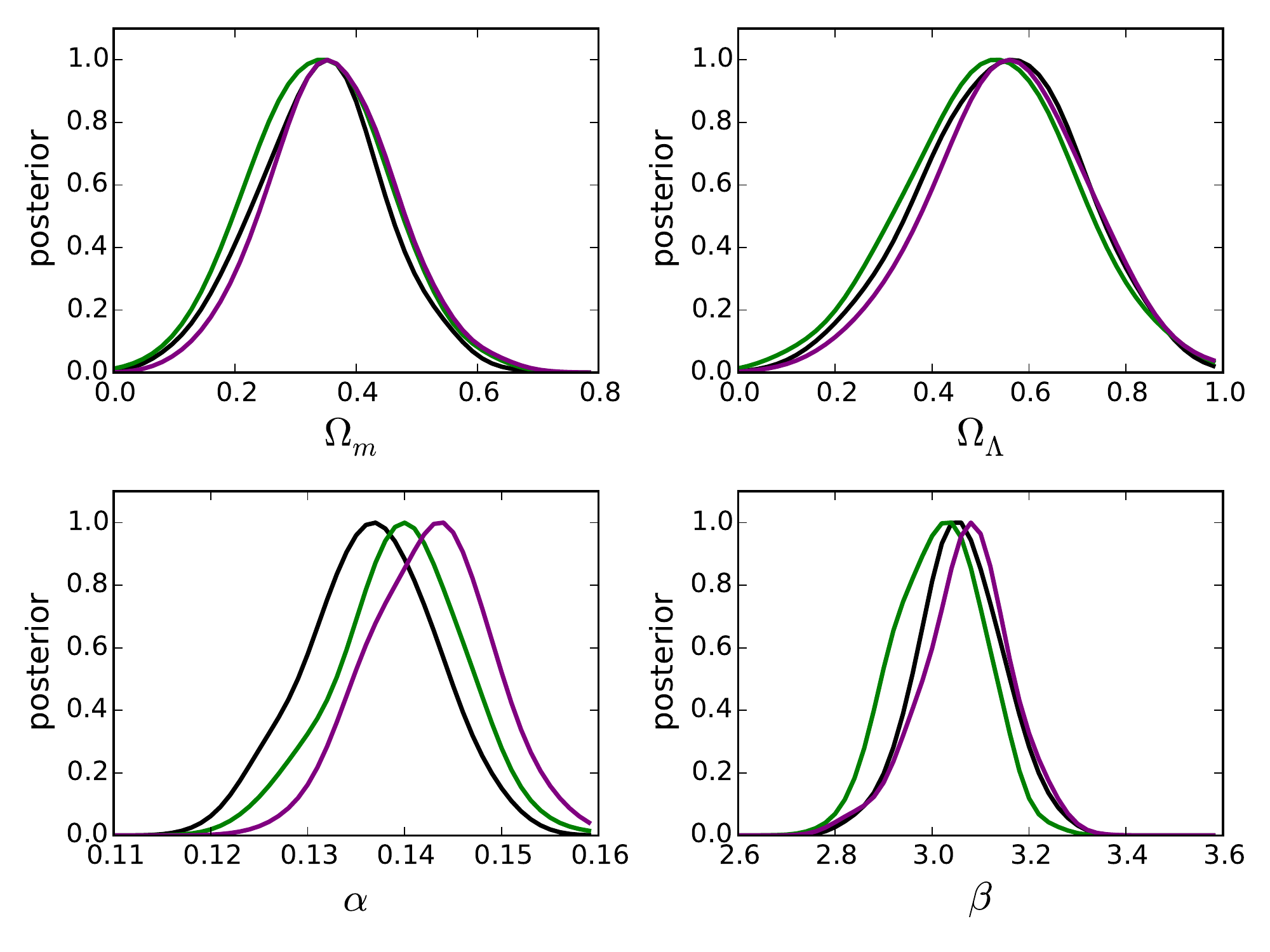} \\
  \includegraphics[width=0.48\linewidth]{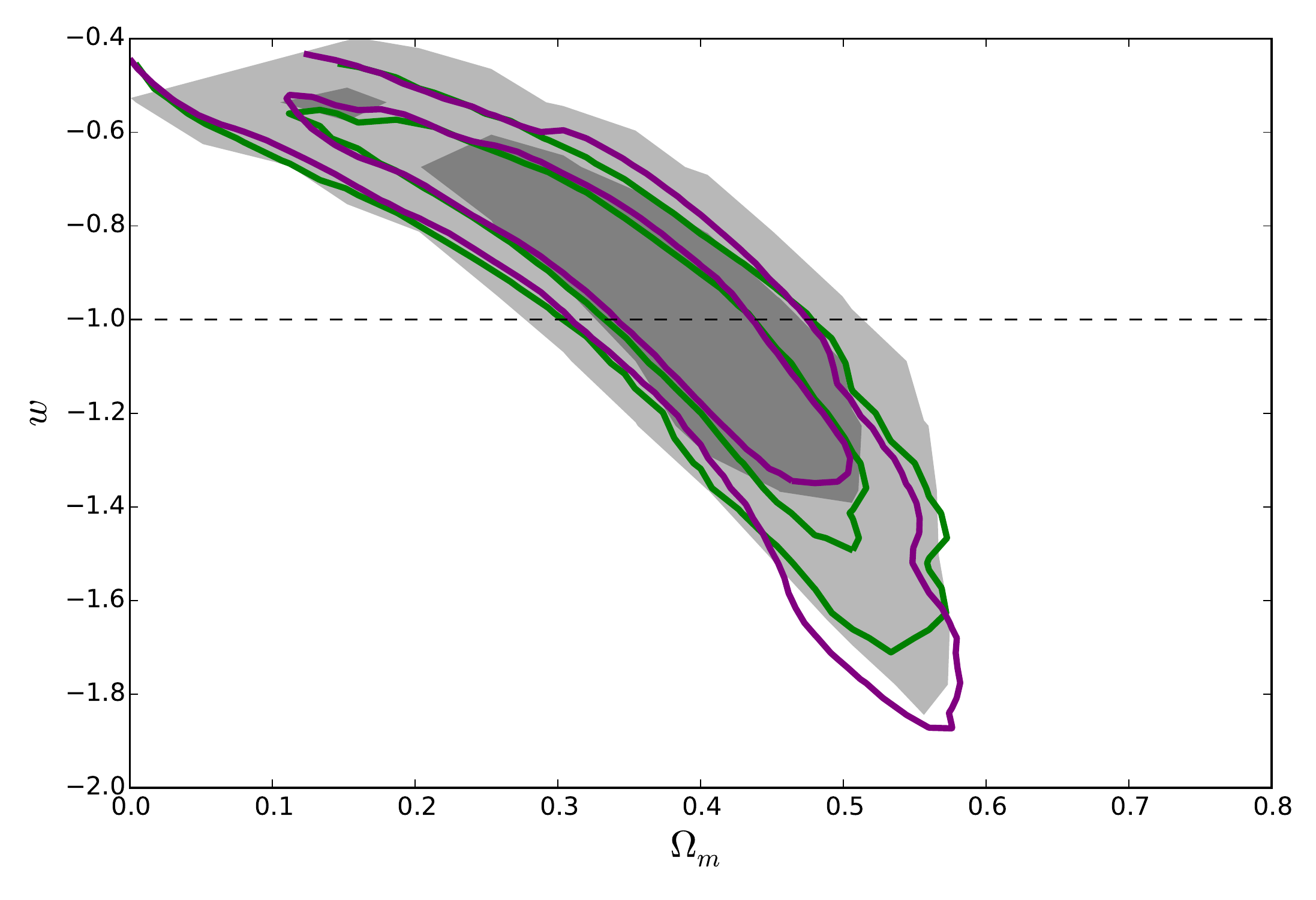} &   \includegraphics[width=0.48\linewidth]{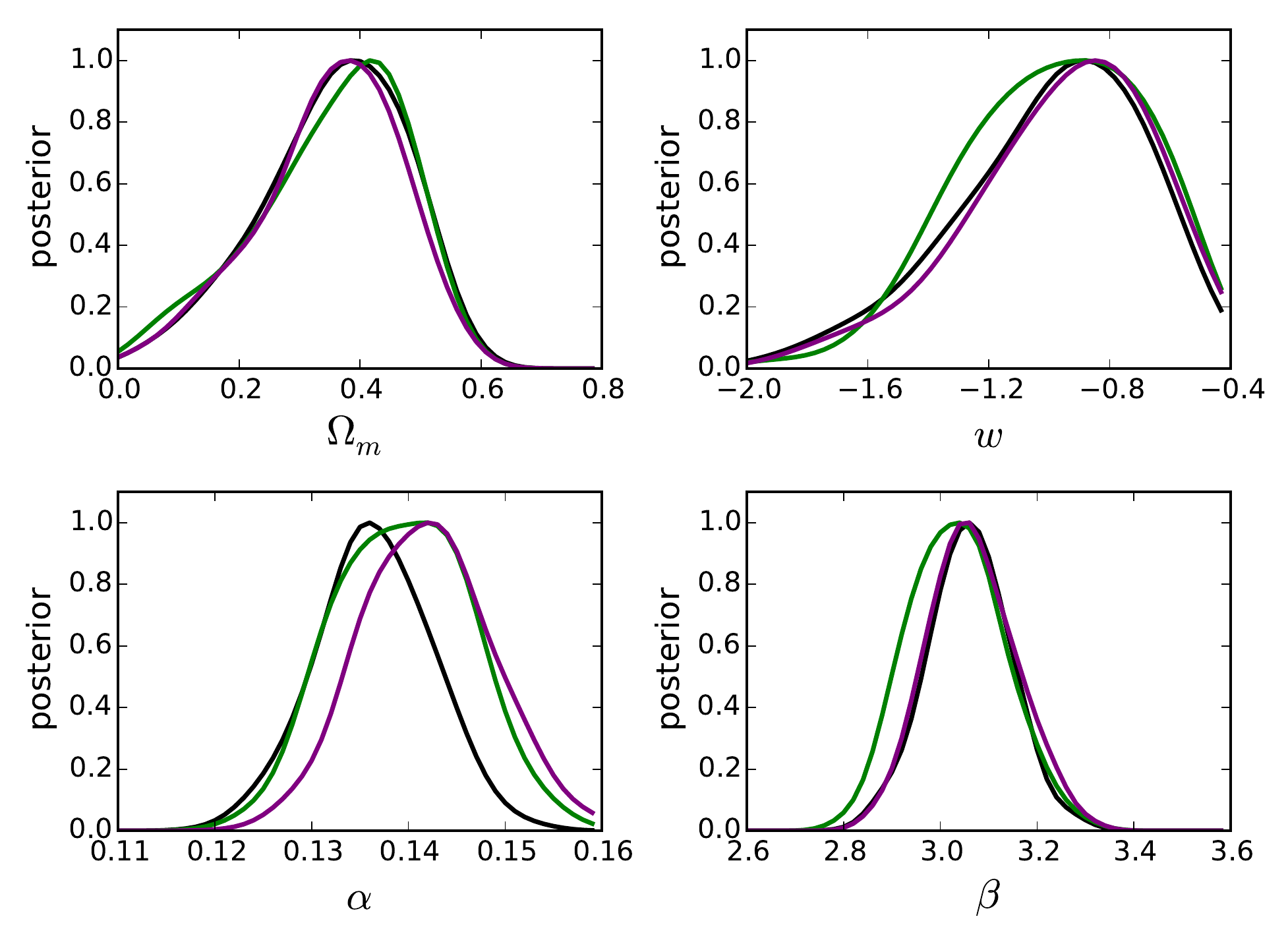} \\
\end{tabular}
  \caption{Comparison of cosmological parameters and standardization parameters with and without host galaxy mass correction (black/shaded: Baseline Model; green: \SC\ Model; purple: \CA\ Model). The result of the \HC\ Model is similar to that of the \SC\ Model and is not shown.
Top panels are fit under the \LCDM, while the bottom panels are fit under the \wCDM. We do not find a significant difference in cosmology when mass information is included in the fit. }
\label{fig:comp_baseline_mgalprobclass}
\end{figure}

\begin{figure}[tbh]
\centering
  \includegraphics[width=0.48\linewidth]{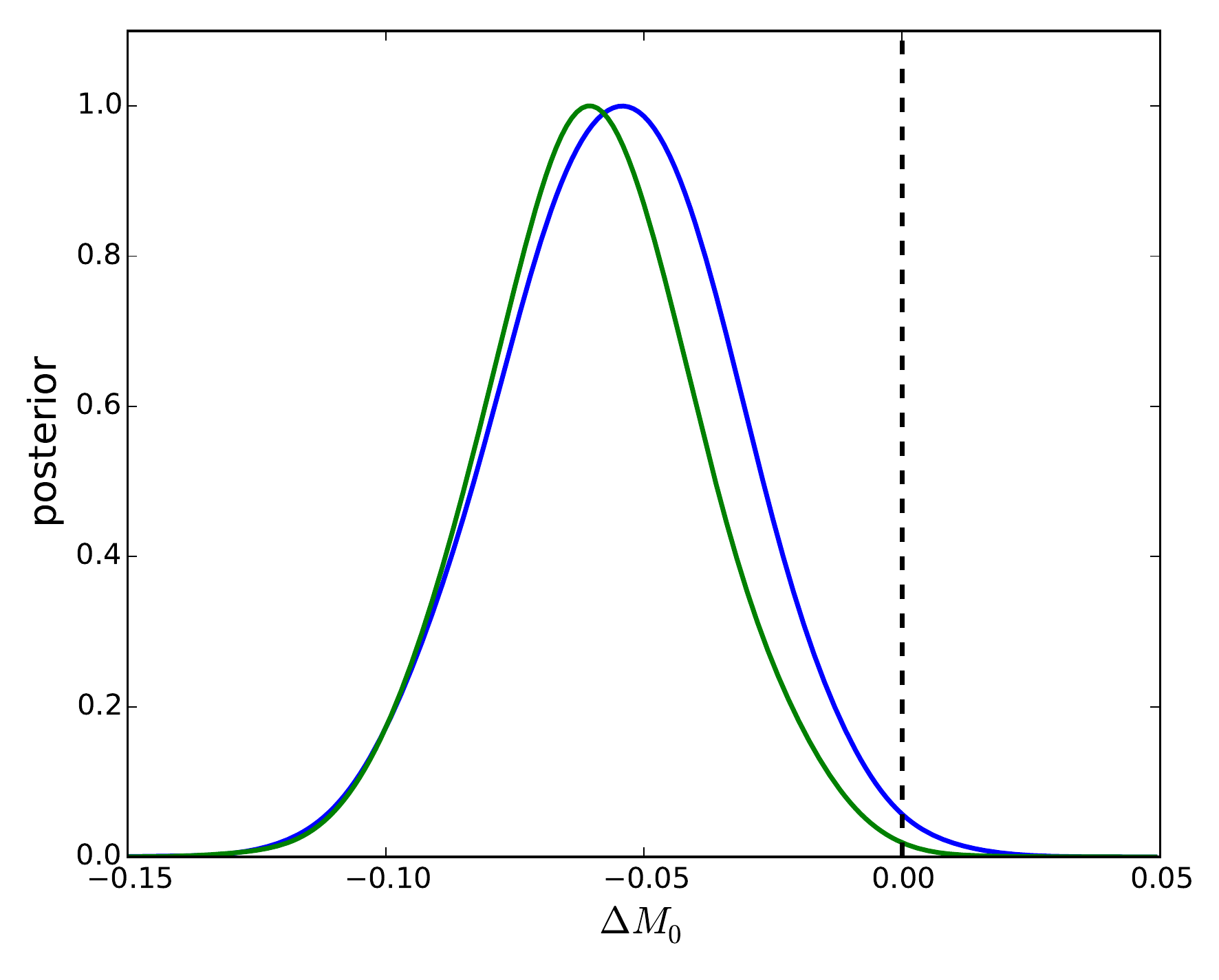}
  \caption{Posterior distribution of $\Delta M_0$, the difference between mean intrinsic magnitudes of SNIa's in high-mass host galaxies ($\Mgal > 10$) and low-mass hosts ($\Mgal<10$).  The blue and green curves correspond to the \HC\ and \SC\ models, respectively.  Under both models, the posterior probability that $\Delta \Mnot<0$ is greater than 95\%, meaning SNIa's in more massive hosts are most probably intrinsically brighter ($\Delta\Mnot < 0$). (This plot assumes a \LCDM\ universe.) }
\label{fig:mean_magnitude_diff}
\end{figure}

The \HC\ model matches exactly the host galaxy mass correction procedure adopted in~\cite{Betoule:2014frx}, hence our results are directly comparable. The only difference is the statistical method adopted in inferring the cosmological parameters from the \salt{} fits. 
For the matter density parameter (assuming \LCDM\ and using JLA data only), we find  $\Omega_m = 0.343 \pm 0.096$ compared to $\Omega_m = 0.295 \pm 0.034$ in~\cite{Betoule:2014frx}. Our posterior uncertainty is about a factor of $\sim 3$ larger, despite the shrinkage effect described above, and the central value is higher by $\sim 0.5 \sigma$.  We find $w=-0.943^{+ 0.363}_{-0.255}$. When compared with the Baseline Model, our cosmological parameter constraints hardly change (see Fig.~\ref{fig:comp_baseline_mgalprobclass})\footnote{Our treatment in the Baseline Model is not fully consistent. While we ignore any host galaxy mass dependence we do include the ``host relation'' term in the systematics covariance matrix. This is however likely to have a negligible effect, since Table 11 in~\cite{Betoule:2014frx} quantifies the contribution to the error budget on $\OmM$ from the host relation uncertainty as a mere 1.3\%. }. 

Despite this, we do detect significant difference (with 95\% probability) between the mean intrinsic magnitude of SNIa in low- and high-galaxy mass hosts. Specifically, we define
\be
\Delta\Mnot\equiv \Mhigh - \Mlow
\ee
as the difference in intrinsic magnitude between the two sub-classes.  The posterior interval for $\Delta\Mnot$ is
\begin{equation}
-0.10 < \Delta\Mnot < 0.00 \quad \text{(95\% equal tail posterior interval)}
\end{equation}
with $\Delta\Mnot = 0$ excluded with 95\% probability. The posterior distribution for $\Delta\Mnot$ is shown in Fig.~\ref{fig:mean_magnitude_diff}, where the result for the \HC\ Model is compared with the \SC\ Model. There is not an appreciable difference in $\Delta\Mnot$ between the \HC\ Model and the \SC\ model. In accordance with previous results~\citep{Kelly:2009iy,2010MNRAS.406..782S,Campbell2015}, we find that SNIa's in more massive galaxies are intrinsically brighter, with our posterior estimate of the magnitude difference being $\Delta\Mnot = -0.055 \pm 0.022$. However, the size of the effect in our study is smaller than previously reported.  For example, \cite{Kelly:2009iy} found (in our notation) $\Delta\Mnot = -0.11$, and \cite{2010MNRAS.406..782S} $\Delta\Mnot = -0.08$, while ~\cite{Campbell2015} reported $\Delta\Mnot = 0.091 \pm 0.045$. 

The residual intrinsic dispersion of the two sub-populations is marginally smaller for the SNIa's residing in more massive hosts: $\sigmareshigh = 0.097 \pm 0.007$; for the lower mass group the residual dispersion is $\sigmareslow = 0.110 \pm 0.009$. (Those values are for the \LCDM\ case, but \wCDM\ is similar.)

\begin{table}[htb]
\scriptsize
\begin{center}
\begin{tabular}{lccccccc}
\hline\hline
& \multicolumn{3}{c}{JLA SNIa only} && \multicolumn{3}{c}{JLA SNIa + Planck 2015}\\
 & Hard & Soft & Covariate &    & Hard & Soft & Covariate  \\[-5pt]
 & Classification & Classification & Adjustment  &&   Classification & Classification  & Adjustment \\ \cline{2-4} \cline{6-8}
&\multicolumn{7}{c}{\underline{ \it Baseline Model parameters}}\\
$\OmM$ &$0.343 \pm 0.096$ & $0.338 \pm 0.107$& $0.361 \pm 0.100$&&$0.423 \pm 0.030$ & $0.400 \pm 0.025$&$0.403 \pm 0.031$\\ 
$\OmL$ &$0.523 \pm 0.144$ &$0.522 \pm 0.165$ & $0.559 \pm 0.151$ &&$0.603 \pm 0.020$ & $0.622 \pm 0.019$&$0.621 \pm 0.023$ \\
$\OmK$ & $0.134 \pm 0.232$&$0.140 \pm 0.263$ & $0.080 \pm 0.244$ &&$-0.026 \pm 0.011$ & $-0.022 \pm 0.008$& $-0.025 \pm 0.010$ \\ 
$\alpha$& $0.141 \pm 0.006$& $0.140 \pm 0.006$& $0.143 \pm 0.006$ && $0.142 \pm 0.006$& $0.142 \pm 0.007$& $0.143 \pm 0.005$\\ 
$\beta$ &$3.058 \pm 0.095$ & $3.014 \pm 0.086$& $3.068 \pm 0.089$  && $3.053 \pm 0.068$& $3.034 \pm 0.060$&  $3.031 \pm 0.086$  \\ 
\cline{2-4} \cline{6-8}
$\Mnot$& n/a & n/a & $-18.837 \pm 0.100$  &&n/a& n/a & $-18.860 \pm 0.096$ \\ 
$\Mlow$& $-19.114 \pm 0.023$ & $-19.110 \pm 0.023$& n/a  &&$-19.111 \pm 0.019$ & $-19.110 \pm 0.021$ & n/a\\ 
$\sigmareslow$ &$0.110 \pm 0.009$  & $0.114 \pm 0.009$  & n/a &&$0.108 \pm 0.006$& $0.113 \pm 0.009$ & n/a\\ 
$\Delta\Mnot$& $-0.055 \pm 0.022$ & $-0.049 \pm 0.022$ & n/a && $-0.062 \pm 0.022$ & $-0.049 \pm 0.019$ & n/a\\ \
$\sigmareshigh$ &$0.097 \pm 0.007$ & $0.096 \pm 0.007$ & n/a &&$0.095 \pm 0.006$& $0.094 \pm 0.006$ & n/a  \\ 
\cline{2-4} \cline{6-8}
$\gamma$  &n/a & n/a&$-0.030 \pm 0.010$ && n/a& n/a & $-0.028 \pm 0.010$ \\
$\sigmares$ &n/a &n/a&$0.101 \pm 0.005$ &&n/a&n/a &$0.102 \pm 0.005$ \\
\hline\hline 
\end{tabular}\caption{Posterior constraints on our model parameters when the host galaxy mass is used as a predictor or a covariate (\LCDM\ case). \HC\ adopts a mass-step correction by splitting the SNIa's according to host galaxy mass into a ``low'' ($\Mgal < 10$) and a ``high'' ($\Mgal>10$) sub-class. \SC\ further accounts for uncertainty due to the host galaxy mass measurement error. \CA\ uses the host galaxy mass as a linear covariate. The quantity $\Delta\Mnot$ is the difference between the mean peak intrinsic magnitudes of the two populations: $\Delta\Mnot \equiv \Mhigh - \Mlow$. \label{tab:1d_constraints_Mgal_wm1} } 
\end{center}
\end{table}

\begin{table}[htb]
\scriptsize
\begin{center}
\begin{tabular}{lccccccc}
\hline\hline
& \multicolumn{3}{c}{JLA SNIa only} && \multicolumn{3}{c}{JLA SNIa + Planck 2015}\\
 & Hard & Soft & Covariate &    & Hard & Soft & Covariate  \\[-5pt]
 & Classification & Classification & Adjustment  &&   Classification & Classification  & Adjustment \\ \cline{2-4} \cline{6-8}
$\OmM$ &$0.342 \pm 0.119$ & $0.343 \pm 0.116$ & $0.348 \pm 0.114$&&$0.343 \pm 0.017$ & $0.350 \pm 0.018$ & $0.347 \pm 0.015$\\ 
$\OmL$ &$0.658 \pm 0.119$ & $0.657 \pm 0.116$ & $0.652 \pm 0.114$&&$0.657 \pm 0.017$ &$0.650 \pm 0.018$ &$0.653 \pm 0.015$ \\
$w$ &$-0.943^{+ 0.363}_{-0.255}$ & $-0.937 ^{+ 0.341}_{-0.213}$ & $-0.958^{+ 0.364}_{-0.271}$ &&$-0.906 \pm 0.043$ & $-0.902 \pm 0.049$& $-0.898 \pm 0.051$ \\ 
$\alpha$ & $0.141 \pm 0.006$ & $0.141 \pm 0.007$ & $0.142 \pm 0.007$  &&$0.135 \pm 0.007$ & $0.142 \pm 0.006$& $0.141 \pm 0.005$ \\ 
$\beta$ & $3.034 \pm 0.078$ & $3.049 \pm 0.085$ & $3.066 \pm 0.087$ && $2.917 \pm 0.092$& $3.054 \pm 0.085$&  $3.057 \pm 0.086$ \\ 
\cline{2-4} \cline{6-8}
$\Mnot$& n/a & n/a & $-18.838 \pm 0.098$ &&n/a& n/a & $-18.846 \pm 0.090$ \\ 
$\Mlow$ & $-19.117 \pm 0.024$ & $-19.111 \pm 0.024$ & n/a&& $-19.126 \pm 0.021$ & $-19.116 \pm 0.020$ & n/a \\ 
$\sigmareslow$ & $0.111 \pm 0.008$ & $0.112 \pm 0.009$ & n/a && $0.110 \pm 0.008$ & $0.112 \pm 0.009$ & n/a\\ 
$\Delta\Mnot$ & $-0.056 \pm 0.021$  & $-0.060 \pm 0.020$ & n/a&& $-0.047 \pm 0.025$ &  $-0.058 \pm 0.020$ &n/a\\ 
$\sigmareshigh$ & $0.098 \pm 0.006$ & $0.094 \pm 0.007$ & n/a & & $0.098 \pm 0.007$ &  $0.094 \pm 0.006$ & n/a\\ 
\cline{2-4} \cline{6-8}
$\gamma$  & n/a& n/a& $-0.030 \pm 0.009$ &&n/a & n/a& $-0.030 \pm 0.009$ \\
$\sigmares$ &n/a &n/a& $0.101 \pm 0.005$ &&n/a&n/a & $0.100 \pm 0.005$\\ 
\hline\hline 
\end{tabular}\caption{As in Table~\ref{tab:1d_constraints_Mgal_wm1}, but for \wCDM. \label{tab:1d_constraints_Mgal_flat}}
\end{center}
\end{table}%

Fig.~\ref{fig:galaxymass_split} shows the posterior estimates of the empirically corrected SNIa's intrinsic magnitudes, $M_i\ep$, as a function of the measured host galaxy mass. Histograms on either side of the graph show the distribution of the posterior mean estimates of $M_i\ep$ for the two populations. The average measurement error of the host galaxy mass is fairly large, especially for low-mass hosts. Therefore, galaxies whose mass is close to the cut-off of $\Mgal=10$ are of uncertain classification, once the measurement error is taken into account. This could influence the estimate of $\Delta\Mnot$ and the ensuing cosmological constraints.   

\begin{figure}[tbh]
\centering
  \includegraphics[width=\linewidth]{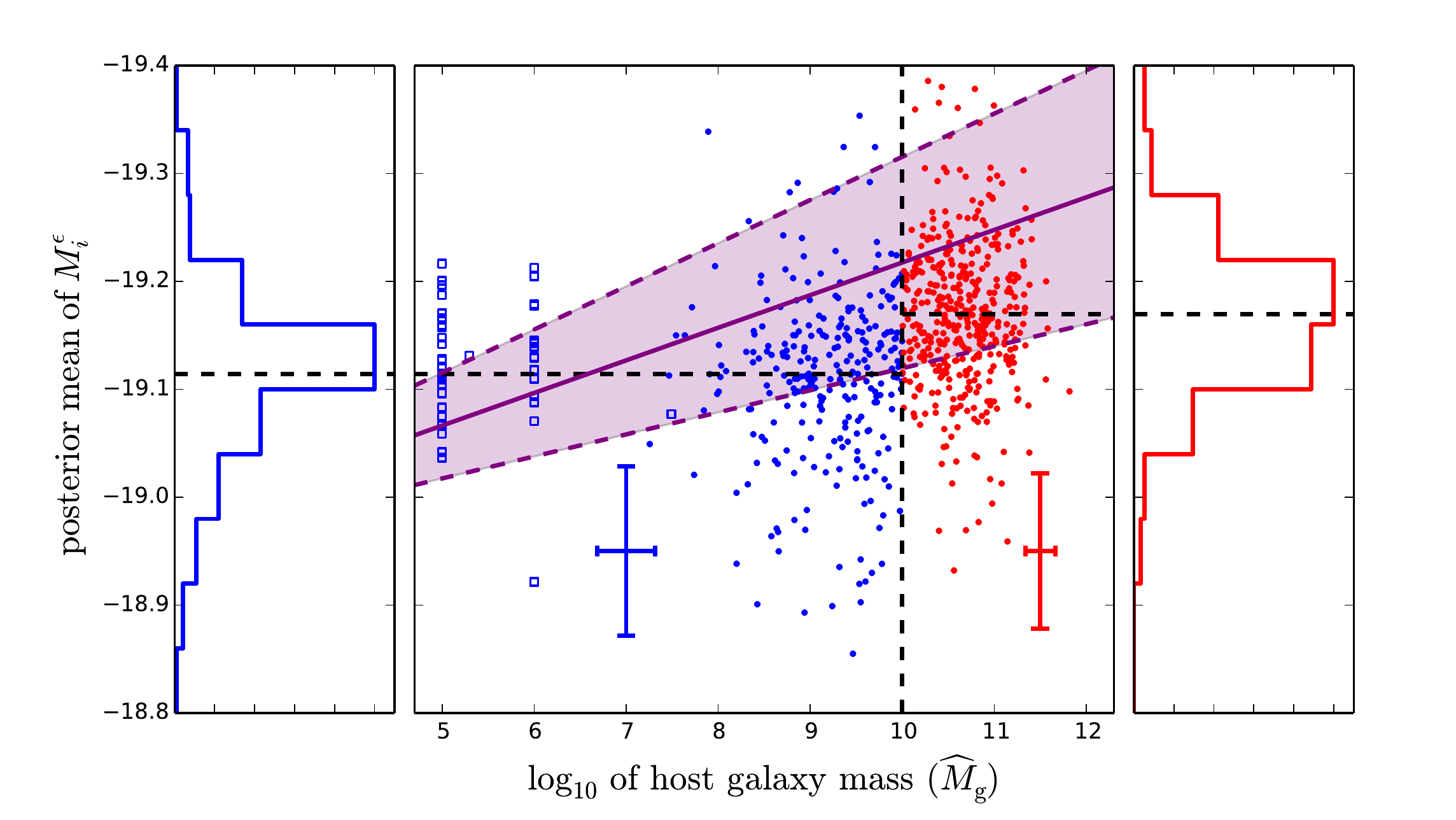}
  \caption{Posterior means and standard deviations for the empirically corrected intrinsic magnitudes of SNIa's in the JLA sample versus measured host galaxy mass. The sample has been divided into two populations, with $\Mgal$ smaller (larger) than $10$, depicted in blue (red). A hollow square represents SNIa whose nominal measurement error on $\Mgal$ is equal to or larger than $5$. The population means of the intrinsic magnitudes are $\Mlow = -19.114 \pm 0.023$ and $\Mhigh= -19.169 \pm 0.022$  (horizontal dashed lines) respectively for the low- and high-host mass classes. The blue and red vertical errorbars represent the average posterior standard deviations of the intrinsic magnitudes  in the low- and high-host mass classes, respectively. The horizontal errorbars represent the average measurement errors of $\Mgal$ in the two classes. The average errorbars exclude the SNIa's represented by hollow squares. The slope of the purple regression line is the posterior mean of $\gamma$ under the \CA\ Model, while the purple shaded area represents the $1\sigma$ credible region for $\gamma$. (The regression line is computed under \LCDM.)}
\label{fig:galaxymass_split}
\end{figure}

To investigate the importance of mass measurement errors, we fit the \SC\ Model which includes indicator variables for each SNIa; recall that $Z_i$ is one if SNIa $i$ belongs to the high-mass host class and zero if it does not. Treating $Z_i$ as an unknown  variable allows us to assess the posterior probability that each SNIa belongs to the high-mass host class. 
In Fig.~\ref{fig:Zi} we plot the posterior means and standard deviations for each $Z_i$. The posterior mean of $Z_i$ is the posterior probability that SNIa $i$ belongs to the high-mass class. 
Although measurement errors in the host galaxy mass are suppressed for clarity in Fig.~\ref{fig:Zi}, the fitted model fully accounts for them.

\begin{figure}[tbh]
\centering
  \includegraphics[width=0.8\linewidth]{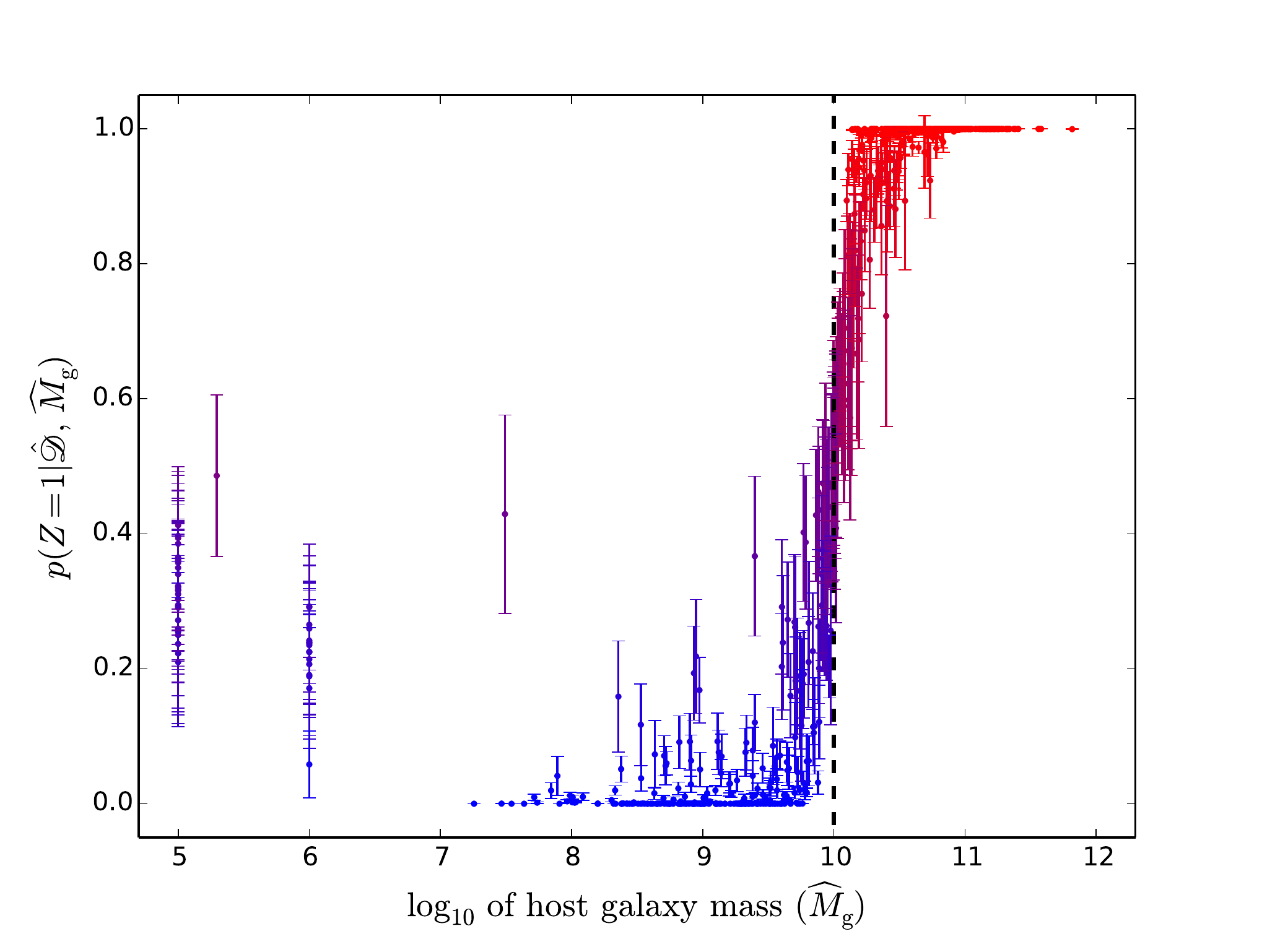}
  \caption{Posterior means and standard deviations of the $Z_i$, the indicator variables for each SNIa belonging to the high-mass host class ($Z_i = 1$, red) versus measured host galaxy mass. If $Z_i=0$ (blue), SNIa belongs to the low-mass host class. The posterior mean of $Z_i$ is the posterior probability that SNIa belongs to the high-mass host class
 Although the horizontal error bars are suppressed for clarity, the model fully accounts for measurement errors in the host galaxy mass. (This plot assumes the \LCDM.)}
\label{fig:Zi}
\end{figure}

The posterior constraints for the \SC\ Model are compared with those under the  Baseline Model in Fig.~\ref{fig:comp_baseline_mgalprobclass}. There is not a significant difference between the cosmological fits or in the fitted nuisance parameters of the Baseline, \HC, or \SC\ Models. 

Finally, the \CA\ Model
includes host galaxy mass as a covariate; the fitted regression line under this model is plotted as a solid purple line in Fig.~\ref{fig:galaxymass_split}. The fitted regression line can be expressed as 
$\mBhat{i}-\mu_i=\text{intercept}+\bar\gamma\ \Mgalobs$, where $\bar\gamma$ is the posterior mean of $\gamma$ and
the intercept is  $(\Mnot-\alpha x_1+\beta c)$ with $\Mnot$, $\alpha$, and $\beta$ replaced by their posterior means, $\bar{\Mnot}$, $\bar{\alpha}$, and $\bar{\beta}$; $x_1$ replaced by $\frac{1}{n}\sum_{i=1}^n\xone{i}$;  and  $c$ replaced with $\frac{1}{n}\sum_{i=1}^n{\chat_i}$. 
The shaded purple region corresponds to a 68\% posterior credible interval of $\gamma$ (with the intercept fixed as described above).
Fig.~\ref{fig:gamma_covariate} plots the posterior distribution for the slope $\gamma$.
We find that the posterior probability that $\gamma< 0$ is 99\%.
The posterior 68\% credible interval for $\gamma$ is $-0.030 \pm 0.010$.
This is qualitatively consistent with previous work, but our slope is shallower. Previous analyses~\citep{Childress:2013wna,Gupta:2011pa,2010MNRAS.401.2331L,Pan:2013cva,Campbell2015} (using various SNIa samples) find values of the slope in the range $\gamma = -0.08$ to $\gamma= -0.04$. 

\begin{figure}[tbh]
\centering
  \includegraphics[width=0.48\linewidth]{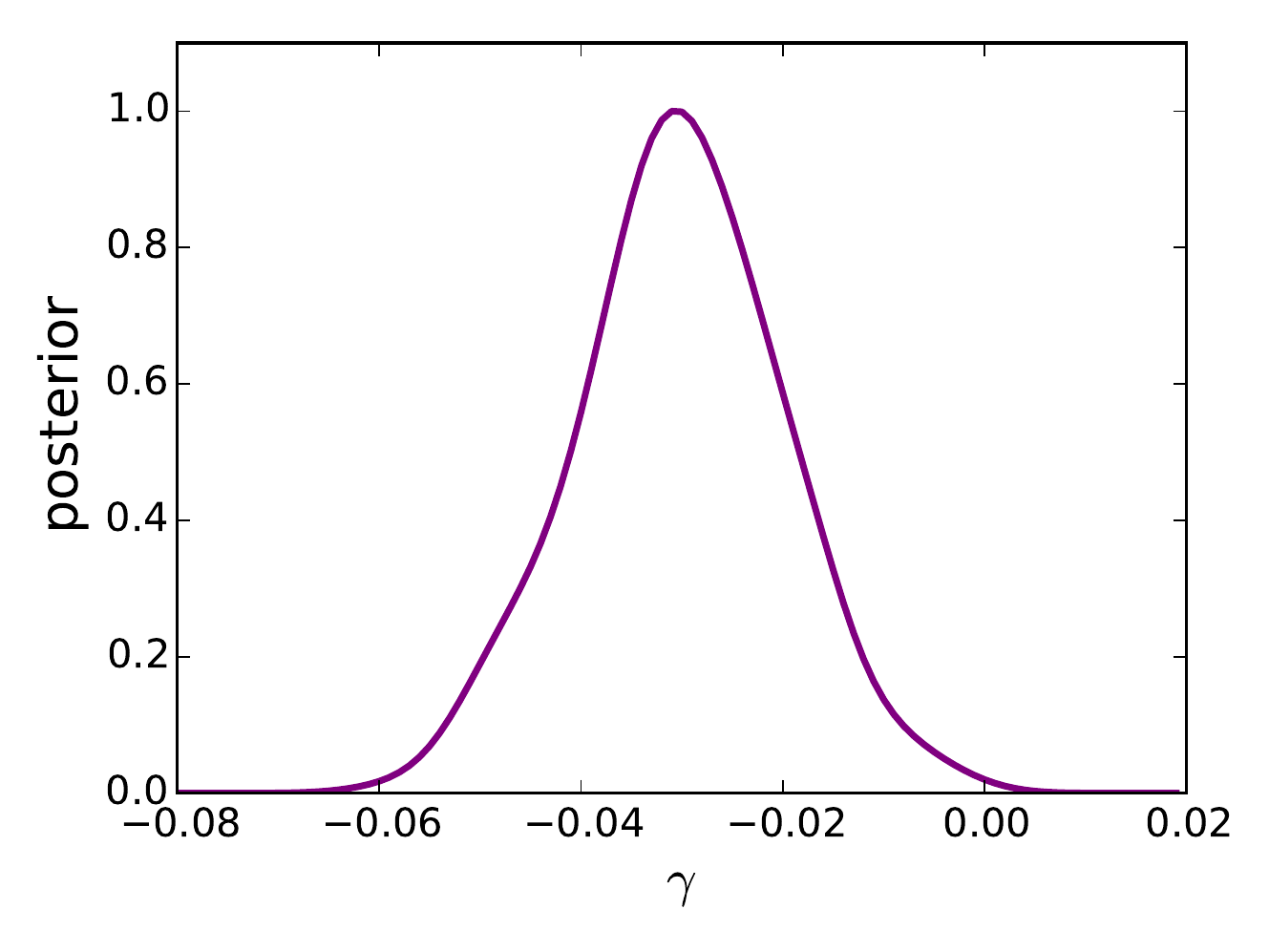}
  \caption{Marginal posterior distribution for $\gamma$, the regression coefficient for $\Mgal$ in the \CA\ Model.
(The model is fit assuming a \LCDM\ universe.) The probability that $\gamma$ is less than zero is $99\%$.}
\label{fig:gamma_covariate}
\end{figure}

Posterior constraints under the \CA\ Model are compared with those under the Baseline Model in Fig.~\ref{fig:comp_baseline_mgalprobclass}.  Despite the fact that the posterior probability that $\gamma <0$ is 99\%, there is not a significant shift in the cosmological parameters or the residual standard deviation, $\sigmares$. Although intuition stemming from standard linear regression suggests that adding a significant covariate should reduce residual variance, the situation is more complicated in Eq. \eqref{eq:covariates_relation_generalized} owing to the measurement errors in both the independent and the dependent variables. While the variances of the left and right sides of \eqref{eq:covariates_relation_generalized} must be equal, there are numerous random quantities whose variances and covariances can be altered by adding a covariate to the model.

\subsection{Redshift Evolution of the Color Correction} 

We now examine possible redshift evolution of the color correction parameter. The posterior distributions of the cosmological parameters under the Baseline, \betalin, and \betastep\ Models are compared in Fig.~\ref{fig:betaw} (\LCDM) and Fig.~\ref{fig:betaflat} (\wCDM). The corresponding marginal posterior constraints are reported alongside the Baseline Model in Tables~\ref{tab:1d_constraints_wm1} and \ref{tab:1d_constraints_flat}.

\begin{figure}[tbh]
\centering
 \includegraphics[width=0.40 \linewidth]{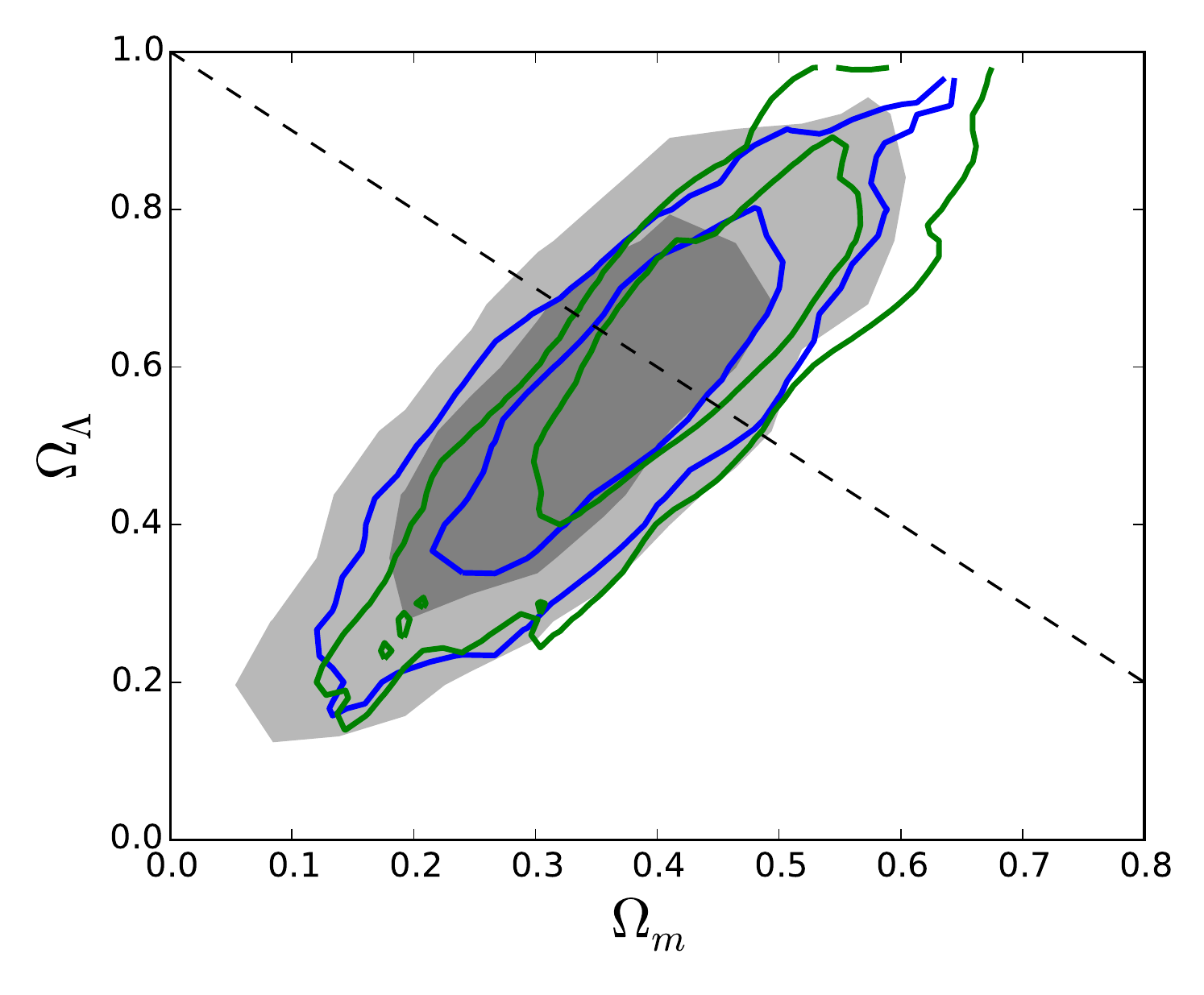}\\
 \includegraphics[width=0.80 \linewidth]{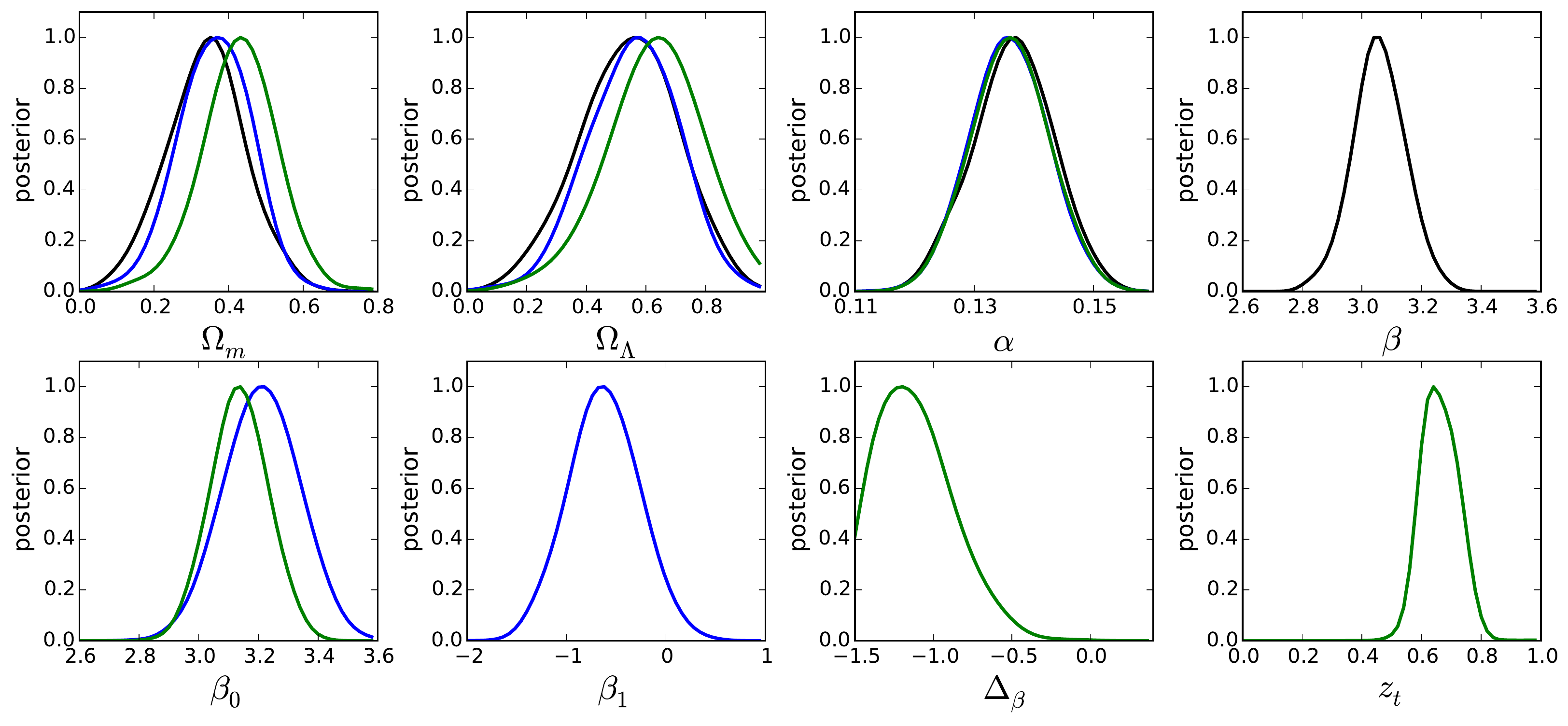} 
  \caption{Comparisons of the posterior distributions for the cosmological parameters and the standardization parameters under different models for the color correction parameter: Black: Baseline Model (no evolution); blue: \betalin\ model; green: \betastep\ model. Posterior are normalized to the peak. (All models are fit assuming a \LCDM\ universe). }
\label{fig:betaw}
\end{figure}

\begin{figure}[tbh]
\centering
 \includegraphics[width=0.40 \linewidth]{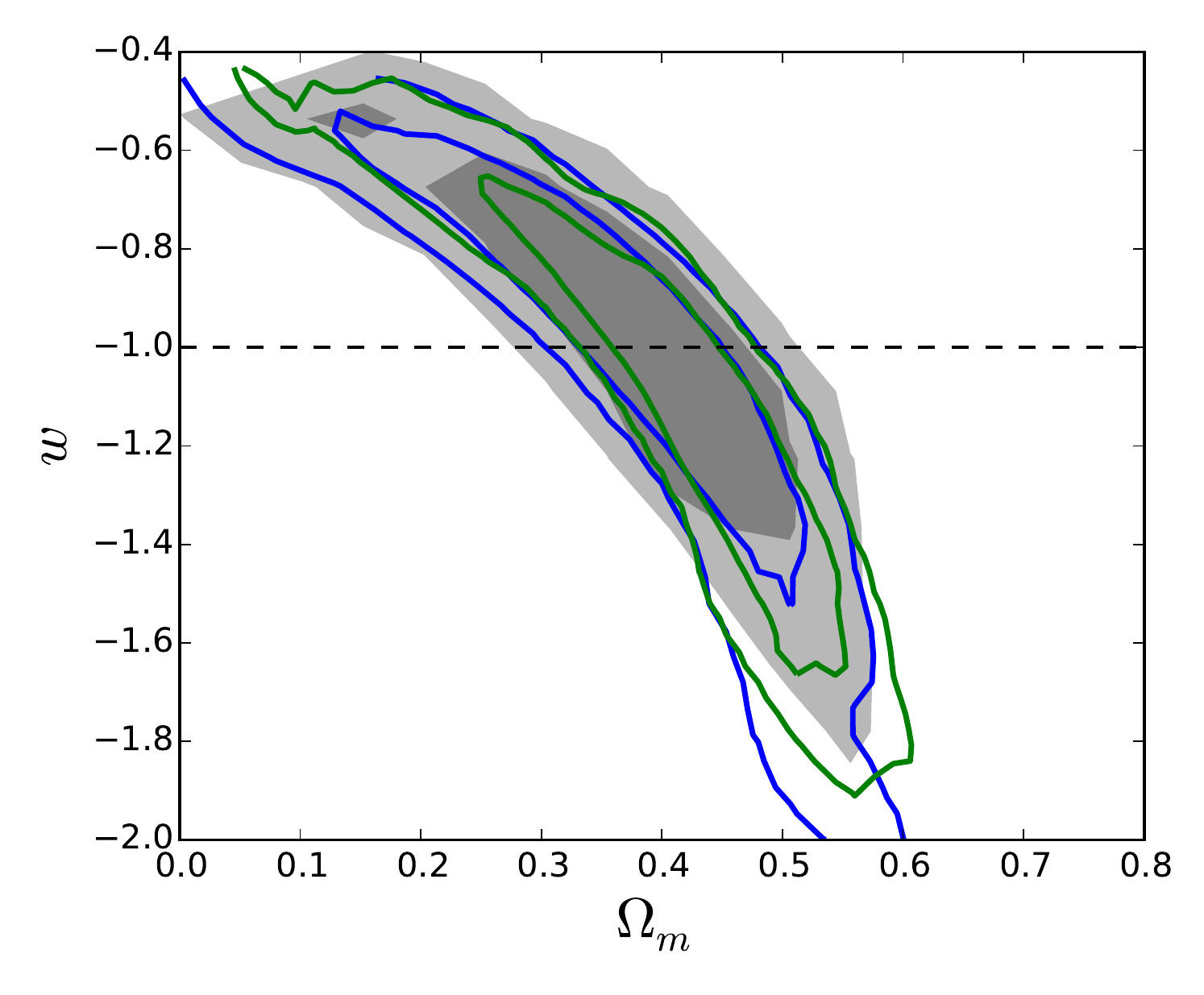}\\
 \includegraphics[width=0.80 \linewidth]{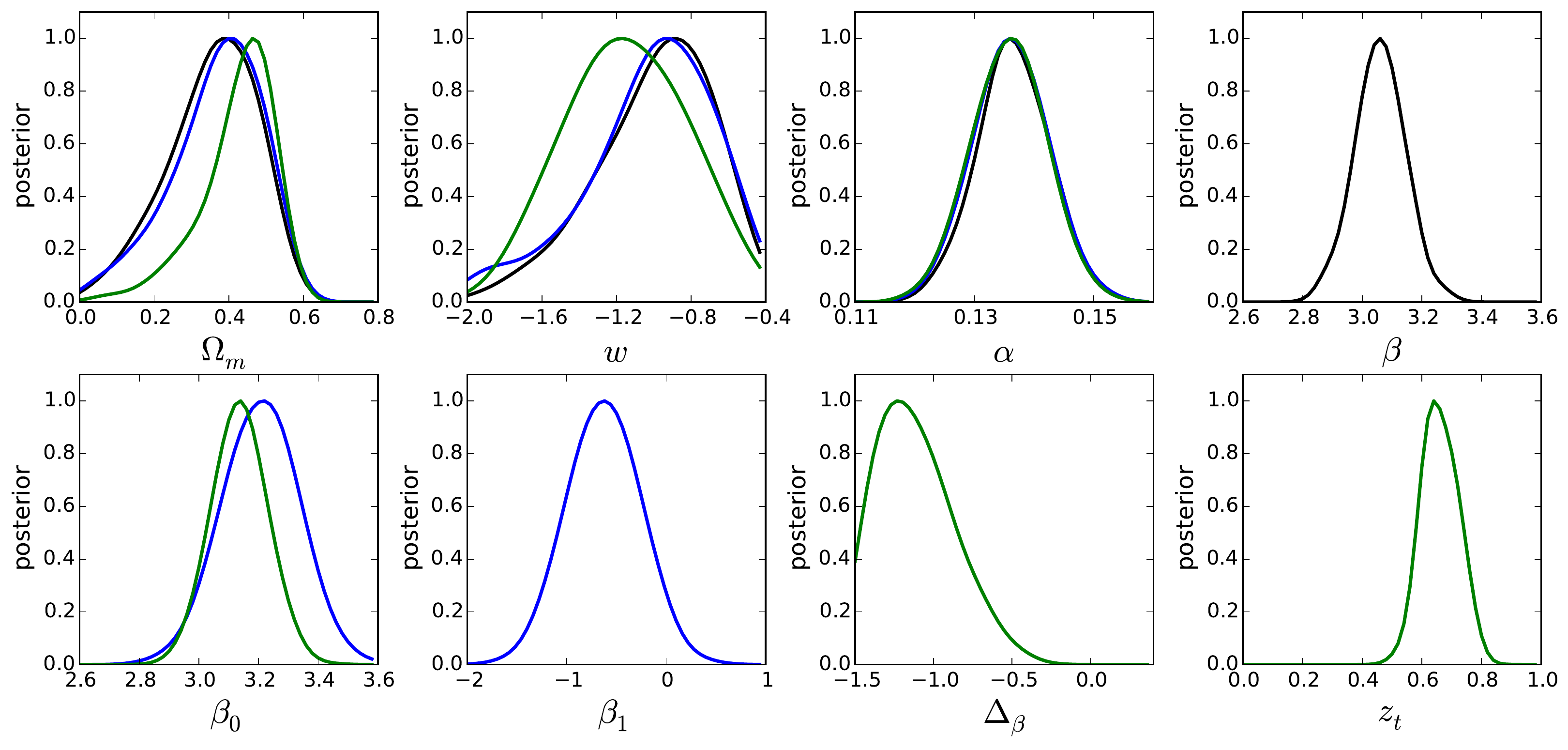}
  \caption{Comparisons of the posterior distributions for the cosmological parameters and the standardization parameters under different models for the color correction parameter: Black: Baseline Model (no evolution); blue: \betalin\ model; green: \betastep\ model. Posterior are normalized to the peak. (All models are fit assuming a \wCDM\ universe.)}
\label{fig:betaflat}
\end{figure}

When evolution that is linear in redshift is allowed (as in the \betalin\ Model), we find that a non-zero, negative linear term $\beta_1$ is preferred with $\sim 95\%$ probability, $\beta_1 = -0.622 \pm 0.342$ (JLA data only). Because the standard deviation of $\chat_i$ is of order $\sim 0.1$, high-redshift SNIa's (at $z \sim 1$) are typically $\sim 0.06$ mag brighter than those nearby. However, there is not a significant shift in the ensuing distributions of the cosmological parameters when compared with the Baseline Model.

When a sharp transition with redshift is allowed  (as in the \betastep\ Model), there is strong evidence for a significant drop in $\beta$ at $z_t = 0.66 \pm 0.06$.  At this redshift, $\beta$ drops from its low-redshift value, $\beta_0 = 3.14 \pm 0.09$ by $\Delta\beta = -1.12 \pm 0.24$, with a nominal significance of approximately $4.6\sigma$. This represents a correction of typically $\sim 0.11$ mag for SNIa's at $z>z_t$.  
The mean value and 1$\sigma$ uncertainty band in the redshift-dependent $\beta(z)$ are shown in Fig.~\ref{fig:betaflat}. This trend is qualitatively similar to what is reported in~\cite{Kessler:2009ys}, which attributed the shift to an unexplained effect in the first-year SNLS data. \cite{Wang:2013tic} also found evidence for evolution of $\beta$ with redshift in the SNLS 3-yr data. The drop however disappears in~\cite{Betoule:2014frx}, after their re-analysis of the (three-years) SNLS data.  The present work however uses identical data to~\cite{Betoule:2014frx}. This is further discussed at the end of this Section.

\begin{figure}[tbh]
\centering
 \includegraphics[width=0.80 \linewidth]{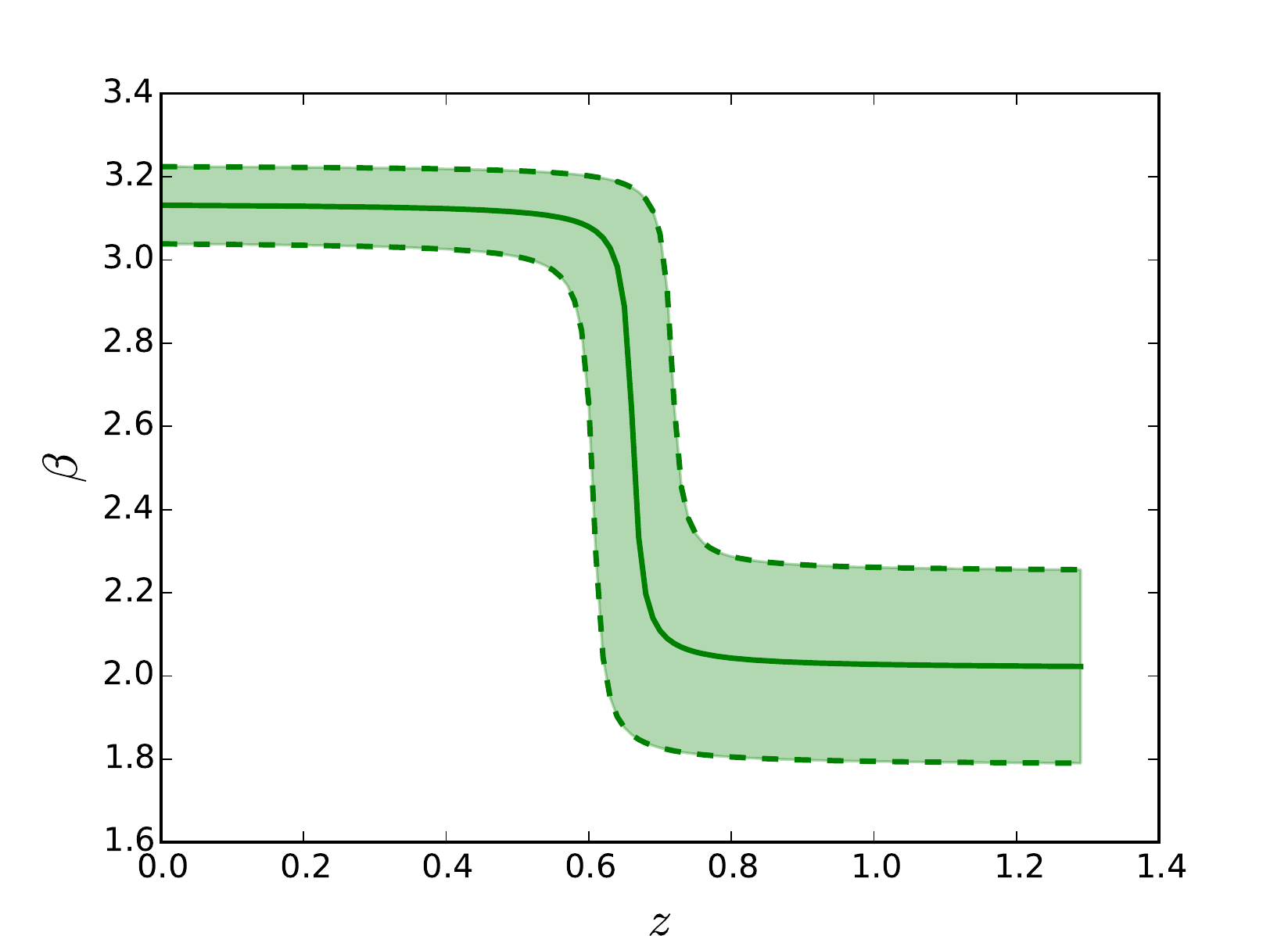}
  \caption{Redshift evolution of the color correction parameter $\beta$, assuming the \betastep\ model. The green line is the posterior mean, while the shaded region represents the $1\sigma$ credible region. 
 (\LCDM\ case). }
\label{fig:betaflat}
\end{figure}

Despite significant evidence for redshift evolution of the color correction, the cosmological parameters are only mildly affected  with respect to the Baseline Model. (Differences between the two fits are one standard deviation or less.)
The posterior distribution of the residual intrinsic scatter also remains unchanged,  giving $\sigmares = 0.103 \pm 0.005$. 

In order to quantify the residual scatter around the Hubble diagram, we consider the difference between the theoretical distance modulus, $\mu(\zhat_i; \Cparams)$, and an estimate based on the observables, $\hat\mu_i(M_0,\alpha, \beta)=\omb-M_0+\alpha\ox-\beta\oc$; that is, we define
\be \label{eq:residual_distance_modulus}
\remu = \hat\mu_i( M_0, \alpha, \beta)-\mu(\zhat_i; \Cparams)
\ee 
and its sample variance,
\begin{equation}
\sremu  = \frac{1}{n-1}  \sum_{i=1}^n {(\remu-\mremu)}^2,
\end{equation}
where $\mremu=\frac{1}{n}\sum_{i=1}^n \remu$.
Notice that both $\hat\mu_i( M_0, \alpha, \beta)$ and $\mu(\zhat_i; \Cparams)$ depend on model parameters and thus for fixed $\saltdata$, we can view $\remu$ and $\sremu$ as functions of the parameters having their own posterior distributions. 

We compare the Hubble diagram residuals, $\remu$, for the Baseline Model, with those for the \betastep\ Model in Fig.~\ref{fig:residuals}. The unknown parameters in $\remu$ are replaced with their posterior means. We only plot SNIa's with $\zhat>0.6$, because the residuals for low-redshift SNIa's are very similar for the two models since the $\beta$ value for $\zhat<0.6$  is similar. The left panel of Fig.~\ref{fig:residuals} shows the Hubble residuals under the Baseline Model; the central panel shows them under the \betastep\ Model; and the right panel compares the two by plotting residuals under the Baseline Model versus residuals under the \betastep\ Model. The scatter is reduced under the \betastep\ Model; it is nearer zero. This indicates that allowing for a sharp transition in $\beta(z)$ improves the standardization of SNIa's. 

\begin{figure}[tbh]
\centering
 \includegraphics[width=0.99\linewidth]{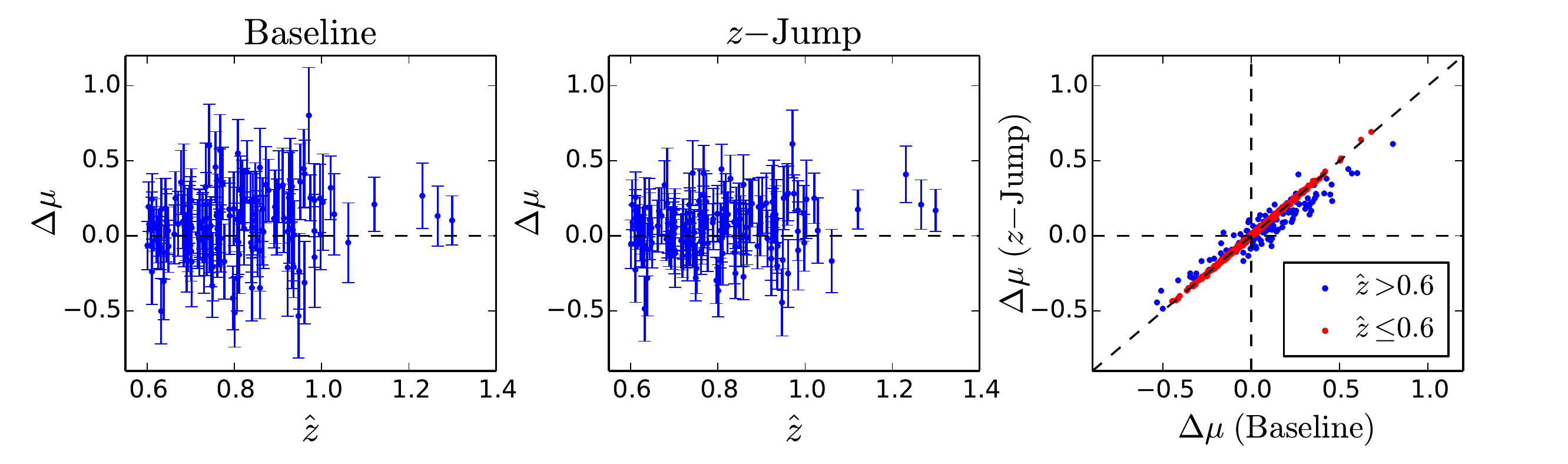}
  \caption{Hubble residuals of the Baseline Model (left, $\beta=\text{constant}$ with redshift), \betastep\ model (centre) and comparison between the two (right). In the left and central panels, only SNIa's with $\zhat>0.6$ are plotted to highlight the difference between the two cases. Errorbars are the posterior standard deviations of $\remu$. In the right panel, SNIa's with $\zhat \leq 0.6$ are plotted in red. This panel shows that the \betastep\ model reduces the scatter around the Hubble diagram noticeably for $\zhat>0.6$, while its Hubble residuals are similar to the Baseline Model for $\zhat \leq 0.6$ (this plot is for \LCDM, and the \wCDM\ case is similar). \label{fig:residuals}}
\end{figure}

We define the cumulative (i.e., summed over redshift) Hubble residual as 
\be
\cumuremun{i} = \sum\limits_{\hatz{j}\le \hatz{i}} |\remun{j}| \quad (1 \leq i \leq \NSN).
\ee
In Fig.~\ref{fig:cumulative_residuals} we use the cumulative residual to highlight the difference in the fit between the Baseline, \betalin\, and \betastep\ Models.  Fig.~\ref{fig:cumulative_residuals} shows the cumulative residual as a function of redshift, where at each redshift the Baseline Model residual has been subtracted to facilitate comparison. For $z \lesssim 0.7$, the Baseline Model offers a slightly better fit than either of the $\beta(z)$ models. But above $z\sim 0.8$ both the \betalin\ and especially the \betastep\ Model provide improved residuals with respect to the Baseline Model. This is shown by their negative values for the relative residual with respect to the Baseline Model. In other words, Fig.~\ref{fig:cumulative_residuals} shows that either of the $\beta(z)$ models improves the fit for high-redshift SNIa's. 
{Although it is beyond the scope of this paper and subject future investigation, formal model comparison should be deployed to weigh the evidence for the evolving color correction model relative to the Baseline model.}

\begin{figure}[tbh]
\centering
 \includegraphics[width=0.85\linewidth]{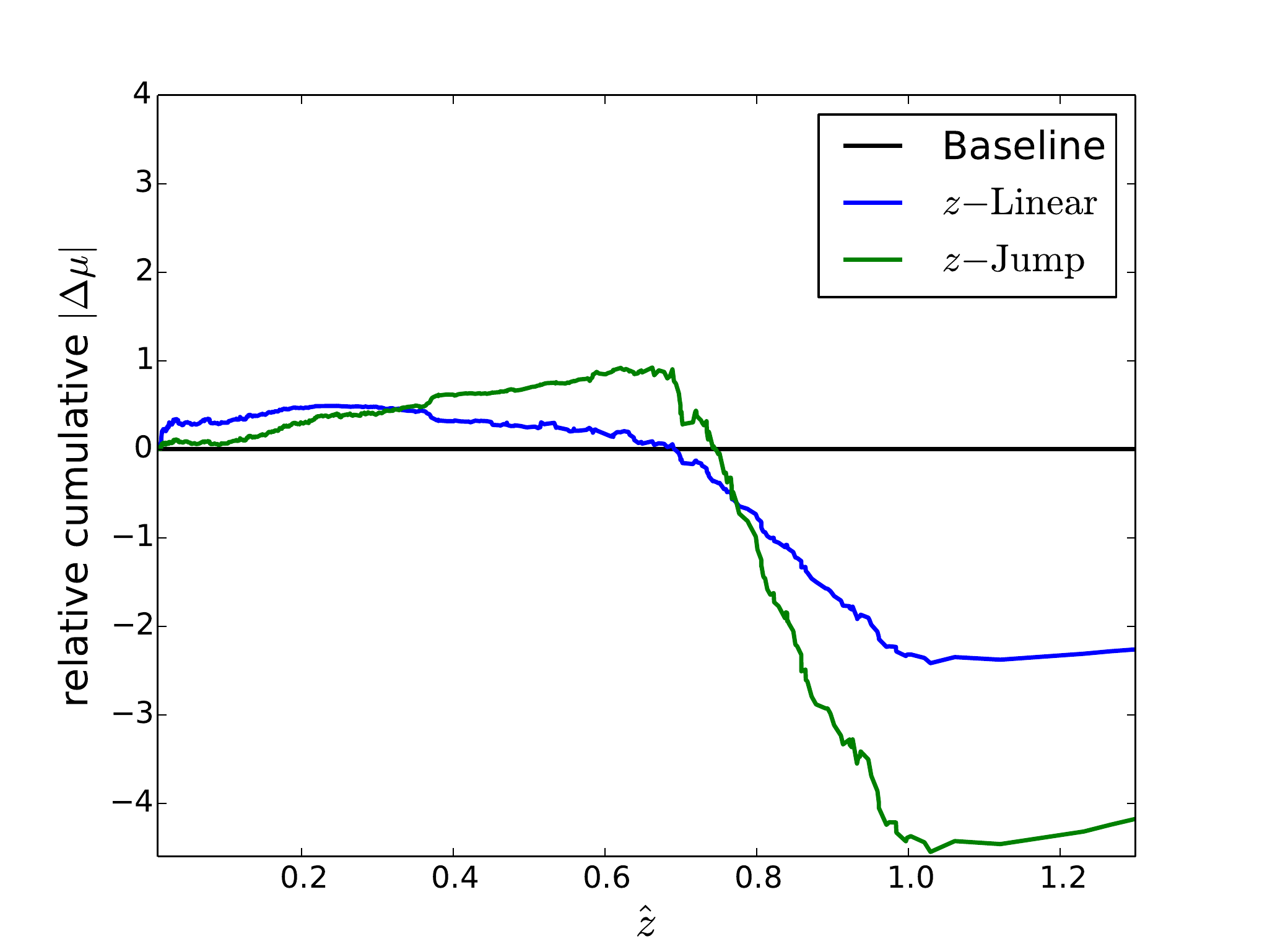}
  \caption{Cumulative Hubble residuals relative to the Baseline Model for the two $\beta(z)$ models considered. For $z \gtrsim 0.8$, both the redshift-dependent models improve the fit with respect to the Baseline Model, which has $\beta = \text{constant}$.  The \betastep\ model shows the largest improvement in the fit. This plot is for \LCDM, but the \wCDM\ case is qualitatively similar. }
\label{fig:cumulative_residuals}
\end{figure}

It is conceivable that the evidence for a step in the evolution of $\beta(z)$ is a spurious consequence of the mass-step correction, which is not included in the above analysis. Since more massive ($\Mgal > 10$) host galaxies are preferentially found at low redshift, and SNIa's in those galaxies are brighter (see Sec.~\ref{sec:galaxymass}), it is possible that such galaxies require on average a smaller color correction than SNIa's in galaxies at high redshift (which are on average less luminous). However, if such a color-mass-redshift interaction were to exist, it could be identified by fitting a model that allows for both a host galaxy mass correction and evolution in the color correction. To investigate this possibility, we fitted a model that included both a mass-step correction (as parameterized in the \HC\ Model) and the \betastep. The posterior constraints on all the model parameters variables change negligibly in this fit compared with the fit of the \betastep\ Model without mass-step correction.

Our result is in stark contrast with~\cite{Betoule:2014frx}, who found no significant departure of $\beta$ from a constant. The dependency of color correction reconstructions on the assumptions of the color scatter model used for \salt{} training has been extensively investigated in~\cite{Mosher:2014gyd}. This study found significant bias (up to $\sim 0.6$) in the reconstructed value for $\beta$ when the underlying color scatter model was misspecified in the reconstruction. However, \cite{Mosher:2014gyd} showed that the reconstructed $\beta$ (constant with redshift) is biased down (in the cases they considered), that is, in the opposite direction of what we observe. This appears to rule out a misspecification of the color scatter model as an explanation for our result. \cite{Mosher:2014gyd} also demonstrated that a color misspecification does not appreciably bias the recovered cosmological parameters. However, they did not investigate a possible $z$-dependence of the recovered $\beta(z)$ value. 
{\cite{Wang:2013yja} analysed the SNLS3 SNIa sample using different parameterisations of the possible redshift dependence of $\beta$, including a linear dependence. They found that $\beta$ increases significantly with redshift, again in contrast to what is seen in our analysis of the JLA data. \cite{Mohlabeng:2013gda} similarly applied a linear $z$-dependence model for $\beta$ using the Union 2.1 SNIa compilation. They found a $7 \sigma$ deviation from a constant $\beta$, with a trend to smaller $\beta$ at larger $z$, similar to our findings.}

The top panel in Fig.~11 in ~\cite{Betoule:2014frx} might suggest that un-modelled selection effects on the color correction at $z\gtrsim 0.6$ could lead to our detection of a drop in the value of $\beta(z)$ in that range. To test this possibility, we have artificially corrected the trend to negative colors (as seen in Fig.~11 of~\cite{Betoule:2014frx}) for $z>\ztrans$, and re-fitted the \betastep\ model. We found that this correction alters the posterior distributions of the cosmological parameters very significantly, while leaving the strong detection of a jump in the value of $\beta(z)$ largely unchanged. This argues against the existence of un-modelled color correction selection effects causing the observed jump in $\beta(z)$ in the \betastep\ Model. By the same token, it is unlikely that our result is driven by the redshift evolution of the color (or stretch) correction, as a consequence of selection effects, as seen e.g.~for SNLS 1-year data in~\cite{Astier:2005qq}. 

In all of our models above, the population mean and variance of the color and stretch corrections are assumed to be redshift-independent. However, the observed color corrections drift towards the blue near the magnitude limit of a survey (i.e., to larger $z$). This happens because intrinsically brighter SNIa's (which are more likely to be observed) are bluer in color. This selection effects thus leads to a $z$-dependency of the observed color correction, even if the underlying color does not change with redshift. 
We allowed the population mean and variance of the color correction to differ for low-redshift ($z < 0.66$) and high-redshift ($z \geq 0.66$) SNIa. (The threshold of $z=0.66$ was chosen because it is the posterior mean of the jump location in the \betastep\ Model.)  With this change, we re-fit both the Baseline Model and the \betastep\ Model. The joint posterior distribution of $\OmM, \OmL$ shifts appreciably toward lower matter and lower cosmological constant values, but the evidence for a drop in $\beta$ persists. This shows that \BAHAMAS\ results are sensitive to the detailed modelling of a potential redshift-dependency (induced by selection effects, or otherwise) of the color correction. However, the model for the redshift dependence of color is not what is driving the shift in the posterior distribution of $\Omega_m$ toward higher values. 
We will further investigate this aspect in future work by including an explicit model of selection effects in \BAHAMAS

\subsection{Influence of the Systematics Covariance Matrix}

To assess the relative importance of the statistical and systematics variance-covariance matrices in our results, we re-fit the Baseline Model with the statistical covariance matrix only, thus omitting $C_{\rm syst}$. The resulting posterior distributions of $(\OmM, \OmL)$ (for \LCDM) and  $(\OmM,w)$ (for \wCDM) are shown in Fig.~\ref{fig:stat_vs_sys}. Fig.~\ref{fig:stat_vs_sys} compares this fit with the previous Baseline Model that includes the systematics covariance matrix. Adding the systematics covariance matrix not only enlarges the size of the contours -- as one expects -- but also significantly shifts the mean value of the posterior distribution of $\OmM$ to larger values, which leads to a smaller $\OmL$ (for \LCDM) and a larger $w$ (for \wCDM). In fact, the posterior means we obtain when neglecting the systematics covariance matrix are broadly compatible with standard results. The Bayesian approach of~\cite{2011MNRAS.418.2308M} is similar to \BAHAMAS\ and produced results comparable to $\chi^2$ fitting on the data set of ~\cite{Kessler:2009ys}; this analysis did not contain the systematic covariance matrix included in JLA. Thus we are led to conclude that the shift in cosmology is driven by some aspect of the systematics error modeling in JLA. 
{The systematics covariance matrix derived by ~\cite{Betoule:2014frx} contains contributions from different sources: calibration uncertainty, Milky Way extinction, light-curve model, bias corrections, host relations, contamination and peculiar velocities. Analysing these individually, shows the main driver shifting $\Omega_m$ towards larger values is the calibration uncertainty.}
The large differences in the fitted values for $\OmM$ and $w$ between \BAHAMAS\ and the standard $\chi^2$ have been observed previously in simulations by~\cite{2014MNRAS.437.3298M}.  These authors showed on simulated SNLS 3-year data that the posterior mean of $\OmM$ tends to be biased high (by $ \sim 0.1$), while the $\chi^2$ fit tends to be biased low (by a similar amount). However, \cite{2014MNRAS.437.3298M} also found that such discrepancies largely disappear when the redshift arm of the SNIa sample is extended to lower and higher $z$. 

\begin{figure}[tbh]
\centering
 \includegraphics[width=0.44\linewidth]{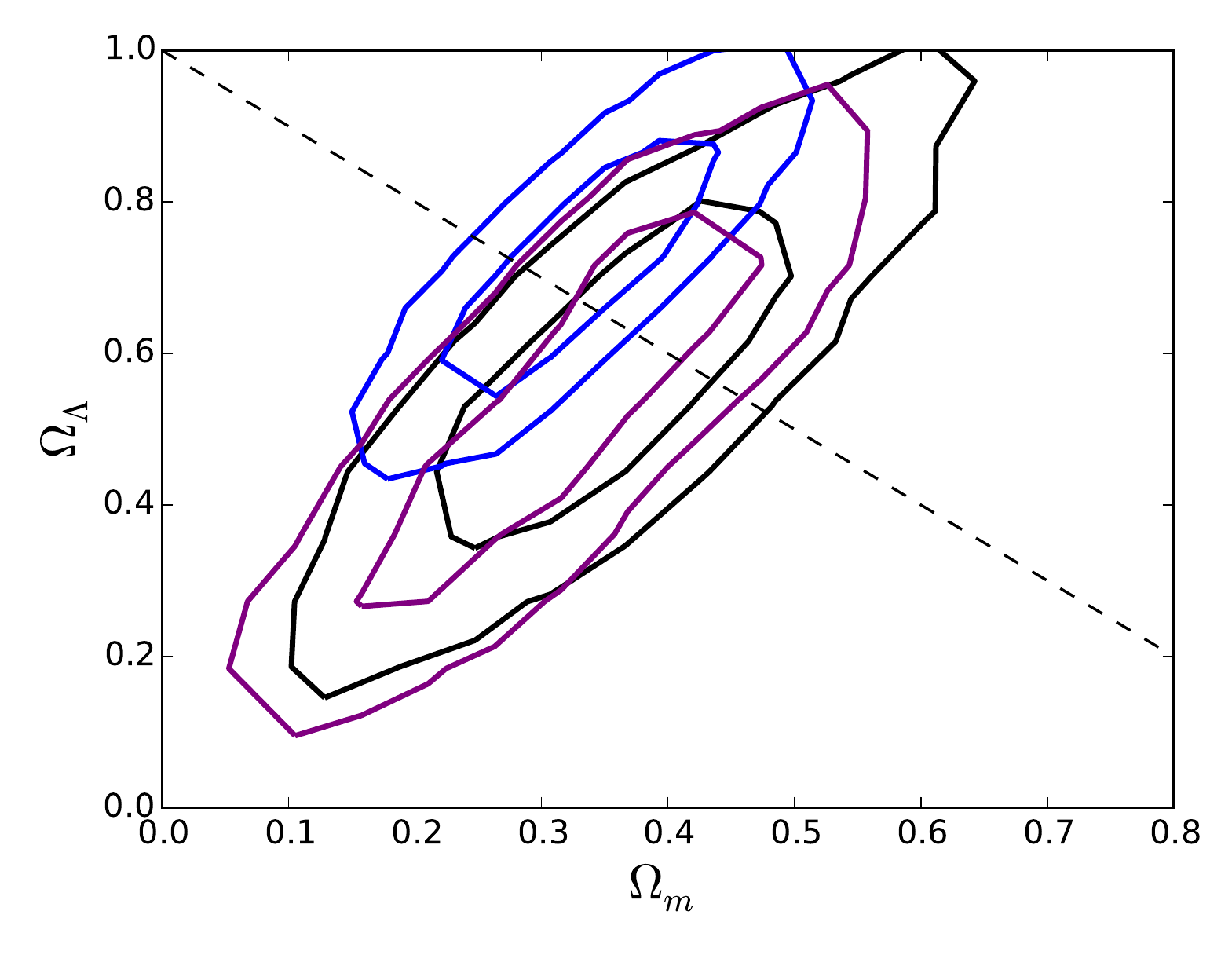}
  \includegraphics[width=0.44\linewidth]{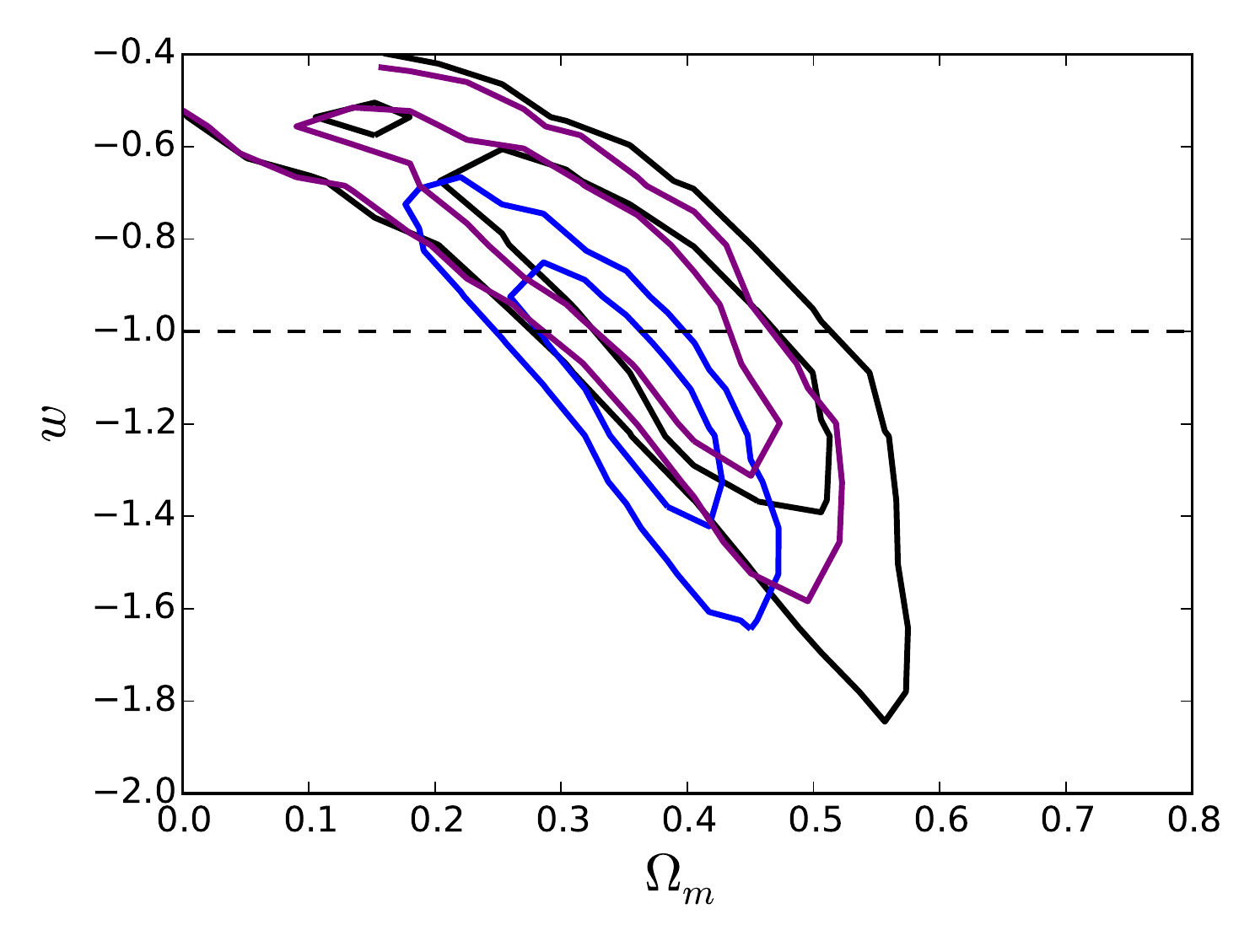}\\
  \caption{Comparison of posterior distributions when including both statistical and systematic errors (black) to the case when the systematics covariance matrix is neglected (blue). Purple: statistics covariance matrix with diagonal errors on $\mB{}$ inflated by the average $\mB{}$ variance from the systematics covariance matrix. Left: \LCDM; Right: \wCDM. }
\label{fig:stat_vs_sys}
\end{figure}

In order to further investigate the origin of the observed shift in the fitted cosmological parameters obtained by \BAHAMAS, we computed the percent increase in the variances of $\mB{}, \xone, c$ when adding the systematics covariance matrix to the statistical covariance matrix, i.e. 
\be
F_i = \frac{1}{\NSN}\sum_{j=1}^{\NSN} \frac{\sigma^{2,\text{syst}}_{i, j}}{\sigma^{2,\text{stat}}_{i, j}}
\ee   
where $i = m$, $x$ or $c$. The quantity $F_i^{1/2}$ is the average percent increase in the standard deviation for observable $i$ when the systematics covariance matrix is added to the statistical covariance matrix (considering diagonal elements only). 
We find $F_{\mB{}}^{1/2} = 2.66$,  $F_{\xone{}}^{1/2} = 0.16$ and  $F_{c}^{1/2} = 0.36$, which shows that the increased error on $\mB{}$ is by far the dominant contribution from the systematics covariance matrix. This is because the dominant source of systematic error in the JLA data is the flux calibration~\citep{Betoule:2014frx}. To check whether the increase in the $\mB{}$ variance is responsible for the fitted cosmological parameters shift,  we multiplied the variance of $\mB{}$ in the statistical covariance matrix by $(1+F_m)$, and refitted (without adding the systematics covariance matrix) our Baseline Model. The resulting cosmological constraints are shown as purple contours in Fig.~\ref{fig:stat_vs_sys}. Comparing with the original Baseline Model fit (black contours), it is clear that most of the shift in the fitted cosmological parameter is due to the large systematic variance of $\mB{}$. If the model were Gaussian and linear, inflating the errors would only enlarge the uncertainty on the parameters, but would not shift the mean of the posterior distribution. Hence we conclude that the cosmology shift is a reflection of the non-Gaussian, non-linear nature of our model, something that is only approximately accounted for in the linear propagation of errors used in standard chi-squared analyses. 

{\subsection{JLA Subsamples}}

{To further investigate the shift in the fitted cosmological parameters and to check for consistency within the JLA SNIa sample, we split the SNIa into a series of sub-samples: low-z+SNLS, SDSS+SNLS, low-z+SNLS+HST, SDSS+SNLS+HST and low-z+SDSS+HST. We did not investigate the low-z+SDSS combination in our analysis as this subsample alone does not have a sufficient redshift range to constrain the  cosmological parameters. In contrast to~\cite{Betoule:2014frx}, we vary both $\Omega_m$ and $\Omega_{\Lambda}$ and do not assume flatness (but we do fix $w=-1$). We compare our results against the entire JLA dataset in~Fig \ref{fig:subsamples}. The left panel shows the results when excluding high-$z$ HST data, while the right panel includes the 9 high-$z$ HST SNIa's.}

\begin{figure}[tbh]
\centering
\includegraphics[width=0.43\linewidth]{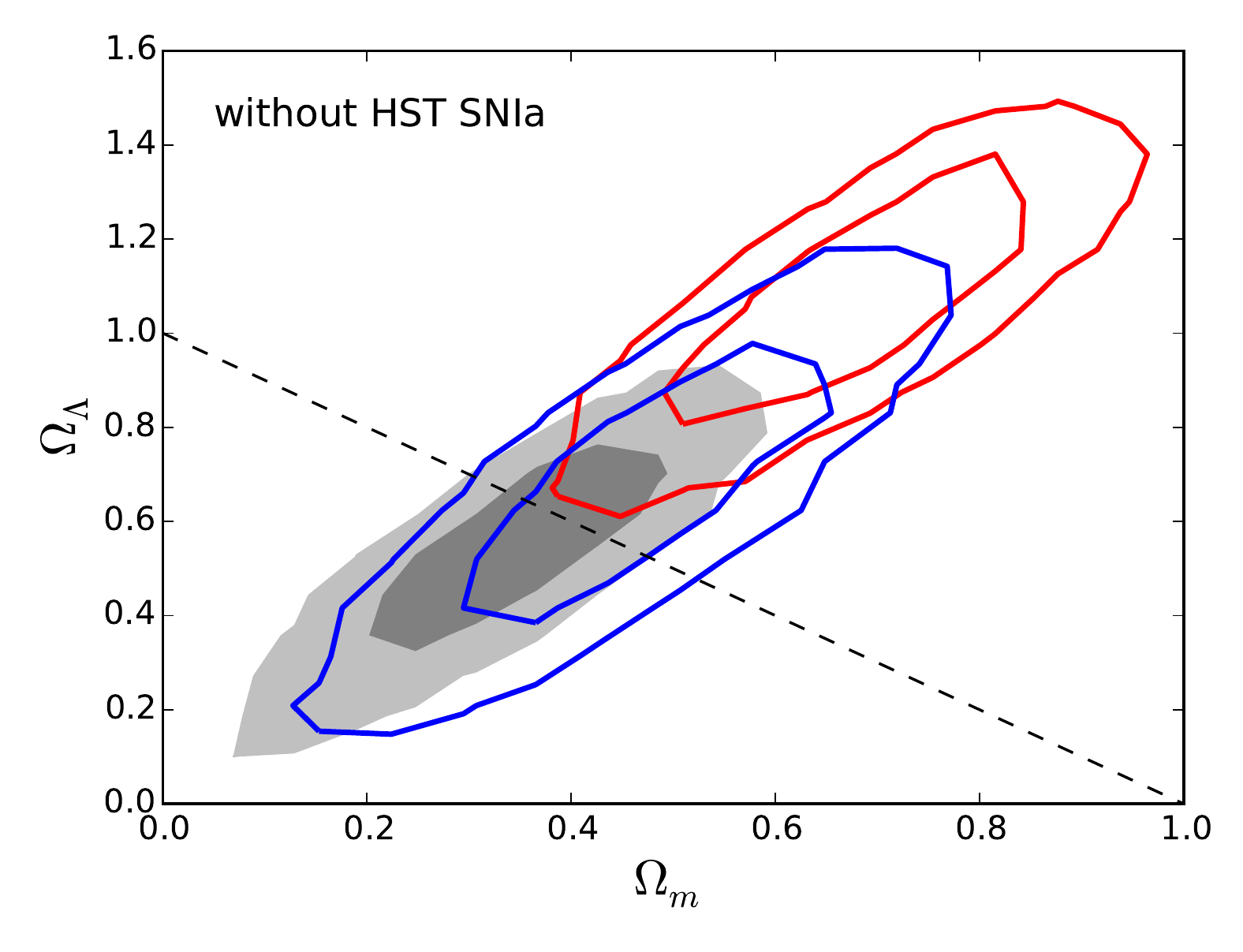}
  \includegraphics[width=0.43\linewidth]{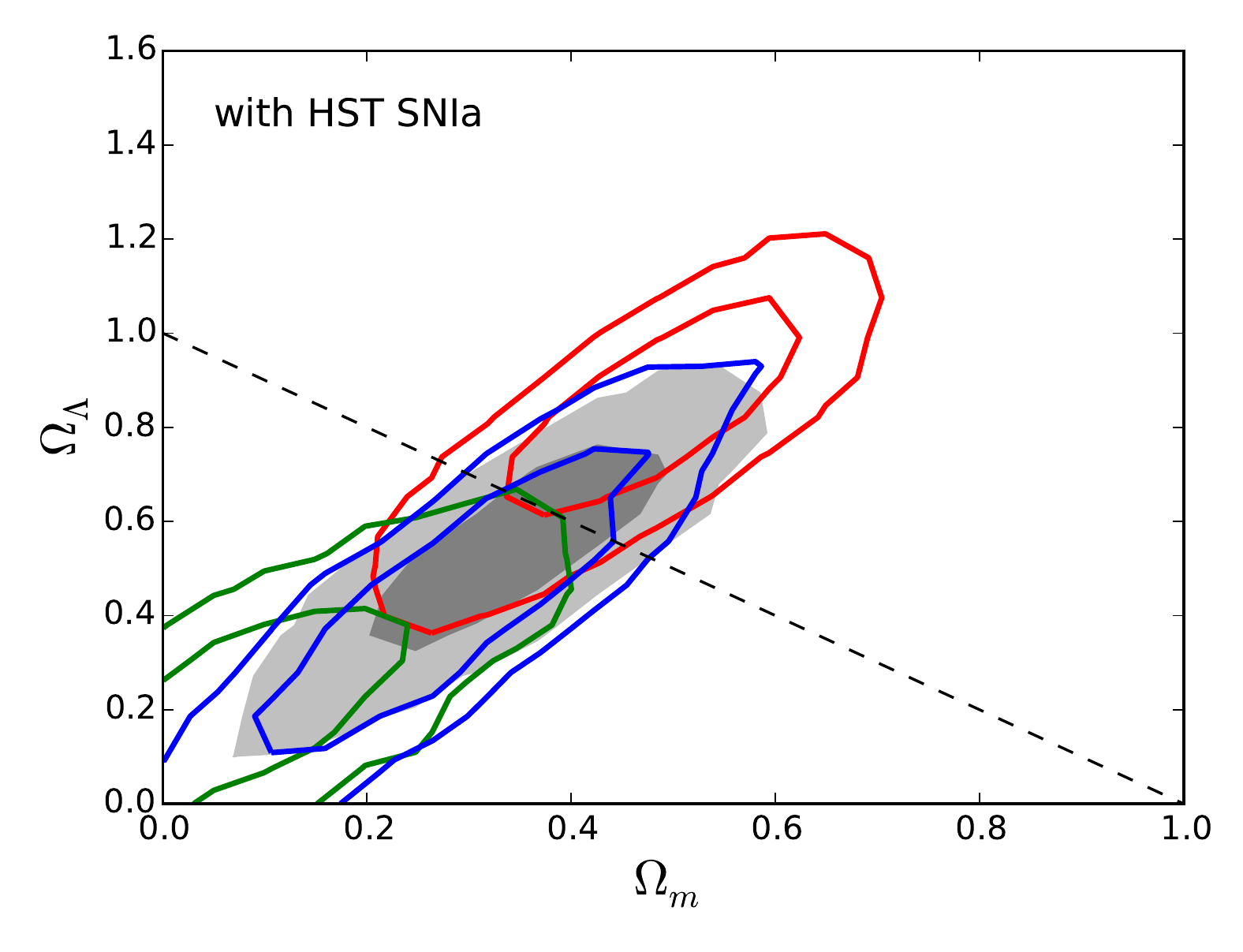}\\
  \caption{{Comparison of posterior distributions for the cosmological parameters in the \LCDM\  case when using different sub-samples of the JLA data, as compared with the result for the entire JLA data set (black/filled). Blue: SDSS+SNLS (613 SNIa, 0.04$<$z$<$1.06), Red: low-z+SNLS (357 SNIa, 0.01$<$z$<$0.08 $\cup$ 0.13$<$z$<$1.06) and Green: low-z+SDSS (492 SNIa, 0.01$<$z$<$0.40). The left plot does not include the HST SNIa (9 SNIa, 0.84$<$z$<$1.30 ) while the right one does.}}
\label{fig:subsamples}
\end{figure}

{In contrast to~\cite{Betoule:2014frx} (see their Table 10), we find significant shifts in the posterior distributions of the cosmological parameters resulting from the different subsamples. The SNLS sample pushes the cosmology toward a closed universe with higher matter and higher dark energy content (an effect previously observed in~\cite{2014MNRAS.437.3298M}) while the HST pulls it in the opposite direction. In particular for the subsample low-z+SNLS (357 SNIa's), including just 9 extra SNIa's from HST shifts the contours very noticeably to much lower values of both $\Omega_m$ and $\Omega_\Lambda$. If we had assumed flatness, as was done in~\cite{Betoule:2014frx}, this effect would have been masked. In all cases in Fig.~\ref{fig:subsamples} if we enforced $ \OmK = 0$, the posterior distribution of  $\Omega_m$ would be similar to the baseline case.
The posterior distributions of the all other parameters for the various subsamples are consistent with each other (hence not shown), except for the low-z+SNLS subsample for which both $\beta$ and $\sigma_{res}$ are smaller. 
This is consistent with the observed redshift dependence of $\beta$, Fig.~\ref{fig:betaflat}.
}

\section{Conclusions}
\label{sec:conclusions}

We have reanalyzed the JLA SNIa data with a principled Bayesian method (\BAHAMAS). As shown in~\cite{2011MNRAS.418.2308M}, our approach has better statistical coverage and smaller mean squared errors than the standard $\chi^2$ method. This paper introduced a series of powerful Gibbs-type samplers that allowed us to explore the posterior distribution of the latent variables associated with SNIa's, such as their empirically corrected intrinsic magnitudes. We have presented a general methodology that can easily incorporate additional standardization variables, over and above the usual stretch and color corrections. We have demonstrated this feature by including host galaxy mass measurements in our fit, fully accounting for the mass measurement uncertainty. 

When the JLA data set is augmented by {\it Planck} CMB data, we find significant discrepancies with the standard $\chi^2$ fit results, in particular in the values of $\OmM$ and $w$.  We measure the average residual dispersion of the post-correction intrinsic magnitudes in the JLA sample to be $\sigmares = 0.104 \pm 0.005$. The magnitude of the host galaxy mass correction is smaller than previously reported. We find significant statistical evidence for a drop in the value of the color correction parameter, $\beta$, at a redshift $\ztrans = 0.66$. While we rule out color-dependent selection effects as responsible for this feature, we cannot trace it back to its origin. Cosmological parameter constraints remain however unaffected by marginalization over this non-standard redshift dependency. 

Future work will incorporate selection effects into our framework (similarly to~\cite{Rubin:2015rza}), include additional covariates (such as star formation rate and metallicity) and test their influence on the recovered cosmology, and allow for the possibility of contamination (as in the BEAMS scenario, ~\cite{Kunz:2006ik,Knights:2012if,Hlozek:2011wq}).

{\it Acknowledgements.} We would like to thank Bruce Bassett, Heather Campbell, Alex Conley, Suhail Dhawan, Josh Frieman, Mike Hobson, Natasha Karpenka, Rick Kessler, Ofer Lahav, Michelle Lochner, Kaisey Mandel, Marisa March, Bob Nichol, Cassio Pigozzo, Mickael Rigault, Mathew Smith, and Mark Sullivan for useful discussions.  We thank an anonymous referee for helpful suggestions. We would like to thank Sakina Ali-Khan for help with the color correction tests, and Cassio Pigozzo for help with checks using {\sc MultiNest}. RT acknowledges partial support from an EPSRC ``Pathways to Impact'' grant and DvD from a Wolfson Research Merit Award (WM110023) provided by the British Royal Society and from Marie-Curie Career Integration (FP7-PEOPLE-2012-CIG-321865) and Marie-Skłodowska-Curie RISE (H2020-MSCA-RISE-2015-691164) Grants both provided by the European Commission. This work was supported by Grant ST/N000838/1 from the Science and Technology Facilities Council.

\newpage
\appendix
\centerline{\Large\bf Appendices} 

\section{Algorithm review}
\label{apsec:algo}
The Gibbs samplers~\citep{gema:gema:84} and Data Augmentation (DA) algorithm~\citep{tann:wong:87}, which is a special Gibbs sampler, are
widely used Markov Chain Monte Carlo (MCMC) methods to sample from highly structured models. Although they are typically easy to implement, they can have slow convergence rates. To improve their convergence, a variety of extensions have been proposed. Among them, the Ancillarity-Sufficiency Interweaving Strategy (ASIS)~\citep{yu:meng:11} is designed to improve the convergence properties of the DA algorithm, and the Partially Collapsed Gibbs (PCG) sampling~\citep{vand:park:08} is a useful tool to improve the convergence of Gibbs samplers. In a Gibbs-type sampler, we may also need the help of Metropolis-Hastings (MH) algorithm~\citep{hast:70,metro:53}, when one of the component conditional distributions is not standard. 

Consider a generic observed data set, $\yobs$, and model parameters, $\theta$, and suppose we wish to sample from the posterior distribution $p(\theta|\yobs)$. When direct sampling is not possible, we may consider introducing a latent variable, $\ymis$, into the model such that the complete-data model $p(\ymis,\yobs|\theta)$ maintains the target posterior, $p(\yobs|\theta)$, as its marginal distribution.
The DA algorithm proceeds by drawing from $p(\ymis|\theta,\yobs)$ and $p(\theta|\ymis,\yobs)$ iteratively. This is a useful strategy when these two distributions are easy to sample and the resulting MCMC is relatively quick to converge.

More generally, when the unknown quantity in a model, $\psi$, consists of two or more components, each of which can be multivariate, that is, $\psi=(\psi_1,\dots,\psi_N)$ with $N\ge 2$, the Gibbs sampler is useful to draw from $p(\psi|\yobs)$. In one iteration of a Gibbs sampler, each component of $\psi$ is sampled from its \emph{complete conditional distribution}, i.e., its distribution conditioning on the current values of all the other components. In this paper we only consider systematic-scan Gibbs samplers \citep{liu:wong:kong:95}, that is, in each complete iteration, the components are updated in a fixed ordering. The DA algorithm is a special case of the Gibbs sampler with two components in $\psi$, i.e., $\psi=(\theta,\ymis)$. 

As mentioned above, although they are easy to implement, in some cases the DA algorithm or Gibbs sampler can be  slow to converge. We now describe two strategies that can significanlty improve their convergence, ASIS and PCG, along with the MH algorithm.  

\paragraph{Ancillarity-Sufficiency Interweaving Strategy.} ASIS improves the convergence of a standard DA algorithm by using a pair of special DA schemes. One is the sufficient augmentation $\symis$, which means the conditional distribution $p(\yobs|\symis,\theta)$ is free of $\theta$. The other is the ancillary augmentation $\aymis$, for which $p(\aymis|\theta)$ does not depend on $\theta$. 
Normally, given the parameter, these two augmentation schemes are related via a one-to-one mapping (but see~\citet{yu:meng:11} for an exception). It is usually the case that if the sampler corresponding to one of these two augmentations is fast, the other is slow. ASIS takes advantage of this ``beauty-and-beast'' feature of the two DA algorithms by interweaving steps of one into the other~\citep{yu:meng:11}. The resulting ASIS sampler can substantially outperform both parent DA samplers in terms of convergence, while the additional computational expense is often fairly small.

\paragraph{Partially Collapsed Gibbs Sampling.} The PCG sampler can be effective in improving the convergence of Gibbs samplers. It achieves this goal by reducing conditioning, that is, by replacing some of the complete conditional distributions of an ordinary Gibbs sampler with the complete conditionals of marginal distributions of the target joint posterior distribution~\citep{vand:park:08}. This generally leads to larger variance of the conditional distribution, and hence bigger jumps. A PCG sampler can be derived from a Gibbs sampler via a three-stage process: 
(i) \emph{marginalization},  (ii) \emph{permutation}, and (iii) \emph{trimming}.
Marginalization can significantly improve the rate of convergence, while permutation typically has a minor effect and trimming has no effect \citep{vand:park:08}. Thus, we generally expect the PCG sampler to exhibit better and often much better convergence properties than its parent Gibbs sampler. In fact, \citet{vand:park:08} already give theoretical arguments and~\citet{park:vand:09} give numerical illustrations of the computational advantage of PCG over ordinary Gibbs samplers. Sometimes, the PCG sampler is simply a blocked or collapsed Gibbs sampler~\citep{liu:wong:kong:94}. However, we are more interested in PCG samplers composed of \emph{incompatible conditional distributions}, that is, there is no joint distribution corresponding to this set of conditional distributions. The incompatibility is introduced by trimming; permuting the order of the steps of a PCG sampler consisting of incompatible conditionals will alter its stationary distribution, see \citet{vand:park:08}. 

\paragraph{Metropolis-Hastings Algorithm.} The MH algorithm is frequently used to obtain a correlated sample from a target distribution, $p(\psi|\yobs)$, for which direct sampling is difficult. Suppose we have sampled ${\psi}^{\cutt}$ and need to generate ${\psi}^{\nextt}$. Instead of sampling from $p(\psi|\yobs)$ directly, we generate a candidate value ${\psi}^{c}$ from a proposal distribution $g(\psi|{\psi}^{\cutt})$ and accept it as ${\psi}^{\nextt}$ with probability ${\rm min}(R, 1)$, where $R =\frac{p({\psi}^{c}|\yobs)g({\psi}^{\cutt}|{\psi}^{c})}
{p({\psi}^{\cutt}|\yobs)g({\psi}^{c}|{\psi}^{\cutt})}$. In this way, we construct a reversible Markov chain, $\{\psi^{\cutt}, t=0,1,\dots\}$ with $p(\psi|\yobs)$ as its stationary distribution.  

\bigskip

To further ease implementation and improve convergence properties, we propose to combine several strategies introduced above into one sampler. \citet{jiao:etal:15} uses a simplified version of the hierarchical model described in Section~\ref{subsec:model} as an example to illustrate the efficiency of both PCG and ASIS in improving the convergence properties of Gibbs-type samplers. They find that combining two strategies into one sampler can produce even more efficient samplers. Thus, we use PCG in each of our samplers to improve the convergence properties of $\Cparams$ or $\regcoeff$. In some samplers, we combine PCG and ASIS for better convergence properties. The general method of combining several strategies into one sampler will appear in \citet{jiao:vandyk:15}.

\section{The Posterior Distribution}
\label{apsec:pos}

\begin{table}[htb]
\spacingset{1.2}
\small
\begin{center}
\begin{tabular}{c p{14cm}} 
\hline\hline 
Symbol  & \multicolumn{1}{c}{Description}\\ 
\hline 

$\dobs$ & Column stacked vector of observed quantities, with apparent magnitude corrected for distance modulus, e.g.,  $\dobs=\{\mBhat{1}-\mu_{1}(\zhat_1, \Cparams), \xhat{1}, \chat_1, \dots, 
\mBhat{n}-\mu_{n}(\zhat_n, \Cparams), \xhat{n}, \chat_n\}^T$ in the Baseline model\\

$\dmis$ & Column stacked vector of latent variables, e.g., $\dmis=\{M_1\ep, \xone{1}, c_1, \dots, M_n\ep, \xone{n}, c_n\}^T$ in the Baseline model\\

$\dmean$  & Vector of population means of the latent variables in $\dmis$, e.g.,  $\dmean=\{\Mnot\ep, \xstar, \cstar\}^T$ in the Baseline model\\

$\dprior$  & Vector of prior means of quantities in $\dmean$, e.g., with priors given in Table~\ref{table:main_params}, $\dprior=\{-19.3, 0, 0\}^T$ in the Baseline model\\

$\sigobs$ & Matrix of variances (uncertainties) of observed quantities in $\dobs$, compiled using $\sigsalt = C_{\rm stat} +C_{\rm syst}$, see Section~\ref{sec:system}\\

$\sigmis$ & Population variance-covariance matrix of latent quantities in $\dmis$. This is a block-diagonal matrix composed of $n$ blocks, i.e., $\sigmis=\diag(S_1, \dots, S_n)$. For example, each $S_i=\diag(\sigmares^2, \Rx^2, \Rc^2)$ in the Baseline model\\

$\sigprior$ & Prior variance-covariance matrix of quantities in $\dmean$,  e.g., with priors given in Table~\ref{table:main_params}, $\sigprior=\diag(2^2, 10^2, 1^2)$ in the Baseline model\\

$\Jmat$ & Top-to-bottom stacked matrix of $n$ matrices, i.e., \par
$\Jmat=\scalefont{0.6}{\left[ \begin{array}{c} J_1\\ \vdots\\ J_n \end{array} \right]}$. In the Hard  and \SC\  models, $J_i=\scalefont{0.6}{\left[
\begin{array}{cccc}
Z_i & 1-Z_i & 0 & 0\\
0 & 0 & 1 & 0\\
0 & 0 & 0 & 1
\end{array}
\right]}$, while in the other models, each $J_i$  is an identity matrix
\\

$\Amat$ & Block-diagonal matrix with $n$ blocks, i.e., $\Amat=\diag(T_1, \dots, T_n)$. Each block is composed of $0$, $1$ and elements of $\regcoeff$, e.g., each $T_i=
\scalefont{0.6}{\left[
\begin{array}{ccc}
1 & -\alpha & \beta\\
0 & 1 & 0\\
0 & 0 & 1
\end{array}
\right]}$ 
in the Baseline model\\

$\sigA$ & $\sigA^{-1}=\Amat^T \sigobs^{-1} \Amat + \sigmis^{-1}$\\

$\sigK$ & $\sigK^{-1}=-\Jmat^T \sigmis^{-1}\sigA \sigmis^{-1} \Jmat + \Jmat^T \sigmis^{-1} \Jmat+ \sigprior^{-1}$\\

$\sigdelta$ & $\sigdelta=\Amat^T \sigobs^{-1} \dobs$\\

$\kvec$ & $\kvec=\sigK (\Jmat^T \sigmis^{-1} \sigA \sigdelta + \sigprior^{-1}\dprior)$\\

$\gapara$ & Parameter in the prior $\invgamma$ distribution of $\sigmares^2$, i.e., $\sigmares^2\sim \invgamma(\gapara,\gapara)$\\
\hline \hline
\end{tabular}
\caption{Unified general notation used in the posterior distributions given in Eq. \eqref{eq:posterior} -- \eqref{eq:marginal}. Here we exemplify the general notation for the Baseline model in terms of the notation used in Section~\ref{sec:model}. These details are given for each of the model extensions in Section~\ref{apsubsec:gibbs}.
} 
 \label{table:list_symbol}
\end{center}
\end{table}%

In this section we give explicit expressions for the posterior distributions of the Baseline model and its extensions listed in Table~\ref{tab:model_extensions}. To this end, we introduce a unified and general notation, see Table~\ref{table:list_symbol}.  We start with an expression that covers all of the models we consider, except the \HC\ and \SC\ models. In particular, this formulation covers the regression model given in Eq.~\eqref{eq:covariates_relation_nonlinear} with the population distributions given in Eq.~\eqref{eq:Mdist} -- \eqref{eq:cdist} and \eqref{eq:gdist} and the systematics covariance matrix described in Section~\ref{sec:system}. Under this extended hierarchical model, the posterior distribution is 
{\small
\begin{eqnarray}
p(\dmis,\  \dmean,\  \sigmis,\ \regcoeff, \ \Cparams  | \ \saltdata) & \propto&
{{| \sigobs \sigmis \sigprior|}^{-\half} \over \Rc^2  \ \Rx^2 }  p(\sigmares^2)
\label{eq:posterior}\\
& \times & {\E} \Big\{
 - \half  \big[{(\dobs-\Amat \dmis)}^{T} \sigobs^{-1} (\dobs-\Amat \dmis) \nonumber\\
& + &\ {(\dmis - J \dmean)}^{T}\sigmis^{-1}(\dmis - J \dmean)
+ {(\dmean - \dprior)}^{T}\sigprior^{-1}{(\dmean - \dprior)}\big]
\Big\},\nonumber
\end{eqnarray}}
where $p(\sigmares^2)$ is the prior distribution of $\sigmares^2$ given in Table~\ref{table:main_params} and
the notation is defined in Table~\ref{tab:model_extensions}. The priors for the cosmological parameters,  $\Cparams = \{\OmM, \OmL, w \}$,  the regression coefficients, $\regcoeff = \{ \alpha, \beta, \beta_1, \deltabeta, \gamma, \ztrans\}$, the latent variables, $\dmis$,  their population means, $\dmean$,  and their variances, $\sigmis$, are given in Table~\ref{table:main_params}.

The posterior distribution under the \HC\ model is formally identical to that in Eq.~\eqref{eq:posterior} except that
\be
p(\sigmares^2) \ \hbox{ is replaced by } \ p\left({\sigmareslow}^2\right) p\left({\sigmareshigh}^2\right), 
\label{eq:hcsub}
\ee
with the prior distributions given in  Table~\ref{table:main_params}. The (assumed known) indicator variables, $Z_i$, for low and high host galaxy masses enter through $J$ and $\sigmis$ using the definitions given in Section~\ref{apsubsubsec:massstep}.

For the \SC\ model, SNIa's are classified on their true (latent) host galaxy masses (rather than on their observed masses as in the \HC\ model). Thus, the indicator variables, $Z_i$, are treated as unknown and the posterior distribution, 
$p(\dmis,\  \dmean,\  \sigmis,\ \regcoeff, \ \Cparams, Z  | \ \saltdata, \Mgaldata)$, 
is formally identical to that in Eq.~\eqref{eq:posterior} except 
\be
p(\sigmares^2) \ \hbox{ is replaced by } \ 
p\left({\sigmareslow}^2\right) p\left({\sigmareshigh}^2\right) \prod_{i=1}^n \pi_i^{Z_i}{(1-\pi_i)}^{1-Z_i},
\label{eq:scsub}
\ee 
with the prior distributions given in  Table~\ref{table:main_params},  $\Mgaldata = \{\Mgalobs, i=1,\ldots, n\}$, and
\be
\pi_i=
\Pr(Z_i=1\mid\Mgalobs)=\Pr(\Mgal<10\mid\Mgalobs)=\int_{-\infty}^{10} \frac{1}{\sqrt{2\pi} \sigMgal }\E\left[-{(\Mgal-\Mgalobs)}^2/(2 \sigMgal^2)\right]{\rm d} \Mgal, 
\ee
for $i=1,\dots,n$. The specific definitions of the unified notation for the \SC\ model are given in Section~\ref{apsubsubsec:prob}.

\section{The MCMC samplers}
\label{apsec:sampler}

To obtain posterior draws of all the variables (including latent variables) of the hierarchical models we use Gibbs-type samplers, sometimes augmented with an MH step. In order to cross-check our sampling results, we have compared the marginal posteriors for the cosmological parameters, the regression coefficients and the population variances obtained with Gibbs-type samplers with those obtained from a pure MH algorithm. The MH algorithm has been used to sample from a marginal posterior with latent variables, $\dmis$, and population mean parameters, $\dmean$, integrated out analytically, akin to what was done in~\cite{2011MNRAS.418.2308M}. 

In our Gibbs-type samplers, we make use of PCG to improve convergence. As detailed below, this involves sampling from conditional distributions of the marginal posterior distribution, 
{\small
\begin{eqnarray} 
\label{eq:margpost}
p(\sigmis,\ \regcoeff, \ \Cparams  | \ \saltdata) &\propto&
{{|\sigobs \sigmis \sigprior|}^{-\half} {|\sigA \sigK|}^\half \over \Rc^2 \ \Rx^2 } p(\sigmares^2)
\label{eq:marginal} \\
&\times &\E \left\{
 - \half \Big[{\dobs}^T \sigobs^{-1} \dobs
 -\sigdelta^T \sigA \sigdelta 
 - \kvec^T \sigK^{-1} \kvec
+ \dprior^T \sigprior^{-1} \dprior\Big]
\right\}, \nonumber
\end{eqnarray}
}
with notation given in Table~\ref{table:list_symbol}. The corresponding marginal posterior distributions for the \HC\ and \SC\ models are obtained using the substitutions in Eq.~\eqref{eq:hcsub} and \eqref{eq:scsub}, respectively.

This section consists of details of sampling steps of these algorithms. 

\subsection{Gibbs-type samplers}
\label{apsubsec:gibbs}
We start with Gibbs-type samplers and consider both the Baseline model and all its extensions discussed in Sections~\ref{subsec:model} and~\ref{subsec:extend}.

\subsubsection{Baseline model}
\label{apsubsubsec:base}
As stated in Table~\ref{table:list_symbol}, in the Baseline model, $\dobs$ is the distance modulus corrected version of $\saltdata$, that is, 
\be
\dobs=\{\dobs_1^T,\dots,\dobs_n^T\}^T, 
\ee
where $\dobs_i=\{\mBhat{i}-\mu_{i}(\zhat_i, \Cparams), \xhat{i}, \chat_i\}^T$. Moreover, $\dmis=\{\dmis_1^T,\dots,\dmis_n^T\}^T$, where $\dmis_i=\{M_i\ep, \covariates_i^T\}^T$ with $\covariates_i=\{\xone{i}, \tc\}^T$; $\dmean=\{\Mnot\ep, \xstar, \cstar\}^T$; $\dprior=\{-19.3, 0, 0\}^T$. For the variance-covariance matrices, $\sigobs=C_{\rm stat} +C_{\rm syst}$; $\sigmis=\diag(S_1, \dots, S_n)$, where each $S_i=\diag(\sigmares^2, \Rx^2, \Rc^2)$; $\sigprior=\diag(2^2, 10^2, 1^2)$. In addition, $\Jmat_{(3n\times 3)}={\left[ \begin{array}{c} J_1\\ \vdots\\ J_n \end{array} \right]}$, where each $J_i$ is a $(3\times 3)$ identity matrix, that is, $J_i=\diag(1, 1, 1)$; $\regcoeff=\{-\alpha, \beta\}^T$ and $\Amat_\dimmat=\diag(T_1,\dots,T_n)$, where
each $T_i=\left[
\begin{array}{ccc}
1 & -\alpha & \beta\\
0 & 1 & 0\\
0 & 0 & 1
\end{array}
\right]$.  

\paragraph{The sampler for the baseline model:}
This is an MH within PCG sampler, that is, we integrate $(\dmis,\dmean)$ out when updating $\Cparams$ and $\regcoeff$. Then the sampling of $\Cparams$ and $\regcoeff$ needs the help of the MH algorithm. While using MH in a Gibbs sampler is a standard strategy, embedding MH into a PCG sampler involves more subtleties. We follow exactly the procedure provided by \citet{vandyk:jiao:15} when deriving an MH within PCG sampler. The steps of the sampler are listed below. We use a prime to indicate the current iteration of a parameter, and $\tk$ to represent the transition function introduced by the MH algorithm.
\begin{description}
\itemsep=0in
\item[Step~1:] $\Cparams\sim\tk(\Cparams|\saltdata,\ \sigmis^\prime,\ \regcoeff^\prime)$:\par
Use MH to sample $\Cparams$ from $p(\Cparams|\saltdata,\ \sigmis^\prime,\ \regcoeff^\prime)$, which is proportional to $p(\sigmis^\prime,\ \regcoeff^\prime, \ \Cparams | \ \saltdata)$, under the constraint imposed by the priors\footnote{The proportionally is a consequence of the relationship: $p(X|Y) = P(X,Y)/P(Y) \propto P(X,Y)$, as a function of $X$.};

\item[Step~2:] $\regcoeff\sim\tk(\regcoeff|\saltdata,\ \sigmis^\prime,\ \Cparams)$:\par
Use MH to sample $\regcoeff$ from $p(\regcoeff|\saltdata,\ \sigmis^\prime,\ \Cparams)$, which is proportional to $p(\sigmis^\prime,\ \regcoeff, \ \Cparams | \ \saltdata)$, under the constraint imposed by the priors;

\item[Step~3:] $(\dmis,\dmean)\sim p(\dmis,\dmean|\saltdata,\ \sigmis^\prime,\ \regcoeff,\ \Cparams)$:\par
This step consists of two sub-steps:
\begin{itemize}
\itemsep=0in
\item Sample $\dmean$ from $\N(\kvec,  \sigK)$, where $\kvec$ and $\sigK$ are defined in Table~\ref{table:list_symbol};
\item Sample $\dmis$ from $\N(\mu_A,\sigA)$, where $\sigA$ is defined in Table~\ref{table:list_symbol} and $\mu_A=\sigA(\sigdelta+\sigmis^{-1}\Jmat\dmean)$;
\end{itemize}

\item[Step~4:] $\sigmares\sim p(\sigmares|\saltdata,\ \dmis,\ \dmean,\ \Rx^\prime,\ \Rc^\prime,\ \regcoeff,\ \Cparams)$:\par
Sample $\sigmares^2$ from 
$\invgamma\left[\frac{n}{2}+\gapara, \frac{\sum_{i=1}^n {(M_i\ep-\Mnot\ep)}^2}{2}+\gapara\right]$, and $\sigmares=\sqrt{\sigmares^2}$;

\item[Step~5:] $\Rx\sim p(\Rx|\saltdata,\ \dmis,\ \dmean,\ \sigmares,\ \Rc^\prime,\ \regcoeff,\ \Cparams)$:\par
Sample $\Rx^2$ from $\invgamma\left[\frac{n}{2},\frac{\sum_{i=1}^{n}{(\xone{i}-\xstar)}^2}{2}\right]$ with $\logr(\Rx)\in [-5,2]$, and $\Rx=\sqrt{\Rx^2}$;

\item[Step~6:] $\Rc\sim p(\Rc|\saltdata,\ \dmis,\ \dmean,\ \sigmares,\ \Rx,\ \regcoeff,\ \Cparams)$:\par
Sample $\Rc^2$ from $\invgamma\left[\frac{n}{2},\frac{\sum_{i=1}^{n}{(\tc-\cstar)}^2}{2}\right]$ with $\logr(\Rc)\in [-5,2]$, and $\Rc=\sqrt{\Rc^2}$.
\end{description}

\subsubsection{\betalin\ model}
\label{apsubsubsec:blin}

In the \betalin\ model, the specification of $\dobs$, $\dmean$, $\dprior$, $\sigobs$, $\sigmis$, $\sigprior$ and $\Jmat$ is identical to that in the Baseline model.  As above, $\dmis= \{\dmis_1^T,\dots,\dmis_n^T\}^T$, where $\dmis_i=\{M_i\ep, \covariates_i^T\}^T$, but under this model, $\covariates_i=\{\xone{i}, \tc, \zhat_i\tc\}^T$. In addition, $\regcoeff=\{-\alpha, \betanot, \beta_1\}^T$; $\Amat_\dimmat=\diag(T_1,\dots,T_n)$, where $T_i=\left[
\begin{array}{ccc}
1 & -\alpha & \betanot+\beta_1 \zhat_i\\
0 & 1 & 0\\
0 & 0 & 1
\end{array}
\right]$.  

\paragraph{The sampler for the \betalin\ model.}
In this sampler, we combine ASIS and MH within PCG algorithms. We integrate $(\dmis,\dmean)$ out when updating $\Cparams$, and use the ASIS algorithm to update $\regcoeff$. The distribution of $\dmis$ conditioning on $\regcoeff$ and other parameters is 
\be
\dmis|\dmean,\sigmis,\regcoeff,\Cparams\sim \N(\Jmat \dmean, \sigmis).
\label{eq:aadist}
\ee
Because this distribution is free of $\regcoeff$, $\dmis$ is an ancillary augmentation for $\regcoeff$ conditioning on other parameters. To derive a sufficient augmentation, we set $\tdmis=\Amat \dmis$. The distribution of $\dobs$ conditioning on $\tdmis$, $\regcoeff$, and other parameters is 
\be
\dobs|\tdmis,\dmean,\sigmis,\regcoeff,\Cparams\sim \N(\tdmis, \sigobs).
\label{eq:sadist}
\ee
Because this distribution is free of $\regcoeff$, $\tdmis$ is the corresponding sufficient augmentation for $\regcoeff$.

We use ``$I$'' in the superscript to indicate intermediate draws that are not part of the final output. The steps of the sampler for the \betalin\ model are: 

\begin{description}
\itemsep=0in
\item[Step~1:] $\Cparams\sim\tk(\Cparams|\saltdata,\ \sigmis^\prime,\ \regcoeff^\prime)$:\par
Use MH to sample $\Cparams$ from $p(\Cparams|\saltdata,\ \sigmis^\prime,\ \regcoeff^\prime)$, which is proportional to $p(\sigmis^\prime,\ \regcoeff^\prime, \ \Cparams | \ \saltdata)$, under the constraint imposed by the priors;

\item[Step~2:] $(\dmis^I,\dmean)\sim p(\dmis,\dmean|\saltdata,\ \sigmis^\prime,\ \regcoeff^\prime,\ \Cparams)$:\par
This step consists of two sub-steps:
\begin{itemize}
\itemsep=0in
\item Sample $\dmean$ from $\N(\kvec,  \sigK)$, where $\kvec$ and $\sigK$ are defined in Table~\ref{table:list_symbol};
\item Sample $\dmis^I$ from $\N(\mu_A,\sigA)$, where $\sigA$ is defined in Table~\ref{table:list_symbol} and $\mu_A=\sigA(\sigdelta+\sigmis^{-1}J\dmean)$;
\end{itemize}

\item[Step~3:] $\regcoeff^I\sim p(\regcoeff|\saltdata,\ \dmis^I,\ \dmean,
\ \sigmis^\prime,\ \Cparams)$:\par
Sample $\regcoeff^I$ from $\N(\zB,\sigmaB)$ (details about this distribution are given below) with constraint 
$\regcoeff^I\in [-1,0]\times [0,4]\times [-4,4]$;\par
Use $\regcoeff^I$ to construct $\Amat^I$. Then set $\tdmis=\Amat^I \dmis^I$;

\item[Step~4:] $\regcoeff\sim p(\regcoeff|\saltdata,\ \tdmis,\ \dmean,
\ \sigmis^\prime,\ \Cparams)$;\par
Sample $\regcoeff$ from $\N(\tzB,\tsigmaB)$ (details about this distribution are given below) with constraint 
$\regcoeff\in [-1,0]\times [0,4]\times [-4,4]$;\par
Use $\regcoeff$ to construct $\Amat$. Then set $\dmis=\Amat^{-1}\tdmis$;

\item[Step~5:] $\sigmares\sim p(\sigmares|\saltdata,\ \dmis,\ \dmean,\ \Rx^\prime,\ \Rc^\prime,\ \regcoeff,\ \Cparams)$:\par
Sample $\sigmares^2$ from 
$\invgamma\left[\frac{n}{2}+\gapara, \frac{\sum_{i=1}^n {(M_i\ep-\Mnot\ep)}^2}{2}+\gapara\right]$, and $\sigmares=\sqrt{\sigmares^2}$;

\item[Step~6:] $\Rx\sim p(\Rx|\saltdata,\ \dmis,\ \dmean,\ \sigmares,\ \Rc^\prime,\ \regcoeff,\ \Cparams)$:\par
Sample $\Rx^2$ from $\invgamma\left[\frac{n}{2},\frac{\sum_{i=1}^{n}{(\xone{i}-\xstar)}^2}{2}\right]$ with $\logr(\Rx)\in [-5,2]$, and $\Rx=\sqrt{\Rx^2}$;

\item[Step~7:] $\Rc\sim p(\Rc|\saltdata,\ \dmis,\ \dmean,\ \sigmares,\ \Rx,\ \regcoeff,\ \Cparams)$:\par
Sample $\Rc^2$ from $\invgamma\left[\frac{n}{2},\frac{\sum_{i=1}^{n}{(\tc-\cstar)}^2}{2}\right]$ with $\logr(\Rc)\in [-5,2]$, and $\Rc=\sqrt{\Rc^2}$.
\end{description}

In Step~3, $\sigmaB^{-1}=\xmat^T \vone^{-1} \xmat$, where $\vone$ is the $(n \times n)$ submatrix of $\sigobs$ after deleting the ${(3i-1)}^\rth$ ($i=1,\dots,n$) and ${(3i)}^\rth$ ($i=1,\dots,n$) rows and columns, and 
$\xmat_{(n\times 3)}={\left[ \begin{array}{c} \covariates_1^T\\ \vdots\\ \covariates_n^T \end{array} \right]}$. Furthermore, 
$\zB=\sigmaB \xmat^T \vone^{-1} (\pxobs-\pxmis-\delxi)$, where $\pxobs=\{\mBhat{1}-\mu_{1}(\zhat_1, \Cparams),\dots,\mBhat{n}-\mu_{n}(\zhat_i, \Cparams)\}^T$, $\pxmis=\{M_1\ep,\dots,M_n\ep\}^T$, and $\delxi=\pvone\rvone^{-1}(\rxobs-\rxmis)$; $\rvone$ is the $(2n \times 2n)$ submatrix of $\sigobs$ after deleting the ${(3i-2)}^\rth$ ($i=1,\dots,n$) rows and columns; $\pvone$ is the $(n \times 2n)$ submatrix of $\sigobs$ after deleting the ${(3i-1)}^\rth$ ($i=1,\dots,n$) and ${(3i)}^\rth$ ($i=1,\dots,n$) rows and the ${(3i-2)}^\rth$ ($i=1,\dots,n$)  columns; $\rxobs=\{\xhat{1},\chat_1,\dots,\xhat{n},\chat_n\}^T$; $\rxmis=\{\xone{1}, c_1,\dots,\xone{n}, c_n\}^T$.

In Step~4, $\tsigmaB^{-1}=(\txmat^T \txmat)/\sigmares^{2^\prime}$, where
$\txmat_{(n\times 3)}={\left[ \begin{array}{c} \txmat_1^T\\ \vdots\\ \txmat_n^T \end{array} \right]}$ with $\txmat_i=\{-\txone{i}, -\ttc, -\zhat_i \ttc\}^T$; 
$\txone{i}$ and $\ttc$ are the ${(3i-1)}^\rth$ and ${(3i)}^\rth$ components of $\tdmis$ respectively. Furthermore, 
$\tzB=\tsigmaB [\txmat^T (\ximnot-\tpxmis)/\sigmares^{2^\prime}]$, where $\ximnot=\{\underbrace{\Mnot\ep,\dots,\Mnot\ep}_n\}^T$ and $\tpxmis=\{\ttM_1\ep,\dots,\ttM_n\ep\}^T$; $\ttM_i\ep$ is the ${(3i-2)}^\rth$ component of $\tdmis$.

\subsubsection{\betastep\ model}
\label{apsubsubsec:btrans}

In the \betastep\ model, the specification of $\dobs$, $\dmean$, $\dprior$, $\sigobs$, $\sigmis$, $\sigprior$ and $\Jmat$ is identical to that in the Baseline model. As above, $\dmis=\{\dmis_1^T,\dots,\dmis_n^T\}^T$, where $\dmis_i=\{M_i\ep, \covariates_i^T\}^T$, but under this model, 
$\covariates_i=\covariates_i(\ztrans)=\left\{\xone{i}, \tc, 
\left({1\over 2} + \frac{1}{\pi} \arctan\left(\frac{\zhat_i-\ztrans}{0.01}\right)\right) \tc \right\}^T$. 
In addition, $\regcoeff=\{-\alpha, \betanot, \Delta\beta\}^T$; $\Amat_\dimmat=\diag(T_1,\dots,T_n)$, where $T_i=\left[
\begin{array}{ccc}
1 & -\alpha & \betanot+\Delta\beta\left({1\over 2} + \frac{1}{\pi} \arctan\left(\frac{\zhat_i-\ztrans}{0.01}\right)\right)\\
0 & 1 & 0\\
0 & 0 & 1
\end{array}
\right]$. 

Because we have an additional unknown parameter, $\ztrans$, under this model, the complete and marginal posterior distributions should be written as $p(\dmis,\  \dmean,\  \sigmis,\ \regcoeff,\ \ztrans, \ \Cparams  | \ \saltdata)$ and $p(\sigmis,\ \regcoeff, \ \ztrans, \ \Cparams  | \ \saltdata)$ respectively, although they are formally identical to (\ref{eq:posterior}) and (\ref{eq:marginal}), respectively.

\paragraph{The sampler for the \betastep\ model.}
As in the sampler for the \betalin\ model, we also combine ASIS and MH within PCG algorithms in this sampler. We integrate $(\dmis,\dmean)$ out when updating both $\Cparams$ and $\ztrans$, and use the ASIS algorithm to update $\regcoeff$. When implementing ASIS, we also regard $\dmis$ as the ancillary augmentation, and $\tdmis=\Amat \dmis$ as the corresponding sufficient augmentation for $\regcoeff$, conditioning on other parameters.

The steps of the sampler for the \betastep\ model are: 

\begin{description}
\itemsep=0in
\item[Step~1:] $\Cparams\sim\tk(\Cparams|\saltdata,\ \sigmis^\prime,\ \regcoeff^\prime,\ \ztrans^\prime)$:\par
Use MH to sample $\Cparams$ from $p(\Cparams|\saltdata,\ \sigmis^\prime,\ \regcoeff^\prime,\ \ztrans^\prime)$, which is proportional to $p(\sigmis^\prime,\ \regcoeff^\prime, \ \ztrans^\prime, \ \Cparams | \ \saltdata)$, under the constraint imposed by the priors;

\item[Step~2:] $\ztrans\sim\tk(\ztrans|\saltdata,\ \sigmis^\prime,\ \regcoeff^\prime,\ \Cparams)$:\par
Use MH to sample $\ztrans$ from $p(\ztrans|\saltdata,\ \sigmis^\prime,\ \regcoeff^\prime,\ \Cparams)$, which is proportional to $p(\sigmis^\prime,\ \regcoeff^\prime, \ \ztrans, \ \Cparams  | \ \saltdata)$, under the constraint $\ztrans\in [0.2,1]$; 

\item[Step~3:] $(\dmis^I,\dmean)\sim p(\dmis,\dmean|\saltdata,\ \sigmis^\prime,\ \regcoeff^\prime,\ \ztrans,\ \Cparams)$:\par
This step consists of two sub-steps:
\begin{itemize}
\itemsep=0in
\item Sample $\dmean$ from $\N(\kvec,  \sigK)$, where $\kvec$ and $\sigK$ are defined in Table~\ref{table:list_symbol};
\item Sample $\dmis^I$ from $\N(\mu_A,\sigA)$, where $\sigA$ is defined in Table~\ref{table:list_symbol} and $\mu_A=\sigA(\sigdelta+\sigmis^{-1}J\dmean)$;
\end{itemize}

\item[Step~4:] $\regcoeff^I\sim p(\regcoeff|\saltdata,\ \dmis^I,\ \dmean,
\ \sigmis^\prime,\ \ztrans,\ \Cparams)$:\par
Sample $\regcoeff^I$ from $\N(\zB,\sigmaB)$ with constraint 
$\regcoeff^I\in [-1,0]\times [0,4]\times [-1.5,1.5]$. The construction of $\zB$ and $\sigmaB$ is identical to that in the \betalin\ sampler;\par
Use $\regcoeff^I$ and $\ztrans$ to construct $\Amat^I$. Then set $\tdmis=\Amat^I \dmis^I$;

\item[Step~5:] $\regcoeff\sim p(\regcoeff|\saltdata,\ \tdmis,\ \dmean,
\ \sigmis^\prime,\ \ztrans,\ \Cparams)$;\par
Sample $\regcoeff$ from $\N(\tzB,\tsigmaB)$ with constraint 
$\regcoeff\in [-1,0]\times [0,4]\times [-1.5,1.5]$. The construction of $\tzB$ and $\tsigmaB$ is identical to that in the \betalin\ sampler, except that under this model, 
$\txmat_i=\left\{-\txone{i}, -\ttc, - \left({1\over 2} + \frac{1}{\pi} \arctan\left(\frac{\zhat_i-\ztrans}{0.01}\right)\right)\ttc\right\}^T$;\par
Use $\regcoeff$ and $\ztrans$ to construct $\Amat$. Then set $\dmis=\Amat^{-1}\tdmis$;

\item[Step~6:] $\sigmares\sim p(\sigmares|\saltdata,\ \dmis,\ \dmean,\ \Rx^\prime,\ \Rc^\prime,\ \regcoeff,\ \ztrans,\ \Cparams)$:\par
Sample $\sigmares^2$ from 
$\invgamma\left[\frac{n}{2}+\gapara, \frac{\sum_{i=1}^n {(M_i\ep-\Mnot\ep)}^2}{2}+\gapara\right]$, and $\sigmares=\sqrt{\sigmares^2}$;

\item[Step~7:] $\Rx\sim p(\Rx|\saltdata,\ \dmis,\ \dmean,\ \sigmares,\ \Rc^\prime,\ \regcoeff,\ \ztrans,\ \Cparams)$:\par
Sample $\Rx^2$ from $\invgamma\left[\frac{n}{2},\frac{\sum_{i=1}^{n}{(\xone{i}-\xstar)}^2}{2}\right]$ with $\logr(\Rx)\in [-5,2]$, and $\Rx=\sqrt{\Rx^2}$;

\item[Step~8:] $\Rc\sim p(\Rc|\saltdata,\ \dmis,\ \dmean,\ \sigmares,\ \Rx,\ \regcoeff,\ \ztrans,\ \Cparams)$:\par
Sample $\Rc^2$ from $\invgamma\left[\frac{n}{2},\frac{\sum_{i=1}^{n}{(\tc-\cstar)}^2}{2}\right]$ with $\logr(\Rc)\in [-5,2]$, and $\Rc=\sqrt{\Rc^2}$.
\end{description}

\subsubsection{\HC\  of host galaxy mass model}
\label{apsubsubsec:massstep}
In this model, we divide the SNIa population into two classes according to host galaxy mass. The specification of $\dobs$, $\dmis$, $\sigobs$ and $\Amat$ is identical to that in the Baseline model. However, the specification of $\dmean$, $\sigmis$, $\sigprior$ and $\Jmat$ is changed to reflect the existence of two host galaxy mass populations. Under this model, $\dmean=\{(\Mnot)_\text{low}, (\Mnot)_\text{high}, \xstar, \cstar\}^T$; $\dprior=\{-19,3, -19.3, 0, 0\}^T$; $\sigmis=\diag(S_1, \dots, S_n)$, where $S_i=\diag[Z_i(\sigmares)_\text{low}^2+(1-Z_i)(\sigmares)_\text{high}^2, \Rx^2, \Rc^2]$; $\sigprior=\diag(2^2, 2^2, 10^2, 1^2)$; $\Jmat_{(3n\times 4)}={\left[ \begin{array}{c} J_1\\ \vdots\\ J_n \end{array} \right]}$, where $J_i=\left[
\begin{array}{cccc}
Z_i & 1-Z_i & 0 & 0\\
0 & 0 & 1 & 0\\
0 & 0 & 0 & 1
\end{array}
\right]$. 
As stated in Section~\ref{apsec:pos}, under this model, $Z=\{Z_1,\dots,Z_n\}$ is assumed known with, 
\be 
Z_i=
\left\{\begin{array}{l}
1\ \text{  if }\Mgalobs<10\\
0\ \text{  otherwise.}
\end{array}\right. 
\ee

\paragraph{The sampler for the \HC\  model.}
This is also an MH within PCG sampler, that is, we integrate $(\dmis,\dmean)$ out when updating $\Cparams$ and $\regcoeff$. The steps of the sampler are listed as follows. 
\begin{description}
\itemsep=0in
\item[Step~1:] $\Cparams\sim\tk(\Cparams|\saltdata, \ \sigmis^\prime,\ \regcoeff^\prime)$:\par
Use MH to sample $\Cparams$ from $p(\Cparams|\saltdata, \ \sigmis^\prime,\ \regcoeff^\prime)$, which is proportional to $p(\sigmis^\prime,\ \regcoeff^\prime, \ \Cparams| \ \saltdata)$, under the constraint imposed by the priors;

\item[Step~2:] $\regcoeff\sim\tk(\regcoeff|\saltdata, \ \sigmis^\prime,\ \Cparams)$:\par
Use MH to sample $\regcoeff$ from $p(\regcoeff|\saltdata, \ \sigmis^\prime,\ \Cparams)$, which is proportional to $p(\sigmis^\prime,\ \regcoeff, \ \Cparams| \ \saltdata)$, under the constraint imposed by the priors;

\item[Step~3:] $(\dmis,\dmean)\sim p(\dmis,\dmean|\saltdata, \ \sigmis^\prime,\ \regcoeff,\ \Cparams)$:\par
This step consists of two sub-steps:
\begin{itemize}
\itemsep=0in
\item Sample $\dmean$ from $\N(\kvec,  \sigK)$, where $\kvec$ and $\sigK$ are defined in Table~\ref{table:list_symbol};
\item Sample $\dmis$ from $\N(\mu_A,\sigA)$, where $\sigA$ is defined in Table~\ref{table:list_symbol} and $\mu_A=\sigA(\sigdelta+\sigmis^{-1}J\dmean)$;
\end{itemize}

\item[Step~4:] $(\sigmares)_\text{low}\sim p((\sigmares)_\text{low}|\saltdata, \ \dmis,\ \dmean,\ (\sigmares^\prime)_\text{high}, \ \Rx^\prime,\ \Rc^\prime,\ \regcoeff,\ \Cparams)$:\par
Sample $(\sigmares)_\text{low}^2$ from 
$\invgamma\left[\frac{\sum_{i=1}^n Z_i}{2}+\gapara, \frac{\sum_{i=1}^n {Z_i(M_i\ep-(\Mnot)_\text{low})}^2}{2}+\gapara\right]$, and $(\sigmares)_\text{low}=\sqrt{(\sigmares)_\text{low}^2}$;

\item[Step~5:] $(\sigmares)_\text{high}\sim p((\sigmares)_\text{high}|\saltdata, \ \dmis,\ \dmean,\ (\sigmares)_\text{low}, \ \Rx^\prime,\ \Rc^\prime,\ \regcoeff,\ \Cparams)$:\par
Sample $(\sigmares)_\text{high}^2$ from 
$\invgamma\left[\frac{\sum_{i=1}^n (1-Z_i)}{2}+\gapara, \frac{\sum_{i=1}^n {(1-Z_i)(M_i\ep-(\Mnot)_\text{high})}^2}{2}+\gapara\right]$, and $(\sigmares)_\text{high}=\sqrt{(\sigmares)_\text{high}^2}$;

\item[Step~6:] $\Rx\sim p(\Rx|\saltdata, \ \dmis,\ \dmean,\ (\sigmares)_\text{low}, \ (\sigmares)_\text{high},\ \Rc^\prime,\ \regcoeff,\ \Cparams)$:\par
Sample $\Rx^2$ from $\invgamma\left[\frac{n}{2},\frac{\sum_{i=1}^{n}{(\xone{i}-\xstar)}^2}{2}\right]$ with $\logr(\Rx)\in [-5,2]$, and $\Rx=\sqrt{\Rx^2}$;

\item[Step~7:] $\Rc\sim p(\Rc|\saltdata, \ \dmis,\ \dmean,\ (\sigmares)_\text{low}, \ (\sigmares)_\text{high},\ \Rx,\ \regcoeff,\ \Cparams)$:\par
Sample $\Rc^2$ from $\invgamma\left[\frac{n}{2},\frac{\sum_{i=1}^{n}{(\tc-\cstar)}^2}{2}\right]$ with $\logr(\Rc)\in [-5,2]$, and $\Rc=\sqrt{\Rc^2}$.
\end{description}

\subsubsection{\SC\  of host galaxy mass model}
\label{apsubsubsec:prob}
In this model, the specification of $\dobs$, $\dmis$, $\sigobs$, $\dmean$, $\sigmis$, $\sigprior$, $\Jmat$ and $\Amat$ is identical to that in the \HC\ model. But here $\Jmat$ is stochastic, since $Z$ is stochastic. 

\paragraph{The sampler for the \SC\  model.}
This is also an MH within PCG sampler, that is, we integrate $(\dmis,\dmean)$ out when updating $\Cparams$ and $\regcoeff$. The steps of the sampler are:
\begin{description}
\itemsep=0in
\item[Step~1:] $Z\sim p(Z|\saltdata, \ \Mgaldata,\ \dmis^\prime,\ \dmean^\prime,\ \sigmis^\prime,\ \regcoeff^\prime,\ \Cparams^\prime)$:

For each $i$, sample $Z_i$ from ${\rm Bernoulli}(\tilde{p}_i)$, where $\tilde{p}_i=\frac{p_{i, {\rm low}}}{p_{i, {\rm low}}+p_{i, {\rm high}}}$, with
\be
p_{i, {\rm low}}=
\frac{1}{(\sigmares^\prime)_\text{low}}
\E \left\{-\frac{{[{(M_i\ep)}^\prime-(\Mnot^\prime)_\text{low}]}^2}{2 (\sigmares^\prime)_\text{low}^2}\right\} \pi_i
\ee
and 
\be
p_{i, {\rm high}}=
\frac{1}{(\sigmares^\prime)_\text{high}}
\E \left\{-\frac{{[{(M_i\ep)}^\prime-(\Mnot^\prime)_\text{high}]}^2}{2 (\sigmares^\prime)_\text{high}^2}\right\} (1-\pi_i);
\ee
$\pi_i$ is defined in Section~\ref{apsec:pos};\par
Use $Z$ to construct $\Jmat$, as in Table~\ref{table:list_symbol};

\item[Step~2:] $\Cparams\sim\tk(\Cparams|\saltdata, \ \Mgaldata,\ \sigmis^\prime,\ \regcoeff^\prime, \ Z)$:\par
Use MH to sample $\Cparams$ from $p(\Cparams|\saltdata, \ \Mgaldata,\ \sigmis^\prime,\ \regcoeff^\prime, \ Z)$, which is proportional to $p(\sigmis^\prime,\ \regcoeff^\prime, \ \Cparams, \ Z| \ \saltdata, \ \Mgaldata)$, under the constraint imposed by the priors;

\item[Step~3:] $\regcoeff\sim\tk(\regcoeff|\saltdata, \ \Mgaldata,\ \sigmis^\prime,\ \Cparams, \ Z)$:\par
Use MH to sample $\regcoeff$ from $p(\regcoeff|\saltdata, \ \Mgaldata,\ \sigmis^\prime,\ \Cparams, \ Z)$, which is proportional to $p(\sigmis^\prime,\ \regcoeff, \ \Cparams, \ Z| \ \saltdata, \ \Mgaldata)$, under the constraint imposed by the priors;

\item[Step~4:] $(\dmis,\dmean)\sim p(\dmis,\dmean|\saltdata, \ \Mgaldata,\ \sigmis^\prime,\ \regcoeff,\ \Cparams, \ Z)$:\par
This step consists of two sub-steps:
\begin{itemize}
\itemsep=0in
\item Sample $\dmean$ from $\N(\kvec,  \sigK)$, where $\kvec$ and $\sigK$ are defined in Table~\ref{table:list_symbol};
\item Sample $\dmis$ from $\N(\mu_A,\sigA)$, where $\sigA$ is defined in Table~\ref{table:list_symbol} and $\mu_A=\sigA(\sigdelta+\sigmis^{-1}J\dmean)$;
\end{itemize}

\item[Step~5:] $(\sigmares)_\text{low}\sim p((\sigmares)_\text{low}|\saltdata, \ \Mgaldata,\ \dmis,\ \dmean,\ (\sigmares^\prime)_\text{high}, \ \Rx^\prime,\ \Rc^\prime,\ \regcoeff,\ \Cparams, \ Z)$:\par
Sample $(\sigmares)_\text{low}^2$ from 
$\invgamma\left[\frac{\sum_{i=1}^n Z_i}{2}+\gapara, \frac{\sum_{i=1}^n {Z_i(M_i\ep-(\Mnot)_\text{low})}^2}{2}+\gapara\right]$, and $(\sigmares)_\text{low}=\sqrt{(\sigmares)_\text{low}^2}$;

\item[Step~6:] $(\sigmares)_\text{high}\sim p((\sigmares)_\text{high}|\saltdata, \ \Mgaldata,\ \dmis,\ \dmean,\ (\sigmares)_\text{low}, \ \Rx^\prime,\ \Rc^\prime,\ \regcoeff,\ \Cparams, \ Z)$:\par
Sample $(\sigmares)_\text{high}^2$ from 
$\invgamma\left[\frac{\sum_{i=1}^n (1-Z_i)}{2}+\gapara, \frac{\sum_{i=1}^n {(1-Z_i)(M_i\ep-(\Mnot)_\text{high})}^2}{2}+\gapara\right]$, and $(\sigmares)_\text{high}=\sqrt{(\sigmares)_\text{high}^2}$;

\item[Step~7:] $\Rx\sim p(\Rx|\saltdata, \ \Mgaldata,\ \dmis,\ \dmean,\ (\sigmares)_\text{low}, \ (\sigmares)_\text{high},\ \Rc^\prime,\ \regcoeff,\ \Cparams, \ Z)$:\par
Sample $\Rx^2$ from $\invgamma\left[\frac{n}{2},\frac{\sum_{i=1}^{n}{(\xone{i}-\xstar)}^2}{2}\right]$ with $\logr(\Rx)\in [-5,2]$, and $\Rx=\sqrt{\Rx^2}$;

\item[Step~8:] $\Rc\sim p(\Rc|\saltdata, \ \Mgaldata,\ \dmis,\ \dmean,\ (\sigmares)_\text{low}, \ (\sigmares)_\text{high},\ \Rx,\ \regcoeff,\ \Cparams, \ Z)$:\par
Sample $\Rc^2$ from $\invgamma\left[\frac{n}{2},\frac{\sum_{i=1}^{n}{(\tc-\cstar)}^2}{2}\right]$ with $\logr(\Rc)\in [-5,2]$, and $\Rc=\sqrt{\Rc^2}$.
\end{description}

\subsubsection{\CA\  of host galaxy mass model}
\label{apsubsubsec:cov}

In this model, since we include $\Mgal$ as an additional covariate, the specification of quantities in the posterior distribution is different from the Baseline model. First, $\dobs$ is the combination of the distance modulus-corrected $\saltdata$ and the host galaxy mass, $\Mgaldata$, that is, $\dobs=\{\dobs_1^T,\dots,\dobs_n^T\}^T$, where $\dobs_i=\{\mBhat{i}-\mu_{i}(\zhat_i, \Cparams), \xhat{i}, \chat_i, \Mgalobs\}^T$. Moreover, $\dmis=\{\dmis_1^T,\dots,\dmis_n^T\}^T$, where $\dmis_i=\{M_i\ep, \covariates_i^T\}^T$ with $\covariates_i=\{\xone{i}, \tc, \Mgal\}^T$; $\dmean=\{\Mnot\ep, \xstar, \cstar, \Mgalstar\}^T$; $\dprior=\{-19.3, 0, 0, 10\}^T$. For the variance-covariance matrices, $\sigobs$ now has the dimension of $(4n \times 4n)$. The $(3n \times 3n)$ submatrix of $\sigobs$, after deleting the ${(4i)}^\rth$ ($i=1,\dots,n$) rows and columns, is $(C_{\rm stat} +C_{\rm syst})$. The ${(4i,4i)}^\rth$ element of $\sigobs$ is $\sigMgal^2$, while the other elements in the ${(4i)}^\rth$ rows and columns are all zero, because we ignore correlations between $\Mgalobs$ and other observed quantites; $\sigmis=\diag(S_1, \dots, S_n)$, where each $S_i=\diag(\sigmares^2, \Rx^2, \Rc^2, \Rgal^2)$; $\sigprior=\diag(2^2, 10^2, 1^2, 100^2)$. In addition, $\Jmat_{(4n\times 4)}={\left[ \begin{array}{c} J_1\\ \vdots\\ J_n \end{array} \right]}$, where each $J_i$ is a $(4\times 4)$ identity matrix; $\regcoeff=\{-\alpha, \beta, \gamma\}^T$ and $\Amat_\dimmat=\diag(T_1,\dots, T_n)$, where
each $T_i=\left[
\begin{array}{cccc}
1 & -\alpha & \beta & \gamma\\
0 & 1 & 0 & 0\\
0 & 0 & 1 & 0\\
0 & 0 & 0 & 1\\
\end{array}
\right]$.  

Because we include host galaxy mass data, $\Mgaldata$, under this model, the complete and marginal posterior distributions should be written as $p(\dmis,\  \dmean,\  \sigmis,\ \regcoeff, \ \Cparams  | \ \saltdata, \ \Mgaldata)$ and $p(\sigmis^\prime,\ \regcoeff^\prime, \ \Cparams  | \ \saltdata, \ \Mgaldata)$ respectively. But they are formally identical to (\ref{eq:posterior}) and (\ref{eq:marginal}), respectively.

\paragraph{The sampler for the \CA\ model.}
This is also an MH within PCG sampler, that is, we integrate $(\dmis,\dmean)$ out when updating $\Cparams$ and $\regcoeff$. Then the sampling of $\Cparams$ and $\regcoeff$ needs the help of the MH algorithm. The steps of the sampler are listed below.  
\begin{description}
\itemsep=0in
\item[Step~1:] $\Cparams\sim\tk(\Cparams|\saltdata, \ \Mgaldata,\ \sigmis^\prime,\ \regcoeff^\prime)$:\par
Use MH to sample $\Cparams$ from $p(\Cparams|\saltdata, \ \Mgaldata,\ \sigmis^\prime,\ \regcoeff^\prime)$, which is proportional to $p(\sigmis^\prime,\ \regcoeff^\prime, \ \Cparams | \ \saltdata, \ \Mgaldata)$, under the constraint imposed by the priors;

\item[Step~2:] $\regcoeff\sim\tk(\regcoeff|\saltdata, \ \Mgaldata,\ \sigmis^\prime,\ \Cparams)$:\par
Use MH to sample $\regcoeff$ from $p(\regcoeff|\saltdata, \ \Mgaldata,\ \sigmis^\prime,\ \Cparams)$, which is proportional to $p(\sigmis^\prime,\ \regcoeff, \ \Cparams | \ \saltdata, \ \Mgaldata)$, under the constraint imposed by the priors;

\item[Step~3:] $(\dmis,\dmean)\sim p(\dmis,\dmean|\saltdata, \ \Mgaldata,\ \sigmis^\prime,\ \regcoeff,\ \Cparams)$:\par
This step consists of two sub-steps:
\begin{itemize}
\itemsep=0in
\item Sample $\dmean$ from $\N(\kvec,  \sigK)$, where $\kvec$ and $\sigK$ are defined in Table~\ref{table:list_symbol};
\item Sample $\dmis$ from $\N(\mu_A,\sigA)$, where $\sigA$ is defined in Table~\ref{table:list_symbol} and $\mu_A=\sigA(\sigdelta+\sigmis^{-1}J\dmean)$;
\end{itemize}

\item[Step~4:] $\sigmares\sim p(\sigmares|\saltdata, \ \Mgaldata,\ \dmis,\ \dmean,\ \Rx^\prime,\ \Rc^\prime,\ \Rgal^\prime, \ \regcoeff,\ \Cparams)$:\par
Sample $\sigmares^2$ from 
$\invgamma\left[\frac{n}{2}+\gapara, \frac{\sum_{i=1}^n {(M_i\ep-\Mnot\ep)}^2}{2}+\gapara\right]$, and $\sigmares=\sqrt{\sigmares^2}$;

\item[Step~5:] $\Rx\sim p(\Rx|\saltdata, \ \Mgaldata,\ \dmis,\ \dmean,\ \sigmares,\ \Rc^\prime,\ \Rgal^\prime,\ \regcoeff,\ \Cparams)$:\par
Sample $\Rx^2$ from $\invgamma\left[\frac{n}{2},\frac{\sum_{i=1}^{n}{(\xone{i}-\xstar)}^2}{2}\right]$ with $\logr(\Rx)\in [-5,2]$, and $\Rx=\sqrt{\Rx^2}$;

\item[Step~6:] $\Rc\sim p(\Rc|\saltdata, \ \Mgaldata,\ \dmis,\ \dmean,\ \sigmares,\ \Rx,\ \Rgal^\prime,\ \regcoeff,\ \Cparams)$:\par
Sample $\Rc^2$ from $\invgamma\left[\frac{n}{2},\frac{\sum_{i=1}^{n}{(\tc-\cstar)}^2}{2}\right]$ with $\logr(\Rc)\in [-5,2]$, and $\Rc=\sqrt{\Rc^2}$;

\item[Step~7:] $\Rgal\sim p(\Rgal|\saltdata, \ \Mgaldata,\ \dmis,\ \dmean,\ \sigmares,\ \Rx,\ \Rc,\ \regcoeff,\ \Cparams)$:\par
Sample $\Rgal^2$ from $\invgamma\left[\frac{n}{2},\frac{\sum_{i=1}^{n}{(\Mgal-\Mgalstar)}^2}{2}\right]$ with $\logr(\Rgal)\in [-5,2]$, and $\Rgal=\sqrt{\Rgal^2}$.
\end{description}

When MH updates are required in the samplers above, we use truncated normal distributions centered at the current draw with variance-covariance matrix adjusted to obtain an acceptance rate of around $40\%$ (univariate) or $25\%$ (multivariate). Truncations are applied according to prior constraints.

PCG and ASIS show significant power in improving the convergence properties of $\Cparams$ and $\regcoeff$.  Although our PCG and ASIS samplers require 30\%--50\% more CPU time per iteration than our ordinary Gibbs samplers, their correlation lengths are smaller. For example, the effective sample size for the components of $\Cparams$ is 5--6 times larger, and for the components of $\regcoeff$ is 3--4 times larger. See \citet{jiao:etal:15} for more numerical illustrations.

\subsection{Metropolis-Hastings samplers}
\label{apsubsec:mh}


We also use MH algorithm to obtain samples of $\sigmis$, $\regcoeff$, $\Cparams$ and (for the \betastep\ model) $\ztrans$ from their combined posterior distribution under all the models, except the \SC\ one, with the purpose of cross-checking the results obtained under the Gibbs-type samplers described above. The proposal distribution of the MH algorithm is a normal distribution centered at the current draw. For the the variance-covariance matrix of the normal proposal distribution we initially choose a diagonal matrix with randomly chosen entries. We then run a preliminary chain and use it to obtain an estimate of the variance-covariance matrix of the parameters. Finally we replace the variance-covariance matrix in the proposal distribution with this estimate and run the MH sampler to obtain posterior samples (ignoring the initial run when plotting the marginal dostributions).

\bibliographystyle{apj}
\bibliography{stats,SNIa}

 \end{document}